\documentclass[12pt]{article}
\usepackage{amsmath}
\usepackage{graphicx}
\usepackage{enumerate}
\usepackage{natbib}
\usepackage{url} 
\usepackage[dvipsnames]{xcolor}
\usepackage{amsmath,amssymb,amsfonts,amsthm}
\usepackage{helvet}
\usepackage{bm,bbm}
\usepackage{graphicx}
\usepackage{enumerate}
\usepackage{enumitem}
\usepackage{natbib}
\usepackage{lscape}
\usepackage{multirow}
\usepackage{rotating}
\usepackage{url}
\usepackage[dvipsnames]{xcolor}
\usepackage{authblk}
\usepackage{IEEEtrantools}
\usepackage{algpseudocode}
\usepackage{algorithm,setspace}
\usepackage{mathrsfs}
\usepackage[mathscr]{eucal}
\usepackage{changepage}
\usepackage{booktabs}   
\usepackage{subcaption}
\usepackage{multirow}
\usepackage{afterpage}
\usepackage{float}
\usepackage[percent]{overpic}

\newcommand{\Q}{\ensuremath{\mathcal{Q}}}
\newcommand{\Qq}{\ensuremath{\Q_q}}
\newcommand{\hQq}{\ensuremath{\hat{\Q}_q}}

\newcommand{\D}{\ensuremath{\mathcal{D}}}
\newcommand{\G}{\ensuremath{\mathcal{G}}}
\newcommand{\hG}{\ensuremath{\hat{\G}}}
\newcommand{\W}{\ensuremath{\mathcal{W}}}
\newcommand{\hW}{\ensuremath{\hat{\W}}}
\newcommand{\Rset}{\ensuremath{\mathcal{R}}}

\newcommand{\M}{\ensuremath{\mathrm{M}}}

\newcommand{\rinf}[1]{\ensuremath{r_{#1}^{\inf}}}
\newcommand{\rsup}[1]{\ensuremath{r_{#1}^{\sup}}}
\newcommand{\loc}{\ensuremath{\mathcal{M}}}
\newcommand{\B}{\ensuremath{\mathcal{B}}}

\newcommand{\Loss}{\ensuremath{\mathcal{L}}}

\newcommand{\SSS}{\ensuremath{\mathbb{S}}}
\newcommand{\s}{\ensuremath{\mathcal{S}}}
\newcommand{\puncS}{\ensuremath{\SSS^{d-1}\backslash S_d}}
\newcommand{\CCC}{\ensuremath{\mathbb{C}}}
\newcommand{\RR}{\ensuremath{\mathbb{R}}}
\newcommand{\PR}{\ensuremath{\mathbb{P}}}
\newcommand{\dd}{\ensuremath{\mathrm{d}}}

\newcommand{\Exc}{\ensuremath{\mathcal{E}}}

\newcommand{\KLD}{\ensuremath{\mathrm{D}_{\mathrm{KL}}}}

\newcommand*\samethanks[1][\value{footnote}]{\footnotemark[#1]}

\newcommand{\blind}{0}

\begin{document}

\def\spacingset#1{\renewcommand{\baselinestretch}%
{#1}\small\normalsize} \spacingset{1}


\if1\blind
{
  \title{\bf Generative modelling of multivariate geometric extremes using normalising flows}
  \author{Lambert {{D}e {M}onte}\thanks{
    The authors gratefully acknowledge \textit{please remember to list all relevant funding sources in the unblinded version}}\hspace{.2cm}\\
    School of Mathematics and Maxwell Institute for Mathematical Sciences, University of Edinburgh, Edinburgh, EH9 3FD, Scotland\\
    and \\
    Rapha\"el Huser \\
    Statistics Program, CEMSE Division, King Abdullah University of Science and Technology (KAUST), Thuwal 23955-6900, Saudi Arabia\\
    and \\
    Ioannis Papastathopoulos \\
    School of Mathematics and Maxwell Institute for Mathematical Sciences, University of Edinburgh, Edinburgh, EH9 3FD, Scotland\\
    and \\
    Jordan Richards\\
    School of Mathematics and Maxwell Institute for Mathematical Sciences, University of Edinburgh, Edinburgh, EH9 3FD, Scotland}
  \maketitle
} \fi

\if0\blind
{
   \title{\bf Generative modelling of multivariate geometric extremes using normalising flows}
   \author{Lambert {{D}e {M}onte}\thanks{
   School of Mathematics and Maxwell Institute for Mathematical Sciences, University of Edinburgh, Edinburgh, EH9 3FD, Scotland},
   Rapha\"el Huser\thanks{
   Statistics Program, CEMSE Division, King Abdullah University of Science and Technology (KAUST), Thuwal 23955-6900, Saudi Arabia},
   Ioannis Papastathopoulos\samethanks[1], and 
   Jordan Richards\samethanks[1]}
   \date{}
   \maketitle
} \fi

\if1\blind
{
  \bigskip
  \bigskip
  \bigskip
  \begin{center}
    {\LARGE\bf Generative modelling of multivariate geometric extremes using normalising flows}
\end{center}
  \medskip
} \fi

\bigskip
\begin{abstract}
  Leveraging the recently emerging geometric
  approach to multivariate extremes and the flexibility of normalising
  flows on the hypersphere, we propose a principled deep-learning-based
  methodology that enables accurate joint tail extrapolation in all directions.\ 
  We exploit theoretical links between intrinsic model parameters 
  defined as functions on hyperspheres to construct models 
  ranging from high flexibility to parsimony, thereby enabling the 
  efficient modelling of multivariate extremes displaying complex dependence 
  structures in higher dimensions with reasonable sample sizes.\ 
  We use the generative feature of normalising flows to perform fast 
  probability estimation for arbitrary Borel risk regions via an efficient
  Monte Carlo integration scheme.\ The good properties of our
  estimators are demonstrated via a simulation study in up to ten dimensions.\ 
  We apply our methodology to the analysis of low and high extremes of wind 
  speeds.\ In particular, we find that our methodology enables probability 
  estimation for non-trivial extreme events in relation to electricity production
  via wind turbines and reveals interesting structure in the underlying data.\  
\end{abstract}

\noindent%
\vfill

\newpage
\spacingset{1.9} 

\section{Introduction}
\label{sec:motivation}
\subsection{Motivation and previous work}
\label{subsec:Motivation}
Extreme events, whether of environmental, biological, or financial
nature, only to name a few examples, pose recurring risks.\ 
Accurate and reliable assessment of their frequency and intensity 
allows for better planning, mitigation, and adaptation efforts.\ The 
mathematical framework of extreme value theory provides a principled
approach to characterise, assess, and quantify the risk of 
extreme events of possibly yet-unobserved magnitudes.\ 

In many real-world applications, a useful and practical approach to 
modelling stochastic phenomena and their extremes is to 
frame the questions of interest in a multivariate setting.\ 
Encoding observables into a finite-dimensional 
random vector, interest lies in estimating the probability that a realisation from the random vector belongs 
to specific extremal subsets.\

Early theoretical work on the geometric interpretation of multivariate extremes
examined conditions under which scaled independent 
draws from the distribution of a random vector converge in 
probability onto compact limit sets with respect to the Hausdorff distance
\citep{geffroy1958,geffroy1959suite,Fisher1069limitconvex,
davis1988almost,kinoshta1991convergence}.\ 
Geometric considerations 
have recently 
found renewed theoretical interest \citep{balkema2010metadensities,
balkema_nolde_2010,balkema_nolde_2012,nolde2014geometric}, with 
\cite{nolde_wadsworth_2022} showing that this general framework unifies
existing multivariate extremes approaches.\ 

This development led 
to new statistical inference methods adopting a radial-angular 
representation in which an extreme event
is defined in terms of the magnitude of radii
given their directions, and hence, enable extrapolation in all directions
of the multivariate support of a random vector.\ 
In particular, \cite{wadsworth2022statistical} propose a parametric approach
to model quantities linked with the limit set of a given random vector.\ 
\cite{papastathopoulos2025statistical} build on the geometric framework  
by establishing the weak convergence of exceedances of 
suitably chosen sequences of functions defined on hyperspheres
to a limit Poisson point process.\ Leveraging the likelihood of 
the limit process, they build a Bayesian statistical model relying on latent
Gaussian processes
which can pool information from the radii and 
the directions of observations to perform extreme event probability 
estimation and to extrapolate probability and return sets in dimensions 2 and 3.\
\cite{Majumder2025semiparametric} adopt a Bayesian approach with 
B{\'e}zier splines to model the limit set.\ 
\cite{murphybarltrop2024deepgauge} extend this line of work via
deep learning methods to perform inference for the limit set 
in up to 8 dimensions, but only enable probability estimation for unbounded 
hyperboxes laying entirely in a quadrant of~$\RR^d$.\
\cite{campbell2024piecewiselinear} 
propose a piecewise-linear representation 
in dimension up~to~5.\ 

Parallel work explores modelling the exceedances 
of high functional thresholds of the univariate radial variable 
given a direction using a generalised Pareto distribution,
motivated by univariate theory; see, \textit{e.g.}, 
\cite{simptawn22bivariate,simpson2024inference}, and \cite{mackay2024modelling}.\ 
An active area of research relates to finding representations for the parameters
of this model to enable statistical inference in higher dimensions
\citep{murphybarltrop2024inference,mackay2024deeplearningjointextremes}.\
Although powerful, current approaches face limitations  
in terms of how the distribution of 
the directional variable is modelled.\  
For instance, they either
resort to sampling from the empirical distribution of the directional variables 
\citep{mackay2024deeplearningjointextremes} or do 
not model it at all \citep{murphybarltrop2024deepgauge}.\ 
Recent work involving generative deep learning methods to model the directional variable
highlights the need for scalable models, see
\cite{lhaut2025wassersteinaitchisonganangularmeasures} and \cite{wessel2025comparisongenerativedeeplearning}.\

In this paper, we adopt a deep learning methodology
based on normalising flows within the geometric extremes framework.\ 
Normalising flows provide a flexible and principled approach to density
estimation and possess generative properties enabling efficient sampling 
from inferred target densities.\ 
A review of normalising flows methods can be found in 
\cite{Kobyzev_2021}, 
and for their application to density estimation, see 
\cite{dinh2017density} and \cite{papamakarios2017masked,Papamakarios2021NF}.\ 
In an extreme value analysis context, normalising flows were 
used by \cite{hickling2023flexible} to model  
the marginal tails of a random vector, but not to model 
its dependence structure; \cite{hu2025gpdflowgenerativemultivariatethreshold}
adopted real-valued non-volume preserving flows to model 
multivariate generalised Pareto distributions 
\citep{Rootzén2006MGPD}.\ 
While typically constructed for hyperrectangular subsets of 
the $d$-dimensional reals, normalising flows
have also been implemented on spheres \citep{Ng2024} and on
hyperspheres and tori \citep{rezende2020normalizing}, 
yet not in the context of multivariate extremes.\

\subsection{Contributions and paper structure}
\label{subsec:contributions_structure}
In this work, we exploit the geometric properties of structural
model parameters to enable semi-parametric statistical inference in higher 
dimensions.\ In particular, using theoretically-justified links 
between model parameters, our methodology allows imposing structure on model
parameters by minimising relevant composite loss functions.\ These
new model formulations bridge parsimony and flexibility, thereby enabling statistical
inference in various dimension settings for varying 
complexity of dependence structures.\  

Adopting the radial-angular representation for random vectors, 
we use normalising flows to model the possibly complex distribution 
of directions of extreme events and the unit-volume shape of 
more general 
positive functions not necessarily integrating to one.\ This yields a new approach to  
perform quantile regression on hyperspherical domains and to enforce theoretically-justified
structure between the parameters of the proposed models.\ 
Additionally, we leverage the generative property of normalising flows to enable 
fast sampling from the fitted models and accurate probability estimation for arbitrary Borel 
subsets of the $d$-dimensional reals via Monte Carlo integration.\ 

The paper is organised as follows.\ In Section~\ref{sec:GE_approach}, 
we provide background on the geometric approach to 
multivariate extremes and formulate specific model structures bridging
parsimony and flexibility based 
on asymptotically-justified arguments.\ In Section~\ref{sec:Stat_inf}, we 
introduce normalising flows and describe the 
gradient descent procedure used to fit our models.\ We discuss model 
regularisation options to reduce over-fitting and increase out-of-sample
generalisation, and detail our probability estimation procedure.\ 
In Section~\ref{sec:Sim_Study}, we conduct a simulation study to demonstrate 
our model's ability to extrapolate and estimate the probability of extreme
events in dimensions up to 10.\ In 
Section~\ref{sec:case_study}, we apply our methodology to jointly 
model low and high extremes of wind speeds in the Pacific Northwest 
region of the United States and assess their impact on electricity production.\ 
Last, in Section 6, we highlight the benefits of our methodology and suggest
possible avenues for future work.\

\section{A geometric approach to multivariate extremes}
\label{sec:GE_approach}

\subsection{Notation and assumptions}
\label{subsec:notation}

To ensure clarity and consistency in the presentation, 
this section defines the notation that will be used throughout 
and states the fundamental assumptions underpining our analysis.\

\noindent\textbf{Commonly used manifolds}:\ 
We denote by $\RR_{\geq0}:=[0,\infty)$ and $\RR_{>0}:=(0,\infty)$
the positive and strictly positive reals, respectively; by $\RR^d$ the $d$-dimensional reals with $d\geq 2$; 
by $\SSS^{d-1}:=\{\bm x \in\RR^d\,:\,\lVert\bm x\rVert = 1\}$
the $(d-1)$-sphere (or hypersphere) with Euclidean norm~$\lVert\, \cdot\, \rVert$; 
by $\CCC^{d-1}:=\SSS^{1}\times(-1,1)^{d-2}$ (for $d>2$, and $\CCC^{1}:=\SSS^{1}$)
the hypercylinder.\ 

\noindent\textbf{Line segments and half-lines}:\ We denote by 
$[\bm x\,:\,\bm y]:=\{(1-t)\bm x +t\bm y\,:\,t\in[0,1]\}\subset\RR^d$ 
the closed line segment between $\bm x\in\RR^d$ and 
$\bm y \in\RR^d$, and by 
$[\bm x\,:\,\bm y):=\{(1-t)\bm x +t\bm y \,:\,t\in\RR_{\geq 0}\}\subset\RR^d$ 
the half-line emanating from $\bm x\in\RR^d$ and passing 
through $\bm y\in\RR^d$.\

\noindent\textbf{Starshaped sets and star-bodies}:\ A set~$\s\subseteq\RR^d$ 
is star-shaped
if there exists a \textit{kernel} subset $\text{Ker}(\s)\subseteq\s$
such that $\forall\,\bm x\in\text{Ker}(\s)$ 
and $\forall\,\bm y\in\s$, $[\bm x:\bm y]\subseteq\s$.\ Further, $\s$ 
belongs to the class of star-bodies, denoted $\s\in\bigstar$, if
it is a compact starshaped set and $\bm 0\in\text{Ker}(\s)$.\ 

\noindent\textbf{Radial functions of star-bodies}:\ 
Any star-body $\s\in\bigstar$ is in one-to-one correspondance with its 
radial function $r_{\s}\,:\,\SSS^{d-1}\to \RR_{\geq0}$ where
$r_{\s}(\bm w)=\sup\{\lambda>0\,:\,\lambda\bm w\in\s\}$, and 
$\partial\s=
\{r_{\s}(\bm w)\bm w\,:\,\bm w\in\SSS^{d-1}\}$ is the boundary of~$\s$.\ A point 
$\bm x\in\RR^d$ belongs to the closure $\overline{\s}$ if $\lVert\bm x\rVert\leq r_{\overline{\s}}(\bm x/\lVert\bm x\rVert)$, 
and to the complement $(\overline{\s})^\prime:=\RR^{d}\backslash\overline{\s}$
of $\overline{\s}$ if $\lVert\bm x\rVert > r_{\overline{\s}}(\bm x/\lVert\bm x\rVert)$.\ 

\noindent\textbf{Normalised functions}:\ We denote by~$f$
any (normalised) function integrating to 1 on its domain.\ A function $f$ is 
understood as a probability density function (PDF) if its subscript is a 
random variable and as a normalised radial function 
if its subscript is a star-body.\ 

\noindent\textbf{Star-bodies as model parameters}:\ Model parameters which 
are positive and bounded functions defined on~$\SSS^{d-1}$ can be 
interpreted as radial functions of star-bodies, providing compact notation.\ A 
model parameter $\Sigma\in\bigstar$ then has value $r_{\Sigma}(\bm w)$ 
at $\bm w\in\SSS^{d-1}$.\ 

\noindent\textbf{Random variables}:\
We write $\bm X\sim \PR_{\bm X}$ for random vector 
$\bm X\in\RR^d$ distributed according to a 
probability measure~$\PR_{\bm X}(\cdot):=\PR(\bm X\in \cdot)$.\ 
Its cumulative distribution function (CDF) $F_{\bm X}$ is given by 
$F_{\bm X}(\bm x)=\PR_{\bm X}((-\infty,x_1]\times\cdots\times(-\infty,x_d])$.\ 
A random variable~$R\in\RR$ has
quantile function $F^{-1}_{R}$ with 
$F^{-1}_{R}(q)=\inf\{r \in \RR_{> 0}\,:\, F_{R}(r)\geq q\}$.
We assume throughout that $\PR_{\bm X}$ 
admits a PDF~$f_{\bm X}$ with respect to the Lebesgue measure 
on~$\RR^d$ and has 
standard Laplace marginals---with CDF 
$F_{X}(x) = \{\exp(x)/2\}^{\mathbbm{1}(x\leq 0)}\{1-\exp(-x)/2\}^{\mathbbm{1}(x> 0)}$.\ 
A random variable $R$ is said to follow a generalised Pareto (GP) distribution if 
$F_R(r)=1-[1+\xi r/\sigma]_+^{-1/\xi}$, 
for $[x]_+=\max\{0,x\}$, $\sigma\in\RR_{>0}$, and $\xi\in\RR$, and a 
standard GP distribution if $\sigma=1$.\

\subsection{Background}
\label{subsec:background}

The geometric approach to multivariate extremes aims to characterise 
the behaviour and to infer the probability of extreme events 
of a random vector $\bm X\in\RR^{d}$ through the consideration of 
star-bodies describing the extremal dependence structure of $\PR_{\bm X}$.\ 
Adopting the practical assumption
that~$\PR_{\bm X}$ admits a PDF~$f_{\bm X}$ 
and has standard Laplace marginal distributions, it follows from the 
work of \cite{nolde_wadsworth_2022} that the scaled sample cloud 
$N_n := \{\bm X_i/\log(n/2)\,:\,i=1,\ldots,n\}$---with $\bm X_i$'s drawn 
independently from $\PR_{\bm X}$---converges in probability
onto a limit star-body~$\G\subseteq[-1,1]^d$ in the sense that the 
Hausdorff distance between $N_n$ and $\G$ 
converges to 0 in probability as $n\to\infty$.\ 
Interest in~$\G$ follows from the rich insight it 
provides into the extremal dependence 
structure of~$\PR_{\bm X}$ and from key connections established 
with known extreme-value frameworks and measures of extremal dependence.\ A sufficient 
condition on~$f_{\bm X}$ for $N_n$ to converge onto $\G$ is that
\begin{IEEEeqnarray}{rCl}
\label{eq:geom_convergence}
   -\frac{\log f_{\bm X}(t\bm x_t)}{t}\to g_{\G}(\bm x), \quad \bm x_t \to \bm x, 
   \text{ as } t\to\infty, \quad \bm x \in \RR^d,
\end{IEEEeqnarray}
for a continuous gauge function $g_{\G}\,:\,\RR^d\to\RR_{\geq 0}$.\ Then, 
$\G\in\bigstar$ and it has radial function $r_{\G}\,:\,\SSS^{d-1}\to\RR_{\geq 0}$ given by
$r_{\G} = 1/g_{\G}$ \citep{nolde_wadsworth_2022}.\ Below, we describe 
connections between the convergence in probability of sample clouds onto 
limit sets 
and a refined weak convergence of exceedances of threshold 
functionals to a limiting multivariate radial generalised Pareto 
(GP) distribution \citep{papastathopoulos2025statistical}.\  

\begin{figure}[t]
    \centering
    \vspace{-0.5em}
    \begin{overpic}[width=0.26\textwidth]{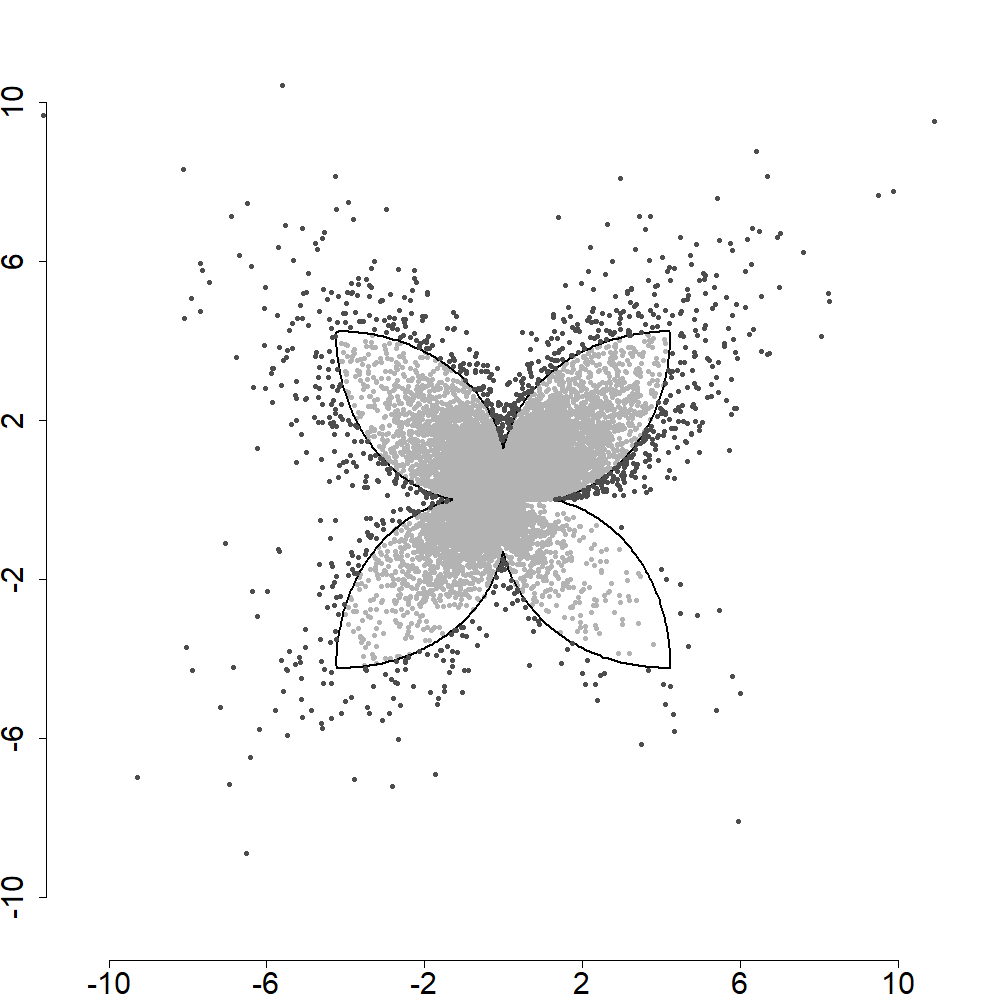}
     \put (46,48) {$\Qq$}
     \put (46,75)  {$\Qq^\prime$}
    \end{overpic}
    \hspace{1em}
    \begin{overpic}[width=0.25\textwidth]{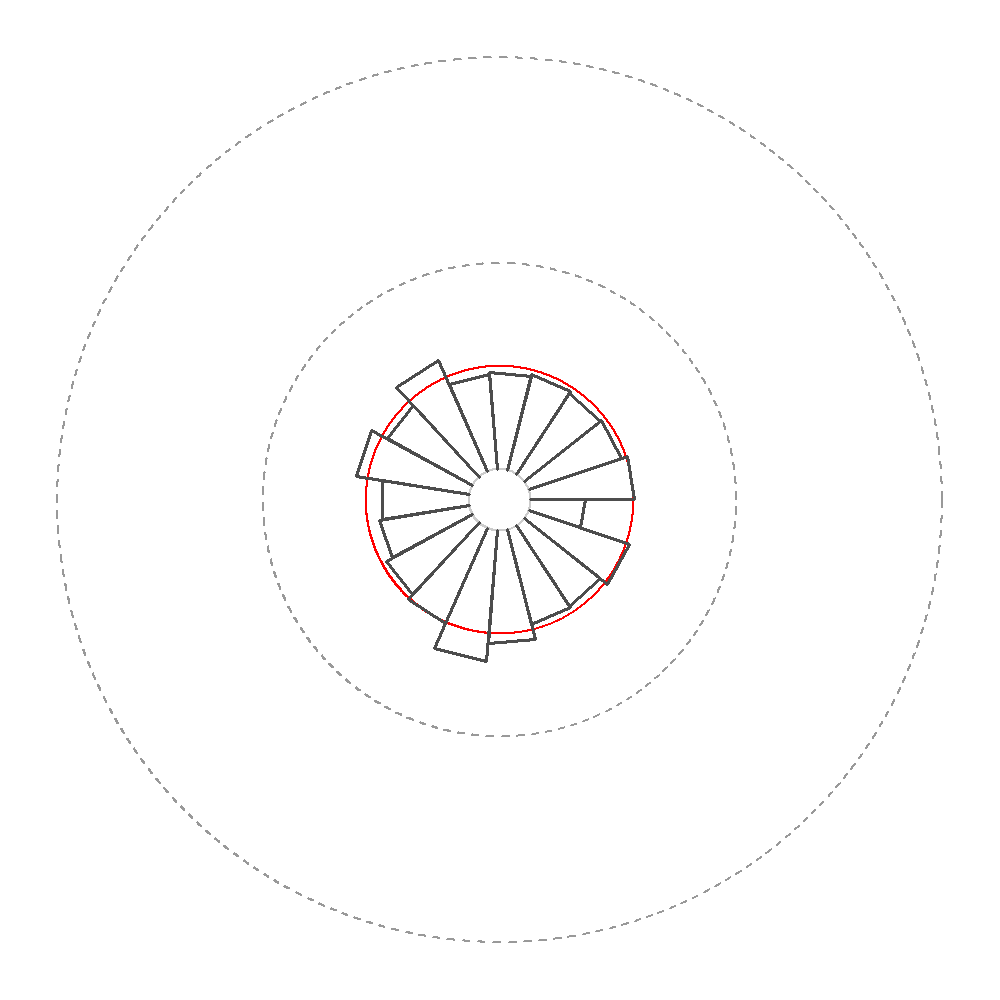}
    \put (48.75,48.5) {\tiny$0$}
    \put (63,32)  {\tiny$0.1$}
    \put (77.5,17.5)  {\tiny$0.2$}
    \end{overpic}
    \hspace{1em}
    \begin{overpic}[width=0.28\textwidth,trim = 0 30 0 30]{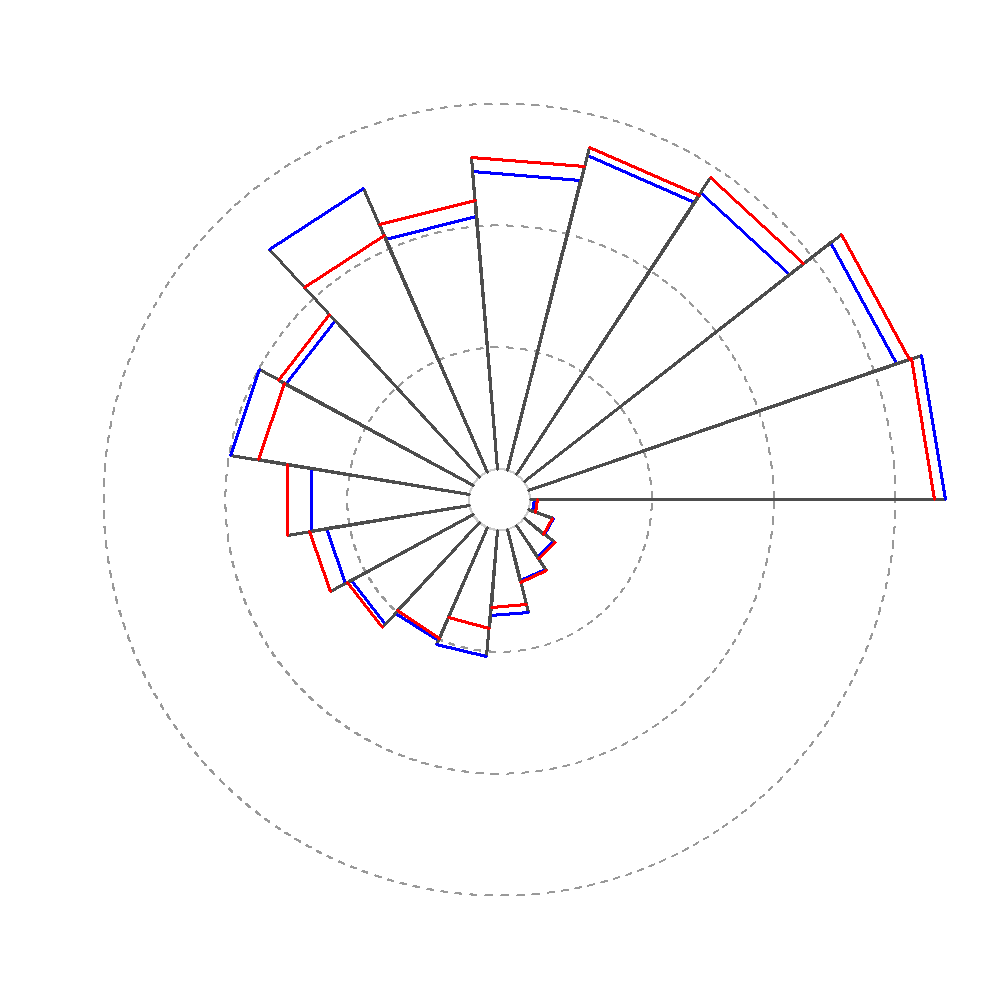}
    \put (48.5,45.8) {\tiny$0$}
    \put (57,36.5) {\tiny$0.1$}
    \put (65.5,27.5)  {\tiny$0.2$}
    \put (74.5,19)  {\tiny$0.3$}
    \end{overpic}
    \caption{\textit{Left}:\ Point cloud (dots) comprising $2\times10^4$ 
    independent samples from a bivariate 
    distribution having quantile set~$\Q_{0.95}$, boundary~$\partial
    \Q_{0.95}$ (solid black line) and complement $\Q_{0.95}^\prime$.\ 
    \textit{Centre}:\ Empirical proportion of exceedances (lying in 
    $\Q_{0.95}^\prime$) binned by angular regions 
    with true exceedance probability $0.05$ (red).\ \textit{Right}:\ Circular 
    histogram of exceedance (blue) and of all sampled (red) directions.\ 
    Concentric circles are density level sets.\ }
    \label{fig:Qq_ex}
\end{figure}
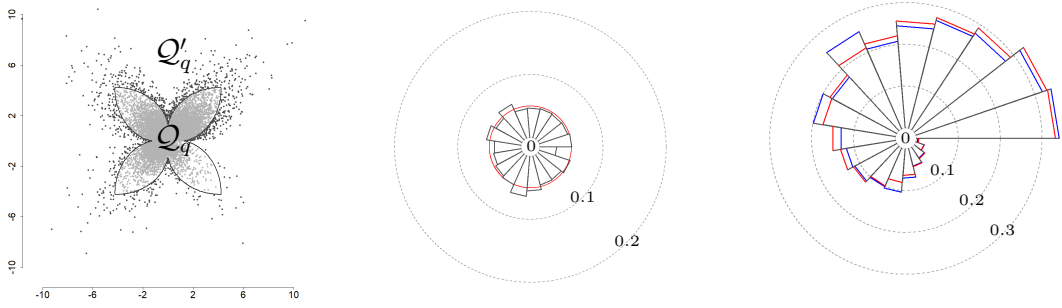

Defining the radius~$R$ and the direction~$\bm W$ of~$\bm X$ 
through
\begin{IEEEeqnarray}{rCl}
\label{eq:rad_ang_decomp}
   (R,\bm W)=(\lVert\bm X\rVert,\bm X/\lVert\bm X\rVert)\in\RR_{>0}\times \SSS^{d-1},\quad \bm X\in\RR^d\backslash \{\bm 0\},
\end{IEEEeqnarray}
allows for the specification 
of a quantile set~$\Qq\in\bigstar$ defined via the $q$th quantile 
of the conditional distribution of $R\mid \{\bm W=\bm w\}$.\ That is, $\Qq$ 
has radial function $r_{\Qq}:\SSS^{d-1}\to\RR_{\geq 0}$ where
$r_{\Qq}(\cdot)=F_{R\mid \bm W}^{-1}(q\mid \cdot)$.\ 
\cite{papastathopoulos2025statistical} show that for any $q\in(0,1)$, 
\begin{IEEEeqnarray}{rCl}
\label{eq:property_Qq}
   \PR(R > r_{\Qq}(\bm W))=\PR(\bm X \in \Qq^\prime)= 1-q \quad\text{and} 
   \quad 
   \{\bm W\mid R>r_{\Qq}(\bm W)\}  \;\smash{\overset{d}{=}}\; \bm W.
\end{IEEEeqnarray}
This is illustrated in Figure~\ref{fig:Qq_ex}.\ 
Since $\PR_{\bm X}$ admits a density by assumption, 
$\PR_{\bm W}$ admits a density $f_{\bm W}$ with respect to the 
($d-1$)-dimensional spherical Lebesgue measure.\
Positivity of
$f_{\bm W}$ on $\SSS^{d-1}$ implies that it describes
a set $\W\in\bigstar$ with radial function $r_{\W}:= f_{\bm W}$.\ 

Further, \cite{papastathopoulos2025statistical} show conditions under 
which there exist a sequence of star-bodies,
$\{\G_{q}\in\bigstar:q\in(0,1)\}$, and a function $\xi:\SSS^{d-1}\to\RR$ 
such that 
\begin{IEEEeqnarray}{rCl}
\label{eq:ass_gp}
   \left(H_{\bm W}\left(\frac{R-r_{\Qq}(\bm W)}{r_{\G_q}(\bm W)}\right), \bm W\right) 
   \,\big |\, \{R>r_{\Qq}(\bm W)\} 
   \overset{d}{\longrightarrow} (U,\bm V),\quad \text{as }q\to1,
\end{IEEEeqnarray}
where $U\sim \mathrm{Uniform}(0,1)$, $\bm V\sim \PR_{\bm W}$, and
$H_{\bm w}$ denotes the CDF of a standard generalised Pareto (GP) distribution with
shape $\xi(\bm w)$.\ Convergence~\eqref{eq:ass_gp} implies a limiting 
distribution for which radial exceedances of $\Qq$ follow a GP distribution
with scale $\G_q$ and shape $\xi$ varying along directions which are distributed according to $\PR_{\bm W}$.\ 

In the important case that~$\bm X$ has standard Laplace marginals, here 
satisfied by assumption, \cite{papastathopoulos2025statistical} show 
that if convergence~\eqref{eq:geom_convergence}
holds uniformly on $\SSS^{d-1}$, then that of~\eqref{eq:ass_gp}
holds with $\G_q=\G$ independent of~$q$ and with $\xi\equiv0$---interpreted as 
$H_{\bm w}(z)=1-\exp(-z)$, $z\geq 0$, the exponential distribution.\ 
Extrapolation is enabled by assuming that the limiting 
distribution 
holds exactly above some fixed level $q\in(0,1)$:\ we onwards assume that 
the excesses~$\bm X\mid \{\bm X\notin\Qq\}$ follow a 
multivariate radial exponential distribution with 
scaling set~$\G$ and directional set~$\W$ exactly---see Section~\ref{sec:MRE} 
of the Supplementary material for a definition.\ 
Then, for every Borel set $\Rset\subseteq \Qq^\prime$, 
the probability that a new draw from~$\bm X$ falls in~$\Rset$ 
is $\PR[\bm X\in\Rset]=(1-q) \PR[\bm X\in\Rset\mid \bm X \in \Qq^\prime]$ where
\begin{IEEEeqnarray}{rCl}
\label{eq:MR_exp}
   \PR[\bm X\in\Rset\mid \bm X \in \Qq^\prime] = 
   \int_{\SSS^{d-1}}\int_{\rho([\bm 0:\bm w)\,\cap\,\Rset)} 
   \frac{1}{r_{\G}(\bm w)}\exp\left\{-\frac{r-r_{\Qq}(\bm w)}{r_{\G}(\bm w)}\right\}
   f_{\bm W}(\bm w)\dd r\,\dd\bm w,\quad
\end{IEEEeqnarray}
$\rho(\cdot) = \{\lVert \bm x \rVert : \bm x \in \cdot\subseteq\RR^d\}$
and 
$[\bm 0:\bm w)$ 
denotes the half-line from~$\bm 0$ through~$\bm w\in\SSS^{d-1}$.\ 

Below, we detail theoretical 
links between the sets~$\Qq$, $\G$, and~$\W$, 
and use them in the next Section~\ref{subsec:proposed_models} to specify new 
models.\ Firstly, \cite{wadsworth2022statistical} show that, under 
mild conditions,~$\Qq$ is asymptotically a scale multiple of~$\G$, 
that is, 
\begin{IEEEeqnarray}{rCl}
\label{eq:Qq_and_G}
   \lim_{q\to1}\frac{r_{\Qq}(\bm w)}{\alpha_q} = r_{\G}(\bm w), 
   \quad\,\bm w \in \SSS^{d-1}, 
\end{IEEEeqnarray}
where $\alpha_q=-\log(1-q)\{1+o(1)\}$.\ 
Secondly, 
while convergence~\eqref{eq:ass_gp} implies that $\G$ 
intervenes in the limit density $f_{R|\{R>r_{\Qq}(\bm W),\bm W\}}$, it
may also more broadly impact the density of exceedances  
$f_{\bm X|\bm X\in\Qq^\prime}(r\bm w) = 
r^{d-1}f_{R|\{R>r_{\Qq}(\bm W),\bm W\}}(r|\bm w)f_{\bm W}(\bm w)$, for~$r>0$
and $\bm w\in\SSS^{d-1}$, via the directional variable $\bm W$.\ 
In particular, \cite{papastathopoulos2025statistical} show that 
if~$f_{\bm X}$ is a composition of a decreasing, 
positive, and continuous function $h_0:\RR_{\geq 0} \to \RR_{\geq 0}$ 
and of the inverse of a \mbox($-1$)-homogeneous function 
$r_{\G}: \RR^d \to \RR_{\geq 0}$ 
describing the shape of a set $\G\in\bigstar$,
that is, $f_{\bm X}(\bm x)=h_0(1/r_{\G}(\bm x))$, $\bm x\in\RR^d$---in which case 
$f_{\bm X}$ is homothetic with respect to~$1/r_{\G}$ 
\citep{balkema_nolde_2010}---then~$\G$ and~$\W$ are indeed linked via
\begin{IEEEeqnarray}{rCl}
\label{eq:G_and_W}
   f_{\bm W}(\bm w)= \frac{r_{\G}(\bm w)^{d}}{\int_{\SSS^{d-1}}r_{\G}(\bm w)^{d}\, \dd \bm w},\quad \bm w \in \SSS^{d-1}.
\end{IEEEeqnarray}
\cite{papastathopoulos2025statistical} show, for known families of 
distributions and real data settings, 
that the use of observed directions may
reduce the uncertainty in estimates of~$\G$.\ 

\subsection{Proposed models}
\label{subsec:proposed_models}
Building upon Section~\ref{subsec:background},
we propose models bridging flexibility and parsimony
to capture complex 
dependence structures in low dimensions and to 
extend semi-parametric estimation to higher dimensions by imposing structure
on the parameters~$\Qq$, $\G$, and~$\W$.\

\afterpage{
\begin{figure}[t!]
   \centering
   \begin{overpic}[width=0.9\textwidth]{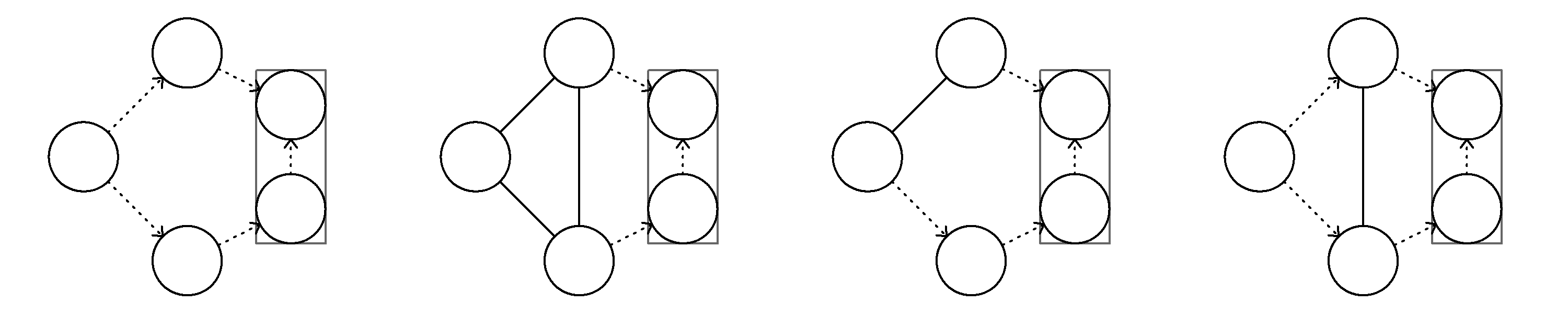}
   \put (17.3,16.5) {{\normalsize $\bm X$}}
   \put (42.3,16.5) {{\normalsize $\bm X$}}
   \put (67.3,16.5) {{\normalsize $\bm X$}}
   \put (92.3,16.5) {{\normalsize $\bm X$}}

   \put (17.3,12.5) {{\normalsize $R$}}
   \put (42.3,12.5) {{\normalsize $R$}}
   \put (67.3,12.5) {{\normalsize $R$}}
   \put (92.3,12.5) {{\normalsize $R$}}

   \put (17,5.6) {{\normalsize $\bm W$}}
   \put (42,5.6) {{\normalsize $\bm W$}}
   \put (67,5.6) {{\normalsize $\bm W$}}
   \put (92,5.6) {{\normalsize $\bm W$}}

   \put (3.5,9.5) {{\normalsize $\Qq$}}
   \put (28.5,9.5) {{\normalsize $\Qq$}}
   \put (53.5,9.5) {{\normalsize $\Qq$}}
   \put (78.5,9.5) {{\normalsize $\Qq$}}

   \put (11,15.8) {{\normalsize $\G$}}
   \put (36,15.8) {{\normalsize $\G$}}
   \put (61,15.8) {{\normalsize $\G$}}
   \put (86,15.8) {{\normalsize $\G$}}

   \put (10.5,2.5) {{\normalsize $\W$}}
   \put (35.5,2.5) {{\normalsize $\W$}}
   \put (60.5,2.5) {{\normalsize $\W$}}
   \put (85.5,2.5) {{\normalsize $\W$}}

   \put (10.5,-1) {{\normalsize $\M_0$}}
   \put (35.5,-1) {{\normalsize $\M_1$}}
   \put (60.5,-1) {{\normalsize $\M_2$}}
   \put (85.5,-1) {{\normalsize $\M_3$}}
   \end{overpic}

   \vspace{1em}

   \begin{overpic}[width=0.9\textwidth]{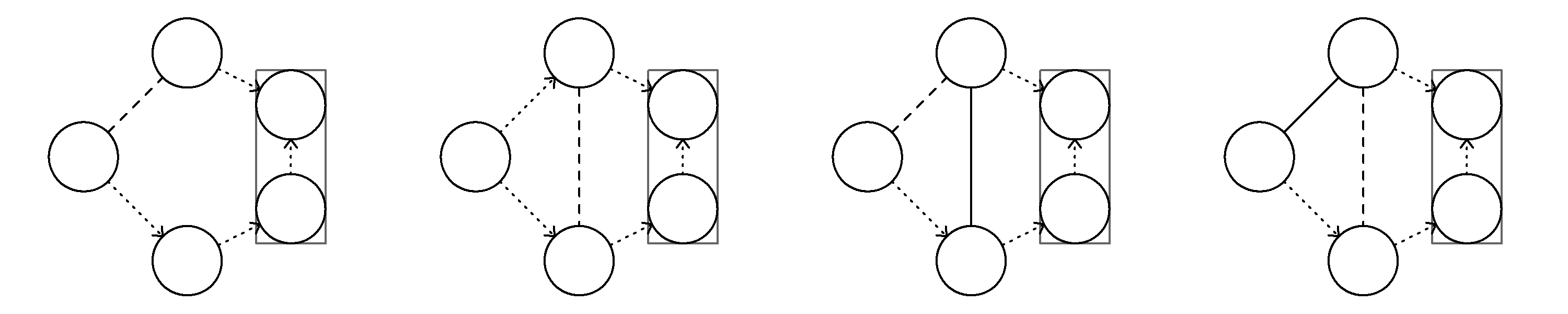}
   \put (17.3,16.5) {{\normalsize $\bm X$}}
   \put (42.3,16.5) {{\normalsize $\bm X$}}
   \put (67.3,16.5) {{\normalsize $\bm X$}}
   \put (92.3,16.5) {{\normalsize $\bm X$}}

   \put (17.3,12.5) {{\normalsize $R$}}
   \put (42.3,12.5) {{\normalsize $R$}}
   \put (67.3,12.5) {{\normalsize $R$}}
   \put (92.3,12.5) {{\normalsize $R$}}

   \put (17,5.6) {{\normalsize $\bm W$}}
   \put (42,5.6) {{\normalsize $\bm W$}}
   \put (67,5.6) {{\normalsize $\bm W$}}
   \put (92,5.6) {{\normalsize $\bm W$}}

   \put (3.5,9.5) {{\normalsize $\Qq$}}
   \put (28.5,9.5) {{\normalsize $\Qq$}}
   \put (53.5,9.5) {{\normalsize $\Qq$}}
   \put (78.5,9.5) {{\normalsize $\Qq$}}

   \put (11,15.8) {{\normalsize $\G$}}
   \put (36,15.8) {{\normalsize $\G$}}
   \put (61,15.8) {{\normalsize $\G$}}
   \put (86,15.8) {{\normalsize $\G$}}

   \put (10.5,2.5) {{\normalsize $\W$}}
   \put (35.5,2.5) {{\normalsize $\W$}}
   \put (60.5,2.5) {{\normalsize $\W$}}
   \put (85.5,2.5) {{\normalsize $\W$}}

   \put (10.5,-1) {{\normalsize $\M_4$}}
   \put (35.5,-1) {{\normalsize $\M_5$}}
   \put (60.5,-1) {{\normalsize $\M_6$}}
   \put (85.5,-1) {{\normalsize $\M_7$}}
   \end{overpic}
   \caption{Structures of the fully decoupled model~$\M_0$, 
   the homothetic model~$\M_1$, the intermediate models~$\M_2$ and~$\M_3$,
   and the constrained models~$\M_4$ to~$\M_7$.\ 
   Solid edges imply an  exact link between parameters.\ Dashed edges imply penalisation from departure of 
   equalities~\eqref{eq:Qq_and_G} or~\eqref{eq:G_and_W}.\ Dotted arrows 
   imply independence in parameter construction.\ }
   \label{fig:models_equalities}
\end{figure}
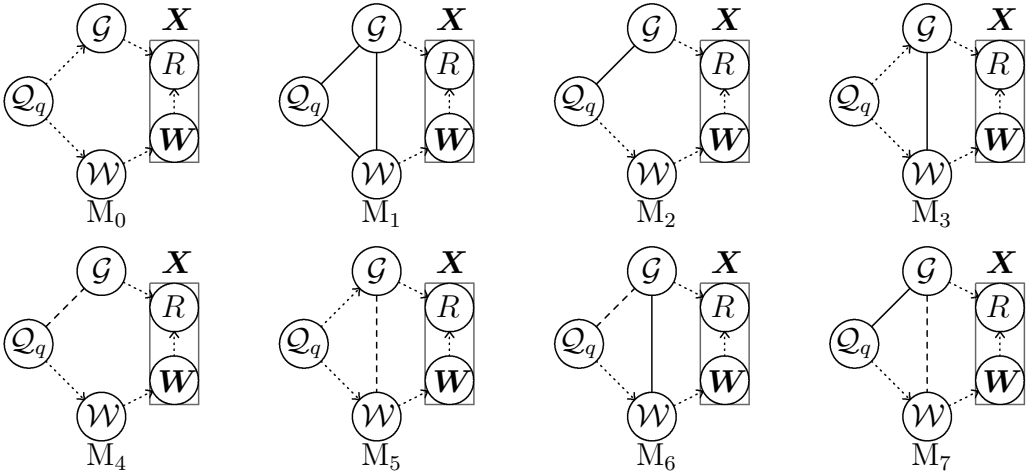
\begin{table}[t!]
   \begin{center}
   \caption{Parameter specification of models~$\M_0$ to~$\M_7$ for scalars
   $\beta_{\Qq},\beta_{\G}>0$, shapes 
   $f_{\Qq},f_{\G},f_{\D}\,:\,\SSS^{d-1}\to\RR_{\geq0}$, and PDF
   of directions $f_{\bm W}\,:\,\SSS^{d-1}\to\RR_{\geq0}$.\ } 
   \begin{tabular}{ccccccccc}
      \toprule \toprule
      \scalebox{.8}{\textbf{Parameter}} & $\M_0$ & $\M_1$ & $\M_2$ & $\M_3$ &
      $\M_4$ & $\M_5$ & $\M_6$ & $\M_7$ \\
      \midrule
      \vspace{0.5em}
      \scalebox{.8}{$r_{\Qq}$} & \scalebox{.8}{$\beta_{\Qq}f_{\Qq}$} &  \scalebox{.8}{$\beta_{\Qq}f_{\bm W}^{\frac{1}{d}}$} & 
                  \scalebox{.8}{$\beta_{\Qq}f_{\G}$} & \scalebox{.8}{$\beta_{\Qq}f_{\Qq}$} &
      \scalebox{.8}{$\beta_{\Qq}f_{\Qq}$} & \scalebox{.8}{$\beta_{\Qq}f_{\D}f_{\G}$} & 
                  \scalebox{.8}{$\beta_{\Qq}(f_{\D}f_{\bm W})^{\frac{1}{d}}$} & \scalebox{.8}{$\beta_{\Qq}(f_{\D}f_{\bm W})^{\frac{1}{d}}$}\\
      \vspace{0.5em}
      \scalebox{.8}{$r_{\G}$} & \scalebox{.8}{$\beta_{\G}f_{\G}$} & \scalebox{.8}{$\beta_{\G}f_{\bm W}^{\frac{1}{d}}$} & 
                  \scalebox{.8}{$\beta_{\G}f_{\G}$} & \scalebox{.8}{$\beta_{\G}f_{\bm W}^{\frac{1}{d}}$} &
      \scalebox{.8}{$\beta_{\G}(f_{\D}f_{\bm W})^{\frac{1}{d}}$} & \scalebox{.8}{$\beta_{\G}f_{\G}$} & 
      \scalebox{.8}{$\beta_{\G}f_{\bm W}^{\frac{1}{d}}$} &\scalebox{.8}{$\beta_{\G}(f_{\D}f_{\bm W})^{\frac{1}{d}}$}\\
      \bottomrule
   \end{tabular}
   \label{tbl:Models}
   \end{center}
\end{table}
}

We first define a base model, $\M_0$, assuming no shared structure between the parameters~$\Qq$, $\G$, and~$\W$,
which can thus have arbitrarily differing geometries.\ 
Among all models proposed below, $\M_0$ can capture the widest range of 
dependence structures and is thus considered the most flexible.\ A visual representation of all model structures 
is given in Figure~\ref{fig:models_equalities};\ their specific 
parameterisations are given in Table~\ref{tbl:Models}.\

To specify more parsimonious models, we leverage theoretical results from
Section~\ref{subsec:background} to motivate links between 
the geometries of the three set parameters.\ 
Concretely,
we say that two sets $\B_1,\B_2\in\bigstar$ with radial functions
$r_{\B_1}$ and~$r_{\B_2}$ are linked if
there exists a bijection $\psi\,:\,\RR_{\geq0}\to\RR_{\geq0}$
such that $r_{\B_1}= \psi\circ r_{\B_2}$.\ 
Without loss and for identifiability purposes, we use the fact that 
the radial function of any set
$\B\in\bigstar$ can be written as 
$r_{\B}=\beta_{\B}f_{\B}$ for
some scalar $\beta_{\B}>0$ and shape function $f_{\B}\,:\,\SSS^{d-1}\to \RR_{\geq0}$ integrating to one.\ 
Also, we use that for any $\B_1,\B_2\in\bigstar$ with 
$r_{\B_1}>0$ and~$r_{\B_2}$, there exists a deformation set~$\D$ 
with radial function
$r_{\D}=\beta_{\D}f_{\D}=r_{\B_2}/r_{\B_1}$ such that $r_{\B_2}=r_{\D}r_{\B_1}$, for $\beta_{\D}>0$ and
$f_{\D}:\SSS^{d-1}\to\RR_{\geq0}$ integrating to one.\ 
Then, $\B_2$ has radial function 
$r_{\B_2}=\beta_{\B_2} f_{\D}f_{\B_1}$ with $\beta_{\B_2}=\beta_{\D}\beta_{\B_1}>0$.\ 
Note that if the shape~$f_{\D}$ is constant, 
$\beta_{\B_2} f_{\D}$ is constant and~$\B_1$ and~$\B_2$ 
have the same shape.\

Therefore, whenever~$f_{\G},f_{\bm W}>0$, we can express~$\Qq$ and~$\G$ in terms of~$\G$ 
and~$\W$, via
\begin{IEEEeqnarray}{rCl}
\label{eq:relax_links}
   r_{\Qq}(\bm w) = \beta_{\Qq} f_{\D_1}(\bm w)f_{\G}(\bm w)
   \quad \text{and}\quad 
   r_{\G}(\bm w) = \beta_{\G}\{f_{\D_2}(\bm w)f_{\bm W}(\bm w)\}^{1/d},\quad \bm w \in \SSS^{d-1},
\end{IEEEeqnarray}
where~$f_{\D_1},f_{\D_2}$
are respectively the shapes of deformation sets $\D_1,\D_2\in\bigstar$, and $\beta_{\Qq},\beta_{\G}>0$.\ 

We specify our most parsimonious model, labelled~$\M_1$, by setting~$f_{\D_1}$ 
and~$f_{\D_2}$ constant on~$\SSS^{d-1}$.\ This corresponds
to assuming that, at a fixed probability level $q\in(0,1)$ and above,~$\Qq$ is exactly a scale 
multiple of~$\G$---that is, $r_{\Qq}\propto r_{\G}$, a practical version of 
assumption~\eqref{eq:Qq_and_G}---and that assumption~\eqref{eq:G_and_W} holds exactly.\ 
For model~$\M_1$, the shapes of~$\Qq$, $\G$, and~$\W$
are thus described by a single shape function~$f_{\bm W}$.\ 

Two models arise from assuming that only one of results~\eqref{eq:Qq_and_G} 
and~\eqref{eq:G_and_W} holds exactly for~$\bm X$ above a finite probability level $q\in(0,1)$.\ 
These models, $\M_2$ and~$\M_3$, correspond to one of~$\D_1$ 
or~$\D_2$ having constant shape, thus linking two parameters via a single shape.\ 
The remaining parameter, either~$\Qq$ or~$\W$, then has independent 
shape $f_{\Qq}$ or~$f_{\bm W}$, respectively.\ 

A third set of models, $\M_4$ to $\M_7$, arises by 
weakening the exact link between
pairs of parameters and constraining them 
from deviating too much from each other.\ 
They are defined using parameterisation~\eqref{eq:relax_links} with non-constant
sets $\D_1$ or $\D_2$---denoted by dashed edges in Figure~\ref{fig:models_equalities}---with 
deformation shapes penalised for departure from a uniform shape.\
Concretely, $\M_4$ bridges $\M_0$ and $\M_2$; $\M_5$ bridges $\M_0$ and $\M_3$;
$\M_6$ bridges $\M_1$ and $\M_3$; $\M_7$ bridges $\M_1$ and $\M_2$.\ Details on 
penalisation are deferred to Section~\ref{sec:penalisation}.\

\section{Statistical inference}
\label{sec:Stat_inf}

\subsection{Marginal model and transformation to Laplace margins}
\label{subsec:Marginal_model}
Let 
$\underline{\bm x}^{\mathrm{o}} = 
\{\bm x_i^{\mathrm{o}} = (x_{i,1}^{\mathrm{o}},\ldots,x_{i,d}^{\mathrm{o}})
\in\RR^d:i=1,\ldots,n\}$ be draws 
from a distribution~$\PR_{\bm{X}^{\mathrm{o}}}$ 
whose marginals $F_{X_j^{\mathrm{o}}}$, $j=1,\ldots,d$, do not necessarily follow the standard Laplace CDF denoted by~$F_L$.\ 
We transform $\underline{\bm x}^{\mathrm{o}}$ to 
$\underline{\bm x} := \big\{\bm x_i = \big(\hat{\varphi}_j(x_{i,j}^\mathrm{o}):j=1,\ldots,d\big):i=1,\ldots,n\big\}$
via the inverse Laplace and approximate probability integral transforms 
$\hat{\varphi}_j = F_L^{-1}\circ\widehat{F}_{X_j^{\mathrm{o}}}$, where 
\begin{equation}
  \widehat{F}_{X_j^{\mathrm{o}}}(x) =
  \begin{cases}
    \Big[1-\left\{1 - \widehat{\xi}_{j, -} \left(\dfrac{x - u_{j,-}}{\widehat{\sigma}_{j,-}}\right)\right\}_+^{-1/\widehat{\xi}_{j,-}}\Big] \widetilde{F}_j(u_{j,-}), \quad& x \leq u_{j,-}\\
    \widetilde{F}_j(x), \quad& u_{j,-} < x \leq u_{j,+}\\
    1-[1-\widetilde{F}_j(u_{j,+})]\left\{1 + \widehat{\xi}_{j,+} \left(\dfrac{x - u_{j,+}}{\widehat{\sigma}_{j,+}}\right)\right\}_+^{-1/\widehat{\xi}_{j,+}}, \quad& x > u_{j,+}
  \end{cases} ,
  \label{eq:PIT_transformation}
\end{equation}
is an estimate of $F_{X_j}$, with
$\widetilde{F}_j(\cdot) = (n+1)^{-1}\sum_{i=1}^n \mathbbm{1}[x_{i,j}^\mathrm{o}
\leq \cdot]$ the empirical distribution function of the $j$th margin, and with
$(\widehat{\sigma}_{j,-}, \widehat{\xi}_{j,-})$ and
$(\widehat{\sigma}_{j,+}, \widehat{\xi}_{j,+})$ the estimated scale and shape
parameters of the GP distribution for the lower and upper
tail fitted to the $j^{\text{th}}$ margin below and above low and high thresholds 
$u_{j,-}$ and $u_{j,+}$, respectively.\ 

As a result, $\underline{\bm x}$ has approximate standard Laplace margins.\ 

\subsection{Deep learning of PDFs and positive functions on $\SSS^{d-1}$}
\label{subsec:NF}

In this section, we detail a method based on normalising flows to infer PDFs 
and positive functions defined almost everywhere (\textit{a.e.}) 
on~$\SSS^{d-1}$ to parameterise the models of
Section~\ref{sec:GE_approach}.\

Formally, a normalising flow is a diffeomorphism---a differentiable 
function with differentiable inverse---mapping a base 
distribution to a target distribution.\ 
If~$\bm Y$ and~$\bm Z$ are random variables 
on probability spaces $(\mathcal{Y},\mathcal{F}_{\mathcal{Y}}, \PR_{\bm{Y}})$ 
and $(\mathcal{Z},\mathcal{F}_{\mathcal{Z}}, \PR_{\bm{Z}})$ 
and admit 
target and base PDFs $f_{\bm Y}$ and $f_{\bm Z}$, 
respectively, the transformation of~$f_{\bm Z}$ 
by a normalising flow~$h$ yields
\begin{IEEEeqnarray}{rCl}
\label{eq:dens_transf}
   f_{\bm Y}(\bm y) = f_{\bm Z}\big\{h^{-1}(\bm y)\big\}\left\lvert
      \frac{\partial h^{-1}(\bm y)}{\partial \bm y}\right\rvert, 
      \quad \bm y \in \mathcal{Y},
\end{IEEEeqnarray}
where $\lvert\partial h^{-1}(\bm y)/\partial \bm y\rvert$ is the 
Jacobian determinant of~$h$.\ In practice, 
$h=h_k\circ\cdots\circ h_1$ is usually taken to be a deep composition 
of~$k$ simple diffeomorphisms $h_1,\ldots,h_k$ with tractable 
Jacobians, thereby enabling flexible transformations of the base 
PDF~$f_{\bm Z}$.\ Details about the normalising flow constructions
are provided in Section~\ref{sec:flow_construction}.\ 
Below, we introduce the method proposed 
by~\cite{rezende2020normalizing} 
to infer a PDF \textit{a.e.\ }on the hypersphere.\ 

\begin{figure}[t]
    \centering
    \vspace{-0.5em}
    \begin{overpic}[width=\textwidth]{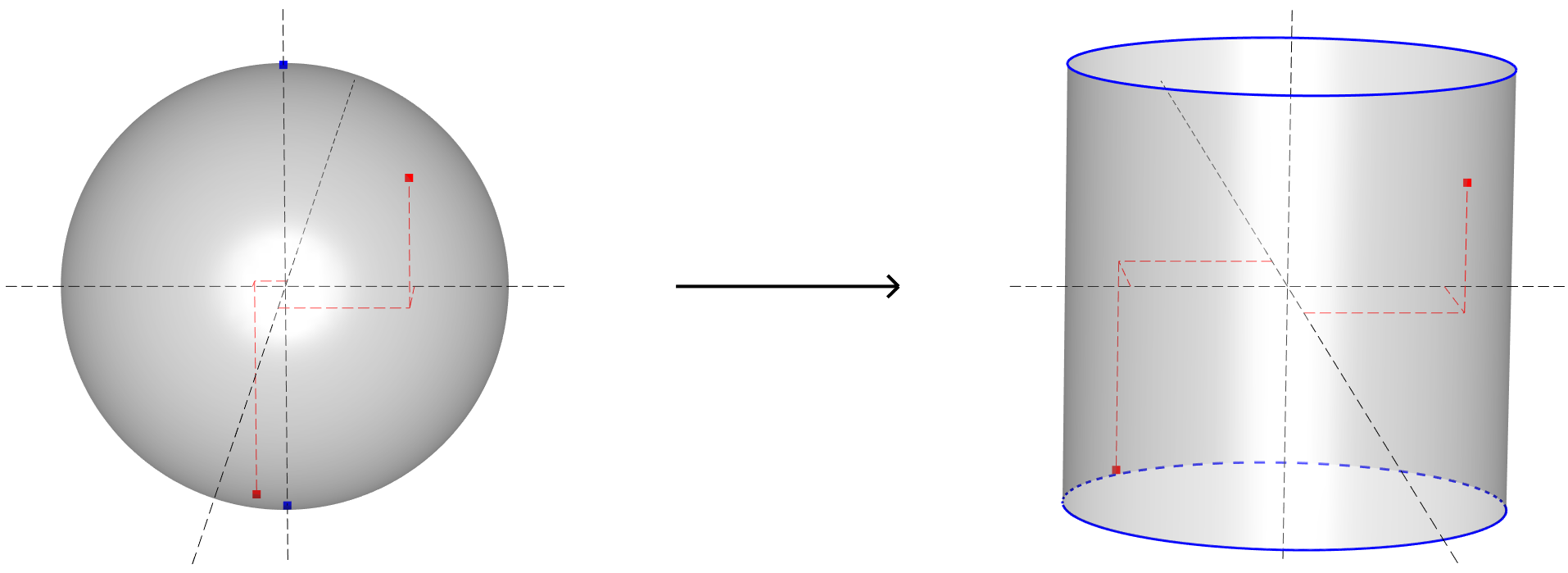}
    \put (34,19) {{\footnotesize $W_1$}}
    \put (9,1.5) {{\footnotesize $W_2$}}
    \put (15,35) {{\footnotesize $W_3$}}
    
    \put (40,21) {{\Large $T:\SSS^2\backslash S_3\to\CCC^{2}$}}
    \put (37.5,14) {{\scriptsize $\bm w\mapsto\bm c = \left(\frac{w_1}{\sqrt{1-w_3^2}},\frac{w_2}{\sqrt{1-w_3^2}},w_3\right)$}}

    \put (97.5,19) {{\footnotesize $C_1$}}
    \put (93.5,0.5) {{\footnotesize $C_2$}}
    \put (80,35) {{\footnotesize $C_3$}}
    
    \put (19,33) {{\footnotesize \color{blue}$\bm e_{3}^{(3)}$}}
    \put (19,2.5) {{\footnotesize \color{blue}$-\bm e_{3}^{(3)}$}}
    
    \put (91,34) {{\footnotesize \color{blue}$\SSS^1\times \{1\}$}}
    \put (68,0) {{\footnotesize \color{blue}$\SSS^1\times \{-1\}$}}
    
    \put (26,26) {{\footnotesize \color{red}$\bm w_1$\color{black}}}
    \put (13.5,5.5) {{\footnotesize \color{red}$\bm w_2$\color{black}}}
    
    \put (93,26) {{\footnotesize \color{red}$\bm c_1$\color{black}}}
    \put (68.5,7) {{\footnotesize \color{red}$\bm c_2$\color{black}}}
    \end{overpic}
    \caption{Visualisation of the map $T:\SSS^{d-1}\backslash\s_{d}\to\CCC^{d-1}$ 
    for the case $d=3$.\ The directions 
    $\bm w_1,\bm w_2\in\SSS^{2}\backslash S_3 = 
    \SSS^{2}\backslash\{\pm \bm e_{3}^{(3)}\}$ are mapped to 
    $\bm c_1=T(\bm w_1),\bm c_2=T(\bm w_2)\in\CCC^{2} = \SSS^{1}\times(-1,1)$.\ }
    \label{fig:T_S_to_C}
\end{figure}

The approach consists in translating the problem of 
learning a PDF \textit{a.e.\ }on the unit hypersphere~$\SSS^{d-1}$ 
to that of learning it on the 
hypercylinder~$\CCC^{d-1}:=\SSS^1\times(-1,1)^{d-2}$, for $d\geq 2$, 
through a recursive, deterministic, and reversible map 
(detailed in Section~\ref{sec:T}),
\begin{IEEEeqnarray}{rCl}
\label{eq:T}
   T:\puncS \to\CCC^{d-1},
\end{IEEEeqnarray}
where $S_d:=\{\pm\,\bm e_{d}^{(3)},\ldots,\pm\,\bm e_{d}^{(d)}\}$ 
is the set of singularities at 
which the map~\eqref{eq:T} is undefined, and 
where~$\bm e_{n}^{(m)}$ denotes the vector
of size~$n$ whose sole non-zero $m$th component is 
equal to~1.\ Figure~\ref{fig:T_S_to_C} illustrates the 
map~$T$ in the case $d=3$ and the two singularities 
$S_3$ of the sphere $\SSS^2$ which are mapped to the 
boundaries $\SSS^1\times\{-1\}$ and $\SSS^1\times\{1\}$ 
of the cylinder~$\CCC^2$.\ Since~$S_d$ has finite cardinality 
$\# S_d=2(d-2)$, it has measure~0 with respect 
to any measure that is absolutely continuous with respect to 
the spherical Lebesgue measure on~$\SSS^{d-1}$.\ 

It follows from the map~\eqref{eq:T} that a target PDF 
$f_{\B}:\puncS\to \RR_{\geq 0}$, describing the shape of a 
star-body $\B\in\RR^d$ \textit{a.e.}, can be written 
as $f_{\B}(\bm w) = f_{\bm Y}(T(\bm w))\lvert\partial T(\bm w)/\partial\bm w\rvert$ 
for $\bm w \in\puncS$ and a PDF~$f_{\bm Y}$ on~$\CCC^{d-1}$.\ 
Using expression~\eqref{eq:dens_transf},~$f_{\B}$ 
can in turn be modelled in terms of a known base PDF 
$f_{\bm Z}:\CCC^{d-1}\to \RR_{\geq 0}$ and a normalising flow~$h_{\B}$ as
\begin{IEEEeqnarray}{rCl}
\label{eq:dens_transf_base_cyl_to_tar_sphere}
   f_{\B}(\bm w) = f_{\bm Z}\big\{h_{\B}^{-1}(T(\bm w))\big\}\left\lvert\frac{\partial h_{\B}^{-1}(T(\bm w))}{\partial T(\bm w)}\right\rvert\left\lvert\frac{\partial T(\bm w)}{\partial \bm w}\right\rvert, \quad \bm w \in \puncS,
\end{IEEEeqnarray}
with Jacobian determinant $\lvert \partial T(\bm w)/\bm w \rvert$ given in Section~\ref{sec:T}.\ 

Equation~\eqref{eq:dens_transf_base_cyl_to_tar_sphere} 
provides a model for~$f_{\bm W}$---by replacing~$\B$ 
with~$\W$---which relies on learning the normalising 
flow~$h_{\W}$.\ 
Further, 
a model for the radial function~$r_\B$ of a general starshaped 
set~$\B$---such as~$\Qq$ or~$\G$---can be obtained via $r_{\B} = \beta_{\B}f_{\B}$ 
where~$f_{\B}$ is as in model~\eqref{eq:dens_transf_base_cyl_to_tar_sphere}, 
and~$\beta_{\B}>0$ is a coefficient to be learned alongside the 
normalising flow~$h_{\B}$.\ 

The invertibility of normalising flows 
enables efficient sampling from a target density~$f_{\B}$ via~$T^{-1}(h_{\B}(\bm Z))$ for $\bm Z\sim f_{\bm Z}$.\
Since~$\bm Z\in\CCC^{d-1}$, a natural base density is 
the uniform density on~$\CCC^{d-1}$.\ 
Since $\bm W$ admits a spherical Lebesgue density by assumption, 
$\PR_{\bm W}(S^d)=0$ and it does not matter that there is no $\bm c \in \CCC^{d-1}$
such that $T^{-1}(h_{\W}(\bm c))\in S^d$.\

\subsection{Statistical inference via loss minimisation}
\label{subsec:loss_minimisation}

To implement the models discussed in Section~\ref{subsec:proposed_models}, 
we use the deep-learning-based methodology described in Section~\ref{subsec:NF} and rely 
on the \textit{normflows} package~\citep{Stimper2023}, a 
\textit{PyTorch}~\citep{paszke2019pytorch} implementation of normalising 
flows via the Python programming language.\ Below, we describe a gradient 
descent approach to model fitting.\ 

\paragraph{Model~$\M_0$:} Model~$\M_0$ can be fitted by sequentially 
obtaining estimates for the parameters~$\Qq$, $\G$, and~$\W$, 
for fixed $q\in(0,1)$.\
To obtain an estimate of~$\Qq$, say~$\hQq$, we seek the 
estimates~$\hat{\beta}_{\Qq}$ and~$\hat{f}_{\Qq}$ minimising 
the quantile loss \citep{Koenker1978}, 
\textit{i.e.,\ }$(\hat{\beta}_{\Qq},\hat{f}_{\Qq})=\mathrm{argmin}_{(\beta_{\Qq}, 
f_{\Qq})}\Loss_{\Qq}(\beta_{\Qq},f_{\Qq}\,;\,\underline{\bm x})$ where,
for $z_i=\lVert \bm x_i \rVert-\beta_{\Qq}f_{\Qq}\left(\bm x_i/\lVert \bm x_i \rVert\right)$,
\begin{IEEEeqnarray}{rCl}
\label{eq:Loss_Qq}
   \Loss_{\Qq}(\beta_{\Qq},f_{\Qq}\,;\,\underline{\bm x}) = 
   \frac{1}{n}\sum_{i=1}^n\max\left\{(1-q)z_i,
   qz_i\right\}. 
\end{IEEEeqnarray}
We minimise~$\Loss_{\Qq}$ via an iterative gradient descent approach 
consisting in a two-step procedure:\ for each gradient descent 
step~$j\in\{1,\ldots,m\}$, we first minimise~$\Loss_{\Qq}$ with 
respect to~$\beta_{\Qq}$ conditionally on the shape~$\hat{f}_{\Qq}^{(j-1)}$ (at step~$j{-}1$) resulting in an 
estimate~$\hat{\beta}_{\Qq}^{(j)}$, and then evaluate the gradient 
of $\Loss_{\Qq}$ at $(\hat{\beta}_{\Qq}^{(j)},\hat{f}_{\Qq}^{(j-1)})$ 
to perform a descent step modifying the flow~$h_{\Qq}$ 
associated with~$\hat{f}_{\Qq}$ conditionally on $\hat{\beta}_{\Qq}^{(j)}$,
resulting in an estimate $\hat{f}_{\Qq}^{(j)}$.\ 
The minimiser $(\hat{\beta}_{\Qq},\hat{f}_{\Qq})$ is 
used to form~$\hQq$ via its radial function 
$r_{\hQq} = \hat{\beta}_{\Qq}\hat{f}_{\Qq}$ and to define 
$\Exc := \{i \in\{1,\ldots,n\}:\lVert\bm x_i \rVert > \smash{r_{\hQq}}(\bm x_i/\lVert \bm x_i \rVert)\}$, 
the indices of 
exceedances of~$\hQq$.\ 

To infer~$\G$, we again use a two-step 
minimisation procedure for~$\beta_{\G}$ and~$f_{\G}$, but with the 
mean negative log-likelihood of the conditional 
radial component, $R\mid \bm W$, as the loss:
\begin{IEEEeqnarray}{rCl}
\label{eq:Loss_G}
   \Loss_{\G}(\beta_{\G},f_{\G}\,;\,r_{\hQq},\underline{\bm x}) {=} 
   {-}\frac{1}{\#\Exc}\sum_{i\in\Exc}\log 
   \left[\left\{\beta_{\G}f_{\G}\left(\frac{\bm x_i}{\lVert\bm x_i\rVert}\right)\right\}^{-1}
   \exp\left\{-\frac{\lVert\bm x_i\rVert {-} r_{\hQq}\left(\bm x_i/\lVert\bm x_i\rVert\right)}
   {\beta_{\G}f_{\G}(\bm x_i/\lVert\bm x_i\rVert)}\right\}\right].\quad~~
\end{IEEEeqnarray}
We note that, conditionally on a shape~$\hat{f}_{\G}^{(j-1)}$ at step~$j{-}1$ 
of gradient descent, the minimiser~$\hat{\beta}_{\G}^{(j)}$ is 
$\hat{\beta}_{\G}^{(j)}= (\#\Exc)^{-1}\sum_{i\in\Exc}
\{\lVert\bm x_i\rVert - r_{\hQq}\left(\bm x_i/\lVert\bm x_i\rVert\right)\}
/\hat{f}_{\G}^{(j-1)}\left(\bm x_i/\lVert\bm x_i\rVert\right)$, 
the closed-form maximum likelihood estimate of the scale 
parameter~$\beta_{\G}$, 
and is therefore fast to compute.\ 

Finally, inferring the directional set~$\W$ is simpler as it solely 
requires minimising the loss~$\Loss_{\W}$ chosen as the negative 
log-likelihood of the observed directions, that is,
\begin{IEEEeqnarray}{rCl}
\label{eq:Loss_W}
   \Loss_{\W}(f_{\bm W}\,;\,\Exc,\underline{\bm x}) = -\frac{1}{\#\Exc}\sum_{i\in\Exc}\log f_{\bm W}(\bm x_i/\lVert \bm x_i \rVert).
\end{IEEEeqnarray}
Owing to property~\eqref{eq:property_Qq}, and assuming 
that~$\Qq$ is well estimated by $\hQq$, we may use all 
observed directions $\{\bm x/\lVert \bm x\rVert : \bm x\in\underline{\bm x}\}$ 
to infer~$\W$ instead of only the exceedances of~$\hQq$;\ 
in this case, the loss~\eqref{eq:Loss_W} can be modified by 
replacing~$\Exc$ with~$\{1,\ldots,n\}$.\
This proves useful as exceedance directions of~$\hQq$ 
can be scarce for~$q$ near~1.\ 

\paragraph{Model~$\M_1$:}Since the negative losses~$-\Loss_{\G}$ and~$-\Loss_{\W}$ 
are mean log-likelihoods of radial exceedances
$\{R-r_{\Qq}(\bm w)\}\mid \{R>r_{\Qq}(\bm w), \bm W=\bm w\}$ and 
of directions of exceedance $\bm W \mid \{R>r_{\Qq}(\bm W)\}$, respectively, their 
sum forms a valid mean log-likelihood for the pair
$(R-r_{\Qq}(\bm W),\bm W)\mid \{R>r_{\Qq}(\bm W)\}$.\ Hence, 
for fixed $q\in(0,1)$, 
model~$\M_1$ can naturally be fitted by minimising the composite loss
\begin{IEEEeqnarray}{rCl}
\label{eq:Loss_M1}
   \Loss_{\Qq,\G,\W}(\beta_{\Qq},\beta_{\G},f_{\bm W}\,;\,\underline{\bm x}) 
   &=& 
   \lambda\Loss_{\Qq}(\beta_{\Qq},f_{\bm W}^{\frac{1}{d}}\,;\,\underline{\bm x})
    + \nonumber\\
   && \quad(1{-}\lambda)\big[\Loss_{\G}(\beta_{\G},f_{\bm W}^{\frac{1}{d}};\beta_{\Qq}f_{\bm W}^{\frac{1}{d}},\underline{\bm x}) 
   +
   \Loss_{\W}(f_{\bm W}\,;\,\Exc,\underline{\bm x})\big]\quad~~
\end{IEEEeqnarray}
with respect to~$\beta_{\Qq}$, $\beta_{\G}$, and~$f_{\bm W}$, and where~$\lambda\in(0,1)$ is a mixing hyperparameter.\ 

In combining the shape of~$\Qq$ with those of~$\G$ 
and~$\W$, the number of exceedances of the estimated $\Qq$
may change at each gradient descent step.\ 
Letting~$\Loss_\G$ and~$\Loss_\W$ be mean negative log-likelihoods instead
of negative log-likelihoods ensures that the sum 
in~$\Loss_{\Qq,\G,\W}$ is 
not minimised by minimising the number of exceedances.\ This same 
argument is applied to~$\M_2$ and~$\M_4$ to~$\M_7$ in which the shape of~$\G$
is also tied to those of~$\Qq$ or~$\W$.\ 

\paragraph{Models~$\M_2$--$\M_3$:}The procedure to fit models~$\M_2$ 
and~$\M_3$ borrows from that of~$\M_0$ and~$\M_1$ in that a pair of their 
parameters is inferred jointly while the third is 
estimated independently.\ Figure~\ref{fig:models_equalities} 
indicates, via solid edges, the losses among~$\Loss_{\Qq}$, $\Loss_{\G}$, 
and $\Loss_{\W}$ that must be combined into a joint loss.\ 
Obtaining estimates~$\hQq$ and~$\hG$ for~$\M_2$ first involves minimising
\begin{IEEEeqnarray}{rCl}
\label{eq:Loss_M2}
   \Loss_{\Qq,\G}(\beta_{\Qq},\beta_{\G},f_{\G}\,;\,\underline{\bm x}) = 
   \lambda\Loss_{\Qq}(\beta_{\Qq},f_{\G}\,;\,\underline{\bm x})+
   (1-\lambda)\Loss_{\G}(\beta_{\G},f_{\G}\,;\,\beta_{\Qq}f_{\G},\underline{\bm x}),
\end{IEEEeqnarray}
and then $\Loss_{\W}$ 
in~\eqref{eq:Loss_W}  
to infer~$\hW$.\ Similarly, model~$\M_3$ can be fitted by first 
obtaining an estimate~$\hQq$ via the minimisation of 
$\Loss_{\Qq}$ in~\eqref{eq:Loss_Qq}, and then~$\hG$ and~$\hW$ by minimising
\begin{IEEEeqnarray}{rCl}
\label{eq:Loss_M3}
   \Loss_{\G,\W}(\beta_{\G},f_{\bm W}\,;\,r_{\hQq},\underline{\bm x}) 
   = 
   \Loss_{\G}(\beta_{\G},f_{\bm W}^{1/d}\,;\,r_{\hQq},\underline{\bm x})+\Loss_{\W}(f_{\bm W}\,;\,\Exc,\underline{\bm x}).
\end{IEEEeqnarray}

\paragraph{Models~$\M_4$--$\M_7$:}Models~$\M_4$ to~$\M_7$ require a different 
treatment as they involve an additional normalising flow~$h_{\D}$ 
associated with a deformation set~$\D$.\ 
The appropriate loss for each of these 
combines the individual losses~$\Loss_{\Qq}$, $\Loss_{\G}$, 
and~$\Loss_{\W}$ according to the 
solid or dashed edges depicted in 
Figure~\ref{fig:models_equalities}.\ That is, model~$\M_4$ shares the 
same composite loss as~$\M_2$, $\M_5$ the same as~$\M_3$, and~$\M_6$ and~$\M_7$ the 
same as~$\M_1$, with the exception that their arguments~$r_{\Qq}$ 
and~$r_{\G}$ are replaced by the appropriate radial functions 
given in Table~\ref{tbl:Models} and that a penalisation 
term for the deformation set~$\D$ is added.\ The penalisation 
term $\lambda_U\overline{\mathrm{D}}_{\mathrm{KL}}[f_U\lVert f_{\D}]$---made 
explicit in Section~\ref{sec:penalisation}, with $\lambda_U\geq0$ 
controlling the strength of penalisation---is a weighted Monte 
Carlo approximation of the Kullback--Leibler (KL) divergence of the uniform 
density~$f_{U}$ on~$\SSS^{d-1}$ from~$f_{\D}$;\ it ensures that~$f_{\D}$ 
is penalised for departure from~$f_{U}$.\ 
This is akin to a penalised complexity prior \citep{Simpson2017PCpriors} 
that puts an exponential prior on the KL divergence to shrink
complex models toward a simpler counterpart.\ 
The models are fitted by alternately minimising, at every 
gradient descent iteration, the appropriate loss with respect 
to~$f_{\D}$ (conditionally on the other model parameters) and to 
the other model parameters (conditionally on~$f_{\D}$).\ 

Fitting the above models requires fixing the hyperparameter~$\lambda$ 
and the penalisation constant $\lambda_U$, both determining
which composite loss components bear more weight.\ 
To avoid significant computational overhead, 
in the following sections we select a $\lambda$ which empirically 
leads to roughly equal contributions of~$\lambda\Loss_{\Qq}$ 
and~$(1-\lambda)(\Loss_{\G}+\Loss_{\W})$ to the 
loss $\Loss_{\Qq,\G,\W}$~\eqref{eq:Loss_M1} of model~$\M_1$, 
and of $\lambda\Loss_{\Qq}$ and $(1-\lambda)\Loss_{\G}$ 
to the loss $\Loss_{\Qq,\G}$~\eqref{eq:Loss_M2} of model~$\M_2$.\
In both cases, however, classical selection methods 
such as cross-validation may be used.\

In practice, we minimise the above losses
via gradient descent using the \textit{PyTorch} 
implementation of the \textit{Adam} 
optimiser~\citep{kingma2017adammethodstochasticoptimization}.\  

\subsection{Model regularisation}
\label{subsec:regularisation}

Due to the possibility of model overfitting, we discuss regularisation techniques
that can be applied to our proposed statistical inference methodology.\ 

A first way to reduce overfitting consists in performing early 
stopping \citep{Prechelt2012}.\ Defining training and validation 
sets~$\underline{\bm x}_{\mathrm{T}}$ and~$\underline{\bm x}_{\mathrm{V}}$ 
by partitioning the observed data~$\underline{\bm x}$, 
early stopping involves halting the gradient descent procedure on a loss evaluated
at~$\underline{\bm x}_{\mathrm{T}}$ 
when the loss at~$\underline{\bm x}_{\mathrm{V}}$ has not 
decreased in a certain number of iterations.\ This prevents 
a decrease of the training loss 
at the detriment of increasing the generalisation error.\

A second type of model regularisation technique, called data mollification, 
has been proposed by~\cite{song2021scorebased}, and \cite{tran2023onelineofcode} 
provided evidence of great improvement to 
density estimation and to the quality of generated samples in likelihood-based generative models 
such as normalising flows.\ Mollification consists in adding 
gradually-vanishing noise to the data at training time to create a transition
from a noisier and simple-to-optimize density to the target density 
of the data.\ To avoid deforming the tail 
properties of the variable $R\mid\bm W$, we only mollify 
the training directions 
$\{\bm x/\lVert \bm x\rVert:\bm x 
\in \underline{\bm x}_{\mathrm{T}}\}$ by adding independent 
von Mises noise with gradually-decreasing dispersion parameter~$\sigma$.\ 
Concretely, at the $j$th of~$J$ gradient descent iterations
of the training period, we use the mollified dataset
\begin{IEEEeqnarray}{rCl}
\label{eq:mollification}
   \underline{\bm x}_{\mathrm{T},j} = 
   \left\{\lVert \bm x_i\rVert\bm w_{i,\varepsilon}\;:\; 
   \bm w_{i,\varepsilon} \sim \mathrm{vonMises}(\bm x_i/\lVert \bm x_i\rVert,\sigma_j),\, 
   \bm x_i \in \underline{\bm x}_{\mathrm{T}}\right\},
\end{IEEEeqnarray}
where vonMises($\bm \mu$,$\sigma$) denotes the von 
Mises distribution with location $\bm \mu\in\SSS^{d-1}$ and dispersion 
$\sigma\in\RR_{>0}$.\ Following the logistic decay proposed 
by~\cite{tran2023onelineofcode}, we impose that $\sigma_j$ decreases to $0$ as~$j$ 
increases to $J$, see Section~\ref{sec:sigma_j} for details.\ 

Combining both regularisation methods yields a sensible procedure which
stops the training when the gradient descent algorithm starts overfitting 
the (decreasingly mollified) data~$\underline{\bm x}_{\mathrm{T},j}$ and 
increases the generalisation error on the validation data~$\underline{\bm x}_{\mathrm{V}}$.\  

\subsection{Rare event probability estimation}
\label{subsec:prob_est}
In this section, we detail a method to 
extrapolate beyond the range of observed 
data $\underline{\bm x}=\{\bm x_1,\ldots,\bm x_n\}$---assumed to form a random 
sample from~$\PR_{\bm X}$---and to estimate
the probability $\PR[\bm X \in\Rset]$ that a new draw~$\bm X$ 
lies in a Borel risk region~$\Rset\subset \RR^d$.\ 

Given estimates~$\hQq$, $\hG$, and~$\hW$, and following the prodecure
outlined in \cite{papastathopoulos2025statistical}, we partition $\Rset$ into 
two disjoint regions $\Rset\cap\hQq^\prime$ and $\Rset\cap\hQq$ to write
\begin{IEEEeqnarray}{rCl}
\label{eq:P_B}
   \PR[\bm X\in\Rset] = \PR[\bm X\in\Rset\mid \bm X \notin \hQq]
   \PR[\bm X \notin \hQq] + \PR[\bm X\in\Rset \cap \hQq].
\end{IEEEeqnarray}
By model assumptions, 
$\Rset\cap\hQq^\prime$ is a region over 
which our threshold exceedance model is defined; 
the method described below is especially suited for risk regions~$\Rset$ 
such that $\Rset\cap\hQq$ is minimal or null.\  
Assuming that excesses above~$\hQq$ follow a multivariate radial exponential 
distribution with scale~$\hG$ and directions $\hW$ exactly, we estimate 
$\PR[\bm X\in\Rset\mid \bm X \notin \hQq]$ 
of equation~\eqref{eq:P_B} via
a Monte Carlo approximation of the outer integral in~\eqref{eq:MR_exp}, \textit{i.e.}, 
\begin{IEEEeqnarray}{rCl}
\label{eq:P_B_noint_Qq}
    \frac{1}{m}\sum_{i=1}^m\int_{\rho([\bm 0:\bm w_i^\star)\,\cap\,\Rset\,\cap\,\hQq^\prime)}
    \frac{1}{r_{\hG}(\bm w_{i}^\star)}
    \exp\left\{-\frac{r-r_{\hQq}(\bm w_{i}^\star)}{r_{\hG}(\bm w_{i}^\star)}\right\}
    \,\dd r, 
\end{IEEEeqnarray}
where directions $\bm w_1^\star,\ldots,\bm w_m^\star\in\puncS$ 
are sampled from~$\hat{f}_{\bm W}$.\ 
Notably, the integral in expression~\eqref{eq:P_B_noint_Qq} 
is solved exactly whenever the set 
$\rho([\bm 0:\bm w)\,\cap\,\Rset\,\cap\,\hQq^\prime)$
is known.\ In the important case 
where~$\Rset$ is a hyperbox, $\rho([\bm 0:\bm w)\,\cap\,\Rset\,\cap\,\hQq^\prime)$ 
is a single interval with explicit lower- and upper-bounds
(see Section~\ref{sec:hyperbox}).\ 
A simple adaptation yields a method to estimate the probability that at 
least~$j\in\{1,\ldots,d\}$ of the components of $\bm X$ fall 
below or above some specifed lower- and upper-bounds
(see Section~\ref{sec:outside_starbodyhyperbox}).\

To estimate the probabilities
$\PR[\bm X\notin\hQq]$ and $\PR[\bm X\in\Rset \cap \hQq]$ of equation~\eqref{eq:P_B}, 
we use their respective unbiased empirical estimates
\begin{IEEEeqnarray}{rCl}
\label{eq:P_not_Qq}
    \frac{1}{n}\sum_{i=1}^n\mathbbm{1}[\bm x_i \notin \hQq] 
   \quad \text{and} \quad 
   \frac{1}{n}\sum_{i=1}^n\mathbbm{1}[\bm x_i \in \Rset \cap \hQq].
\end{IEEEeqnarray}
Their combination with estimate~\eqref{eq:P_B_noint_Qq} via 
equation~\eqref{eq:P_B} yields an estimate of 
$\PR[\bm X \in \Rset]$.\

In practice, the sample of directions 
$\bm w_1^{\star},\ldots,\bm w_m^\star$ required to evaluate 
expression~\eqref{eq:P_B_noint_Qq} is obtained efficiently via the 
generative property of the normalising flow even for 
large~$m$, and the calculation of the estimates~\eqref{eq:P_not_Qq} 
requires a single iteration through~$\underline{\bm x}$.\

\section{Simulation study}
\label{sec:Sim_Study}
In this section, we apply the 
methodology described in Section~\ref{sec:Stat_inf} to 
demonstrate the ability of our models from Section~\ref{sec:GE_approach} to accurately 
estimate the probability of extreme events.\ 

For each $d\in\{3,5,7,10\}$, we draw independent samples 
from the $d$-dimensional multivariate normal 
distribution with zero mean and a randomly generated 
covariance matrix,
resulting in four different simulation study settings.\ 
We transform the standard normal marginal distributions 
to the standard Laplace distribution via the 
probability integral and inverse transforms---using the exact standard 
normal and inverse standard Laplace CDFs---to 
assess the performance of the joint model only, as the marginal model
proposed in Section~\ref{subsec:Marginal_model} is 
standard and widely adopted.\ 

For each setting, we define three risk regions, 
$\Rset_1,\Rset_2,\Rset_3\in\RR^d$, 
for which we can reliably calculate the true probability that a draw from
the specified multivariate distributions lies in.\ 
In all four settings, $\Rset_1$
corresponds to a risk region where all variables are simultaneously large;\ 
$\Rset_2$ and $\Rset_3$ are arbitrary hyperbox extremal regions, some 
corresponding to marginalisation over a subset of the $d$-dimensional vector.\ 
These hyperboxes are explicitly 
stated in the relevant result figures.\
For brevity, we only present results from models~$\M_0$, $\M_1$, and~$\M_2$, 
since model~$\M_3$ was already studied by 
\cite{papastathopoulos2025statistical}.\ 
More results on $\M_4$ to $\M_7$ and visualisations of 
deformation sets are found in 
Section~\ref{sec:Supp_simulation_study}.\ 

To highlight the wide applicability and 
flexibility of our modelling framework, we use the same normalising flow
architecture in all four settings---namely, of 5 stacked rational-quadratic spline 
flows with six knots on each 
interval of the hypercylinder, see 
Section~\ref{sec:flow_construction}.\ More complex 
architectures were tested in the same settings and yielded 
very similar results.\  
To make our probability estimation procedure robust against overfitting, we rely 
both on early stopping, 
with a 0.7--0.3 proportion split for the training and validation
sets, and on mollification (recall Section~\ref{subsec:regularisation}).\ 
Further, to ensure roughly equal weights of 
loss components (see Section~\ref{subsec:loss_minimisation}), 
we fix the mixing hyperparameter $\lambda$ to $0.8$.\

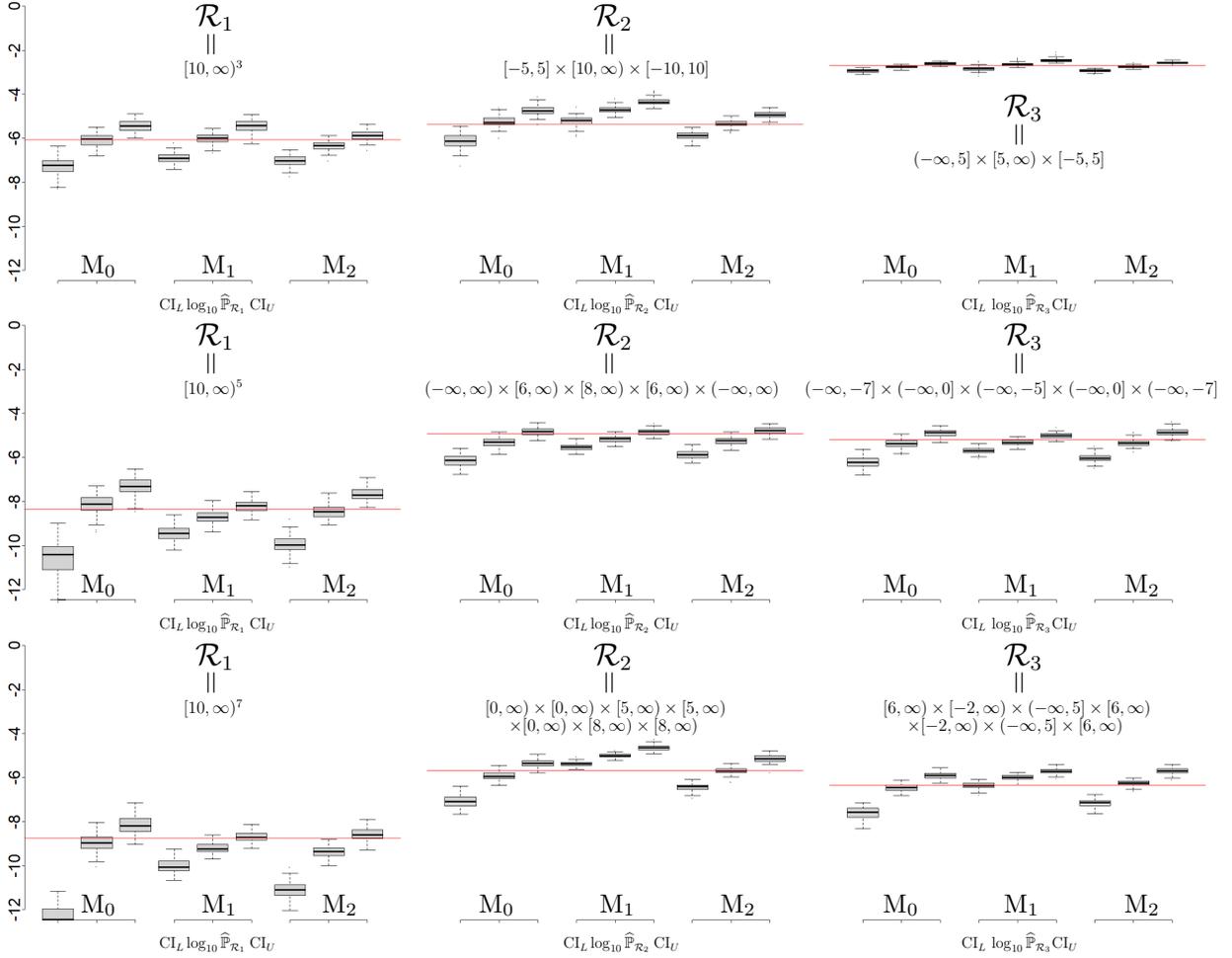
\begin{figure}[t!]
   \centering
   \begin{overpic}[width=\textwidth]{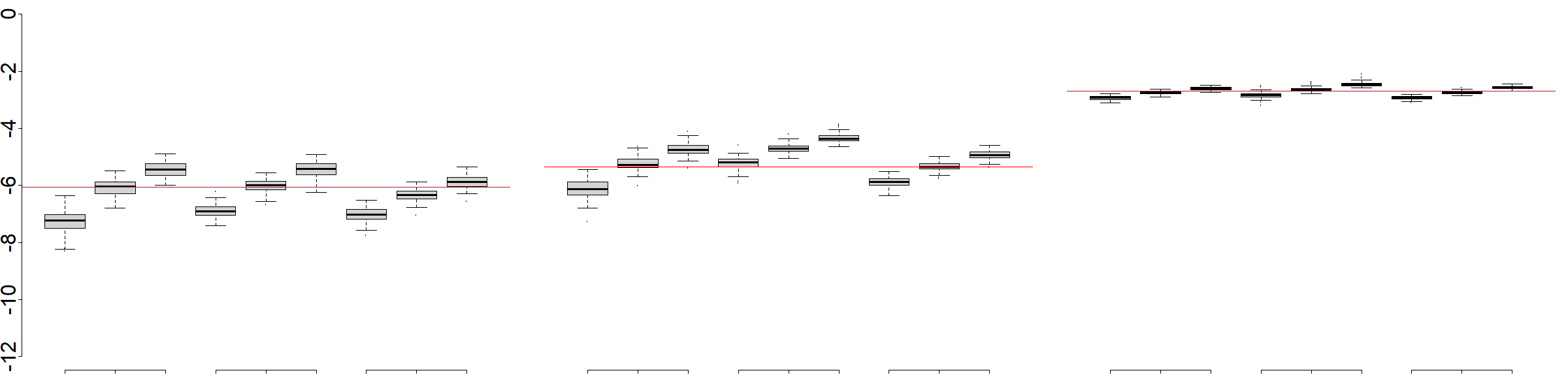}
      \put (12.5,-1) {\scalebox{0.5}{$\mathrm{CI}_{L}$}}
      \put (46,-1) {\scalebox{0.5}{$\mathrm{CI}_{L}$}}
      \put (79,-1) {\scalebox{0.5}{$\mathrm{CI}_{L}$}}
      
      \put (14.75,-1) {\scalebox{0.5}{$\log_{10}\widehat{\PR}_{\Rset_1}$}}
      \put (48.25,-1) {\scalebox{0.5}{$\log_{10}\widehat{\PR}_{\Rset_2}$}}
      \put (81.5,-1) {\scalebox{0.5}{$\log_{10}\widehat{\PR}_{\Rset_3}$}}
      
      \put (20,-1) {\scalebox{0.5}{$\mathrm{CI}_{U}$}}
      \put (53.5,-1) {\scalebox{0.5}{$\mathrm{CI}_{U}$}}
      \put (86.5,-1) {\scalebox{0.5}{$\mathrm{CI}_{U}$}}
      
      \put (6,2) {\footnotesize $\M_0$}
      \put (16,2) {\footnotesize $\M_1$}
      \put (26,2) {\footnotesize $\M_2$}
      \put (39,2) {\footnotesize $\M_0$}
      \put (49,2) {\footnotesize $\M_1$}
      \put (59,2) {\footnotesize $\M_2$}
      \put (72.5,2) {\footnotesize $\M_0$}
      \put (82.5,2) {\footnotesize $\M_1$}
      \put (92.5,2) {\footnotesize $\M_2$}
      
      \put (15.5,22.5) {$\Rset_1$}
      \put (16.5,20) {\rotatebox{90}{$=$}}
      \put (14.5,18.5) {\scalebox{.55}{$[10,\infty)^3$}}
      \put (48.5,22.5) {$\Rset_2$}
      \put (49.5,20) {\rotatebox{90}{$=$}}
      \put (41,18.5) {\scalebox{.55}{$[-5,5]\times[10,\infty)\times[-10,10]$}}
      \put (82.5,15) {$\Rset_3$}
      \put (83.5,12.5) {\rotatebox{90}{$=$}}
      \put (75,11) {\scalebox{.55}{$(-\infty,5]\times[5,\infty)\times[-5,5]$}}
   \end{overpic}

   \vspace{0.5em}
   \centering
   \begin{overpic}[width=\textwidth]{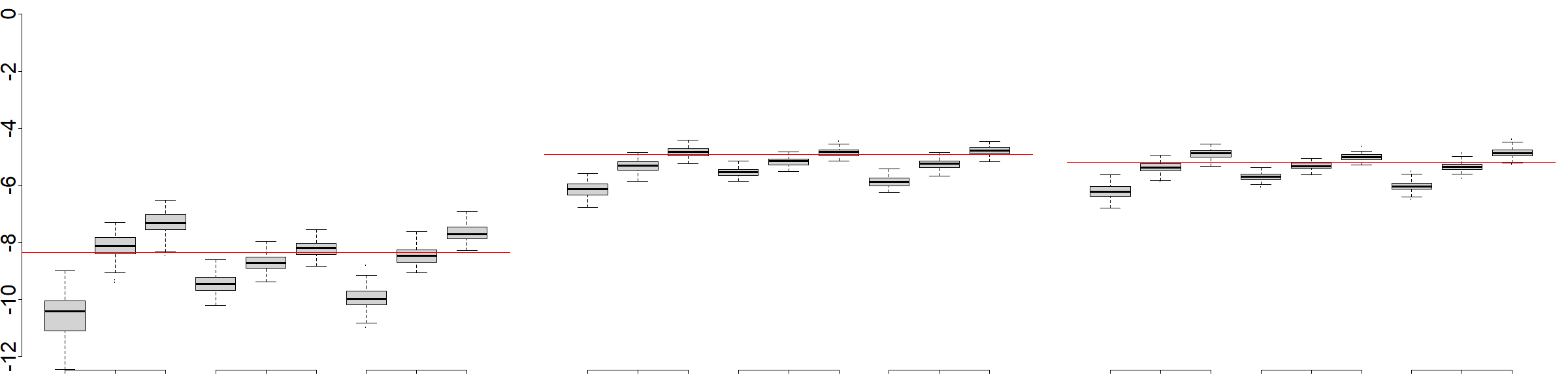}
      \put (12.5,-1) {\scalebox{0.5}{$\mathrm{CI}_{L}$}}
      \put (46,-1) {\scalebox{0.5}{$\mathrm{CI}_{L}$}}
      \put (79,-1) {\scalebox{0.5}{$\mathrm{CI}_{L}$}}
      
      \put (14.75,-1) {\scalebox{0.5}{$\log_{10}\widehat{\PR}_{\Rset_1}$}}
      \put (48.25,-1) {\scalebox{0.5}{$\log_{10}\widehat{\PR}_{\Rset_2}$}}
      \put (81.5,-1) {\scalebox{0.5}{$\log_{10}\widehat{\PR}_{\Rset_3}$}}
      
      \put (20,-1) {\scalebox{0.5}{$\mathrm{CI}_{U}$}}
      \put (53.5,-1) {\scalebox{0.5}{$\mathrm{CI}_{U}$}}
      \put (86.5,-1) {\scalebox{0.5}{$\mathrm{CI}_{U}$}}
      
      \put (6,2) {\footnotesize $\M_0$}
      \put (16,2) {\footnotesize $\M_1$}
      \put (26,2) {\footnotesize $\M_2$}
      \put (39,2) {\footnotesize $\M_0$}
      \put (49,2) {\footnotesize $\M_1$}
      \put (59,2) {\footnotesize $\M_2$}
      \put (72.5,2) {\footnotesize $\M_0$}
      \put (82.5,2) {\footnotesize $\M_1$}
      \put (92.5,2) {\footnotesize $\M_2$}
      
      \put (15.5,22.5) {$\Rset_1$}
      \put (16.5,20) {\rotatebox{90}{$=$}}
      \put (14.5,18.5) {\scalebox{.55}{$[10,\infty)^5$}}
      \put (48.5,22.5) {$\Rset_2$}
      \put (49.5,20) {\rotatebox{90}{$=$}}
      \put (34.5,18.5) {\scalebox{.55}{$(-\infty,\infty)\times[6,\infty)\times[8,\infty)\times[6,\infty)\times(-\infty,\infty)$}}
      \put (82.5,22.5) {$\Rset_3$}
      \put (83.5,20) {\rotatebox{90}{$=$}}
      \put (66,18.5) {\scalebox{.55}{$(-\infty,-7] \times(-\infty,0] \times (-\infty,-5] \times (-\infty,0] \times (-\infty,-7]$}}
   \end{overpic}

   \vspace{0.5em}
   \centering
   \begin{overpic}[width=\textwidth]{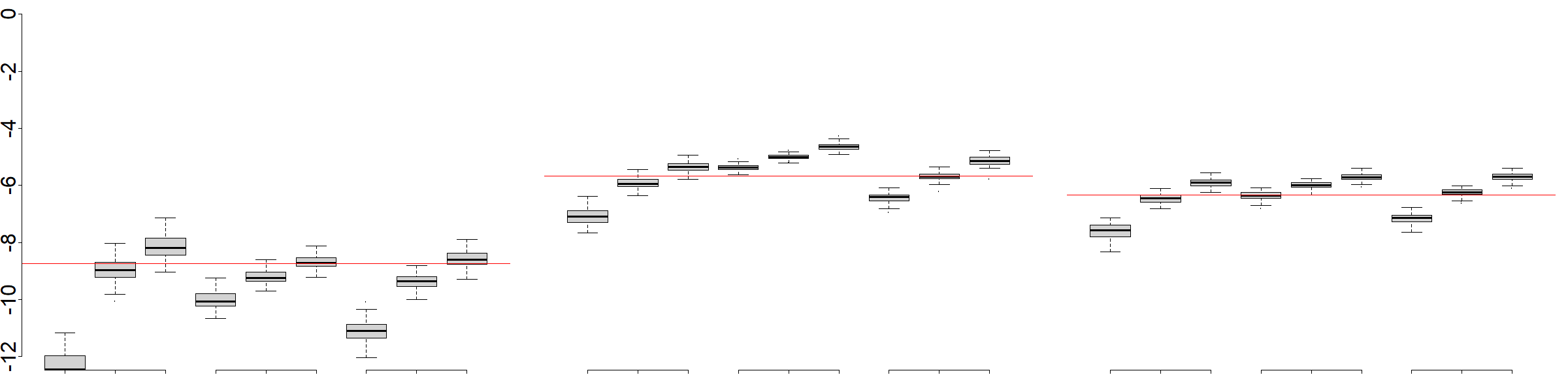}
      \put (12.5,-1) {\scalebox{0.5}{$\mathrm{CI}_{L}$}}
      \put (46,-1) {\scalebox{0.5}{$\mathrm{CI}_{L}$}}
      \put (79,-1) {\scalebox{0.5}{$\mathrm{CI}_{L}$}}
      
      \put (14.75,-1) {\scalebox{0.5}{$\log_{10}\widehat{\PR}_{\Rset_1}$}}
      \put (48.25,-1) {\scalebox{0.5}{$\log_{10}\widehat{\PR}_{\Rset_2}$}}
      \put (81.5,-1) {\scalebox{0.5}{$\log_{10}\widehat{\PR}_{\Rset_3}$}}
      
      \put (20,-1) {\scalebox{0.5}{$\mathrm{CI}_{U}$}}
      \put (53.5,-1) {\scalebox{0.5}{$\mathrm{CI}_{U}$}}
      \put (86.5,-1) {\scalebox{0.5}{$\mathrm{CI}_{U}$}}
      
      \put (6,2) {\footnotesize $\M_0$}
      \put (16,2) {\footnotesize $\M_1$}
      \put (26,2) {\footnotesize $\M_2$}
      \put (39,2) {\footnotesize $\M_0$}
      \put (49,2) {\footnotesize $\M_1$}
      \put (59,2) {\footnotesize $\M_2$}
      \put (72.5,2) {\footnotesize $\M_0$}
      \put (82.5,2) {\footnotesize $\M_1$}
      \put (92.5,2) {\footnotesize $\M_2$}
      
      \put (15.5,22.5) {$\Rset_1$}
      \put (16.5,20) {\rotatebox{90}{$=$}}
      \put (14.5,18.5) {\scalebox{.55}{$[10,\infty)^7$}}
      \put (48.5,22.5) {$\Rset_2$}
      \put (49.5,20) {\rotatebox{90}{$=$}}
      \put (39.5,18.5) {\scalebox{.55}{$[0, \infty)\times[0, \infty)\times[5, \infty)\times[5, \infty)$}}
      \put (41.5,17) {\scalebox{.55}{$\times[0, \infty)\times[8, \infty)\times[8, \infty)$}}
      \put (82.5,22.5) {$\Rset_3$}
      \put (83.5,20) {\rotatebox{90}{$=$}}
      \put (72.5,18.5) {\scalebox{.55}{$[ 6, \infty)\times[-2, \infty)\times(-\infty,  5]\times[ 6, \infty)$}}
      \put (74.5,17) {\scalebox{.55}{$\times[-2, \infty)\times(-\infty,  5]\times[ 6, \infty)$}}
   \end{overpic}
   \caption{Boxplots of 100 estimated probabilities $\widehat{\PR}_{\Rset_j}$, $j=1,2,3$, 
   and of the lower- ($\mathrm{CI}_L$) and 
   upper-bounds ($\mathrm{CI}_U$) of their 95\% bootstrap confidence intervals (on $\log_{10}$ 
   scale) for the sets 
   $\Rset_1,\Rset_2,\Rset_3$ using models $\M_0$, $\M_1$, and $\M_2$ fitted on 
   samples of size $n=10^4$ from multivariate normal distributions with $d=3$ (top row),
   $d=5$ (middle row), and $d=7$ (bottom row).\ 
   The true $\log_{10}$-probabilities are denoted by red lines.\ }
   \label{fig:logPr_d357_n10000}
\end{figure}

For dimensions $d\in\{3,5,7\}$, we draw 100 independent samples of size
$n=10^4$ and set the quantile set level to~$q=0.9$ in all settings to 
assess the effect of the dimension~$d$ on the model fitting and 
probability estimation procedures.\ We fit models~$\M_0$, $\M_1$, and~$\M_2$ to each
sample from each setting as well as to 250 bootstraped samples from 
each original sample, allowing us to perform probability estimation and 
build 95\% equal-tailed confidence intervals for the probability of hitting 
the risk regions~$\Rset_1$, $\Rset_2$, and~$\Rset_3$.\ 
Figure~\ref{fig:logPr_d357_n10000} displays
boxplots of the base-10 logarithm of 100 probability estimates and of lower- and 
upper-bounds of the 95\% equal-tailed bootstrap confidence intervals.\  
The results illustrate the bias--variance trade-off entailed by the  
parsimony--flexibility range of models~$\M_0$, $\M_1$,
and those that bridge them.\ Indeed, it is apparent from 
Figure~\ref{fig:logPr_d357_n10000}
that $\M_0$ has less bias than $\M_1$ and that $\M_1$ has narrower confidence intervals 
than $\M_0$; $\M_2$ appears to be a sensible compromise with accurate probability
estimates.\ Further, as expected, the bootstrap confidence
intervals for the probability estimates widen with increasing $d$,
reflecting the sparsification of data entailed by increasing dimension.\ 

Figure~\ref{fig:logPr_d7_n20000} in Section~\ref{sec:Supp_simulation_study} 
of the Supplementay material displays
the results of the same experiments for model $\M_2$ 
with $d\in\{3,5,7\}$, and alternative sample sizes of $n=5000$,
$20000$, and $20000$, as well as probability levels $q=0.8$, $0.95$, $0.95$.\ 
It illustrates that the method works well in lower dimensions on an 
extrapolation task corresponding to events way beyond the range of 
observed data, especially for~$\Rset_1$.\ 

Last, we apply our methodology to $d=10$ dimensions.\ 
Figure~\ref{fig:logPr_d10_n200000} 
of Section~\ref{sec:Supp_simulation_study} shows the results 
using samples sizes $n=5\times10^4$, $10^5$, and $2\times10^5$ 
and probability levels $q=0.9$, $0.95$, and $0.9625$.\ 
While results are not as accurate as in lower dimensions, they are still very decent
given the difficulty of the estimation task.\ 
Furthermore, we observe that the bias decreases when increasing 
$n$ and $q$.\ 

\section{Data application to low and high wind extremes}
\label{sec:case_study}

\subsection{Data and scientific problem}
\label{subseq:data_and_aims}

We apply our methodology to model low and high wind speed data
and estimate the probability of multivariate extreme 
events that may have significant impact on electricity production.\ 
The method allows modelling the extremal dependence of events at distinct sites 
in an unstructured manner.\ 
We show that 
our method can be used to select new sites at which to build wind farms---to 
minimise or maximise electricty production risk functionals---or 
to guide management of operations at existing stations 
especially when terrain or 
landmarks weaken the belief in spatial modelling assumptions.\ 

\begin{figure}[t!]
\centering
\begin{overpic}[width=0.8\textwidth]{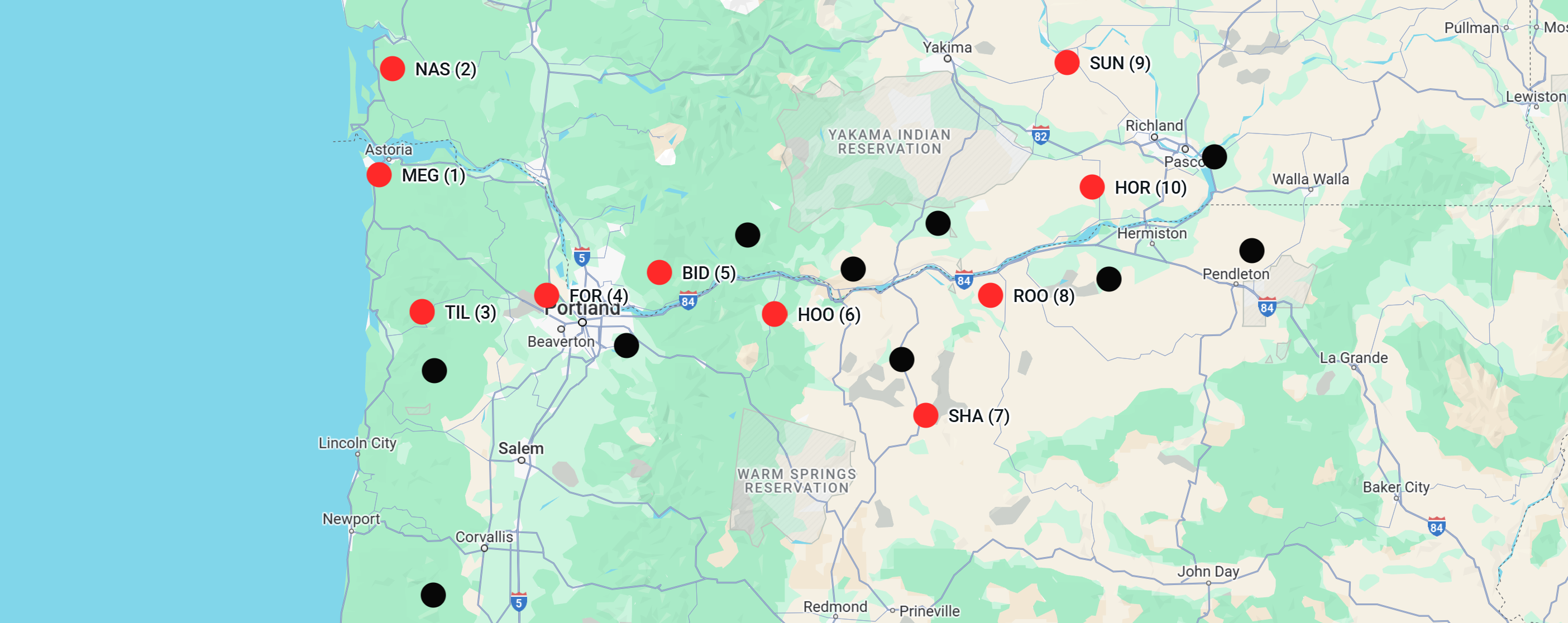}
\end{overpic}
\caption{
Stations (dots) over the Pacific Northwest region, US, 
between the Oregon and Washington states.\ 
Sites without missing data are in red, numbered from west to east.\ }
\label{fig:Data}
\end{figure}

The wind speeds are recorded hourly in meters per second (m/s) 
between 9:00 AM on January 1, 2012, and 8:00 AM on January 1, 2015, at 20
stations near the Oregon and Washington states' border.\ Figure~\ref{fig:Data}
displays the station locations over the region of the Pacific Northwest, United States (US), 
which has large wind energy resources~\citep{Hering2010Wind,Kazor2015wind}.\ 
The data were analysed 
by \cite{Huser2017bridgingGaussian} and \cite{CastroCamilo2019GenGamma}, 
adopting spatial approaches to model extremal dependence.\

We consider the GE 1.5 megawatt (MW) wind turbine---a baseline 
in the study of
wind speed forecasting by \cite{Hering2010Wind} over the same 
region---to define low and high extremes of wind speeds.\ 
According to the GE 1.5 MW power curve, wind speeds (m/s) can be partitioned 
into four intervals $I_1=[0,3.5)$, $I_2=[3.5,13.5)$, $I_3=[13.5,25)$, and 
$I_3=[25,\infty)$, with~$I_1$ and~$I_4$ corresponding to too low 
and too high wind speeds leading to no electricity production
and $I_3$ corresponding to maximal power being produced.\ 

\subsection{Data preprocessing}
\label{subseq:data_preprocessing}

To avoid missingness complications---whose treatment lies outside the scope of 
this exposi\-tion---we consider the 10 stations at which there 
are no missing observations over the whole time period.\ These are 
coloured red and numbered from west to east in Figure~\ref{fig:Data}; their acronyms
are detailed in Table~\ref{tbl:stations} of Section~\ref{sec:Supp_case_study}.\ 
We denote by $\underline{\bm x}^{\mathrm{o}} = 
\{\bm x_i^{\mathrm{o}} = (x_{i,1}^{\mathrm{o}},\ldots,x_{i,10}^{\mathrm{o}})
\in\RR^{10}_{\geq 0}:i=1,\ldots,n\}$ the collection assumed to have been observed from 
the climatology random vector $\bm X^{\mathrm{o}}\in\RR^{10}$.\ 

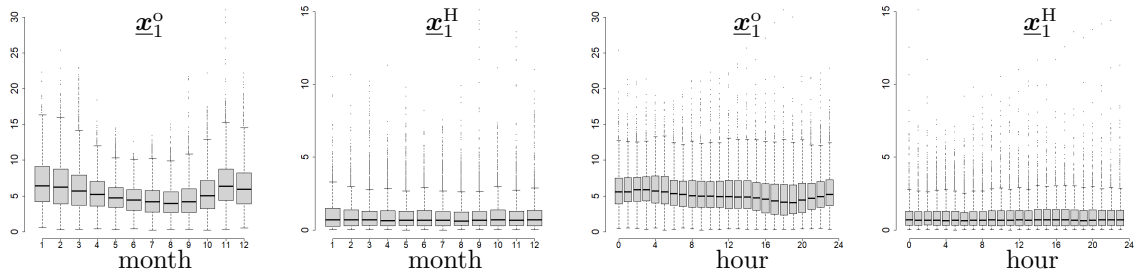
\begin{figure}[t!]
   \centering
   \begin{overpic}[width=0.95\textwidth]{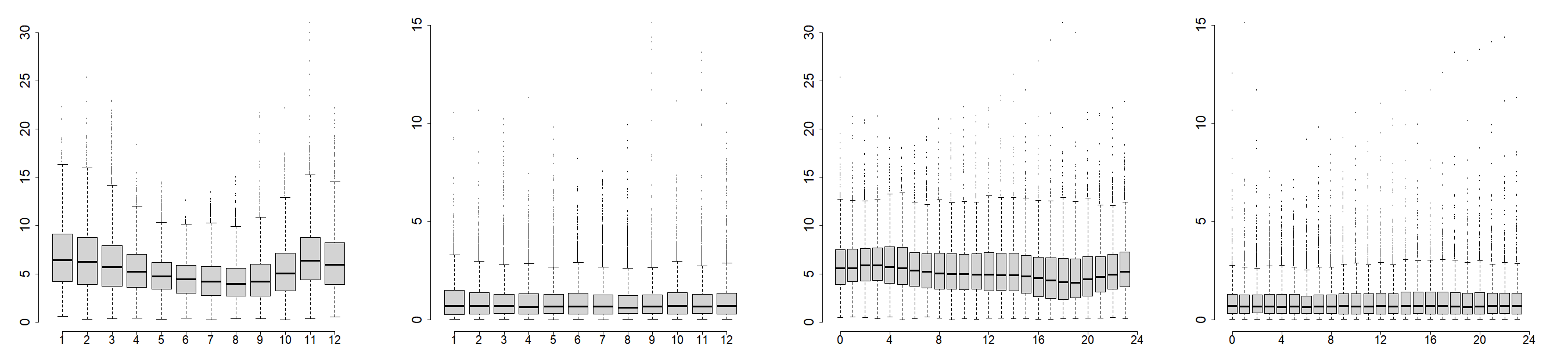}
      \put (10.5,-1.5) {{\footnotesize month}}
      \put (35.5,-1.5) {{\footnotesize month}}
      \put (62,-1.5) {{\footnotesize hour}}
      \put (87,-1.5) {{\footnotesize hour}}
   
      \put (12,19) {{\footnotesize $\underline{\bm x}^{\mathrm{o}}_1$}}
      \put (37,19) {{\footnotesize $\underline{\bm x}^{\mathrm{H}}_1$}}
      \put (63.5,19) {{\footnotesize $\underline{\bm x}^{\mathrm{o}}_1$}}
      \put (88.5,19) {{\footnotesize $\underline{\bm x}^{\mathrm{H}}_1$}}
   \end{overpic}
   \caption{Boxplots of the monthly- and hourly-aggregated original 
   wind speeds (m/s)~$\underline{\bm x}^{\mathrm{o}}_1$ (pannels~1-3) and of the 
   homogenised data~$\underline{\bm x}^{\mathrm{H}}_1$ (pannels~2-4), respectively, for site~MEG.}
   \label{fig:gumbel_detrending}
   \end{figure}

The data display seasonality patterns which can be observed on monthly and 
hourly scales, see Figure~\ref{fig:gumbel_detrending}.\ 
We adopt a site-wise modelling of wind speeds using the generalised additive model
\citep[GAM;][]{Wood2011GAM} framework.\ 
Following classical approaches \citep{Elliott2004WindWeibull}, we 
model the site-wise wind speed random variable $X_{j,m,h}^{\mathrm{o}}$,  
at station $j$ in month $m$ of the year and hour $h$ of the day,
under the working assumption that it follows a Weibull distribution 
with scale~$\lambda_{j,m,h}>0$ and shape~$\kappa_{j,m,h}>0$.\ More precisely, 
\begin{IEEEeqnarray}{rCl}
\label{eq:gumbel_loc_scale}
X_{j,m,h}^{\mathrm{o}} \sim 
\text{Weibull}\big(\lambda_{j,m,h}=s_{j,1}(m)+s_{j,2}(h)\,,\,\kappa_{j,m,h}=s_{j,3}(m)+s_{j,4}(h)\big),
\end{IEEEeqnarray}
where $s$ denotes a cubic cyclic spline on $m\in\{1,\ldots,12\}$ or 
$h\in\{0,\ldots,23\}$.\ 

We fit model~\eqref{eq:gumbel_loc_scale} using the \texttt{evgam} package \citep{Youngman2022evgam}
from the \textsf{R} programming language and obtain estimates~$\hat{\lambda}_{j,m,h}$ 
and~$\hat{\kappa}_{j,m,h}$ 
for all months~$m$
and hours~$h$.\ We obtain homogenised data 
$\underline{\bm x}^{\mathrm{H}} = 
\{\bm x_i^{\mathrm{H}} = (x_{i,1}^{\mathrm{H}},\ldots,x_{i,10}^{\mathrm{H}})
:i=1,\ldots,n\}$ via $x^{\mathrm{H}}_{i,j}:= 
(x^{\mathrm{o}}_{i,j}/\hat{\lambda}_{j,m,h})^{\hat{\kappa}_{j,m,h}}$
for the month~$m$ and hour~$h$ of~$\bm x_{i}$.\  
Figure~\ref{fig:gumbel_detrending} shows monthly- and 
hourly-aggregated boxplots of the homogenised $\underline{\bm x}^{\mathrm{H}}_1$ at site~MEG,
where $\underline{\bm x}^{\mathrm{H}}_j:=\{x^{\mathrm{H}}_{i,j}\,:\, 
i=1,\ldots,n\}$, 
being much more homogeneous
in time than their original-scale counterparts.\ 
Autocorrelation function plots in Figure~\ref{fig:ACF} of 
Section~\ref{sec:Supp_case_study} also support this:\ 
the $\underline{\bm x}^{\mathrm{H}}$ remain autocorrelated 
for up to approximately four days, considerably less than $\underline{\bm x}^{\mathrm{o}}$.\  

Since wind speeds may not follow a Weibull distribution exactly---with 
misspecification particularly pronounced
in the lower and upper tails \citep{Gunturu2012WindWeibull}---we 
apply the marginal model detailed in Section~\ref{subsec:Marginal_model}
to the homogenised observations 
$\underline{\bm x}^{\mathrm{H}}$.\
Due to the known marginal lower-bound of 0 m/s windspeeds, 
we only model the upper-tails and set the threshold~$u_{j,+}$ 
to the 0.995 empirical quantile
of $\underline{\bm x}^{\mathrm{H}}_j$, respectively, for $j=1,\ldots,10$.\ 
Figure~\ref{fig:MarginalGP} in Section~\ref{sec:Supp_case_study}
displays Quantile-Quantile plots of all upper-tail fits, which suggest that the marginal fits 
are all satisfactory.\ We thus obtain  
$\underline{\bm x}:=\{\bm x_{i}:=
(\hat{\varphi}_1(x^\mathrm{H}_{i,1}),\ldots,\hat{\varphi}_d(x^\mathrm{H}_{i,d}))
\,:\,i=1,\ldots, n\}$.\

\afterpage{
   \begin{figure}[t!]
      \centering
      \begin{overpic}[width=0.48\textwidth]{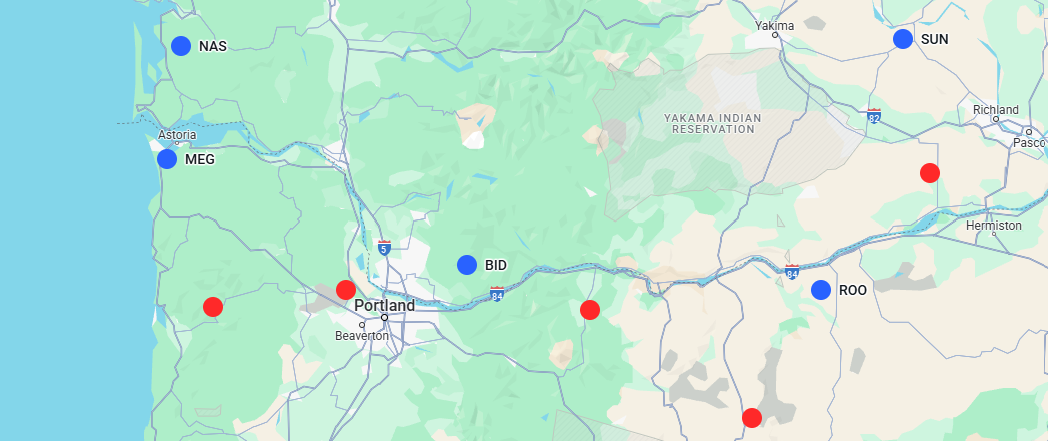}
      \end{overpic}
      \begin{overpic}[width=0.482\textwidth]{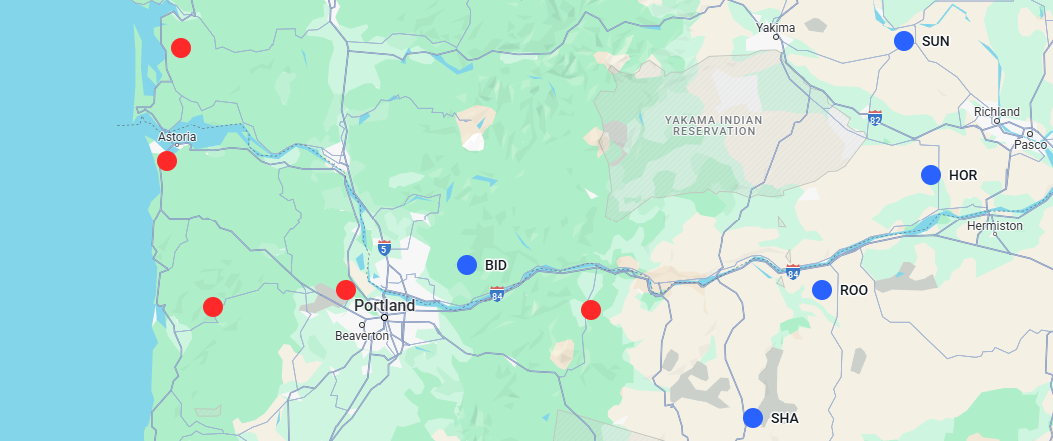}
      \end{overpic}
      \caption{Configurations of five stations (blue) minimising the risk of no production (left) 
      and maximising the probability of full production (right) for month $m=1$ and hour $h=18$.\ }
      \label{fig:config_min_max}
   \end{figure}
   \begin{table}[t!]
      \begin{center}
      \caption{Probability estimates and 95\% block bootstrap confidence intervals of the 
      three configuration of five stations minimising the probability of no 
      production and maximising that of full production for month $m=1$ and hour $h=18$.\ } 
      \begin{tabular}{ccccc}
         \toprule \toprule
         \textbf{Event} & \textbf{Rank} & $\widehat{\PR}_{\bm X_J}(E_{1,18}^{(e,\mathcal{I})})$ & $\text{CI}_{0.95}$ &  \textbf{Stations $\mathcal{I}$} \\
         \midrule
         No production 
         & \scalebox{.8}{1} & \scalebox{.8}{0.0298} & \scalebox{.8}{[0.0233, 0.0354]} & \scalebox{.8}{BID, MEG, NAS, ROO, SUN} \\
         ($E_{1,18}^{(1,\mathcal{I})}$) & \scalebox{.8}{2} & \scalebox{.8}{0.0313} & \scalebox{.8}{[0.0254, 0.0370]} & \scalebox{.8}{BID, NAS, ROO, SHA, SUN} \\
         & \scalebox{.8}{3} & \scalebox{.8}{0.0332} & \scalebox{.8}{[0.0283, 0.0376]} & \scalebox{.8}{BID, HOO, NAS, ROO, SUN} \\
         \midrule
         Full production 
         & \scalebox{.8}{1} & \scalebox{.8}{$1.82{\times}10^{-4}$} & \scalebox{.8}{$[4.64{\times}10^{-5}, 3.73{\times}10^{-4}]$} & \scalebox{.8}{BID, HOR, ROO, SHA, SUN} \\ 
         ($E_{1,18}^{(2,\mathcal{I})}$)& \scalebox{.8}{2} & \scalebox{.8}{$1.14{\times}10^{-4}$} & \scalebox{.8}{$[3.99{\times}10^{-5}, 2.41{\times}10^{-4}]$} & \scalebox{.8}{HOR, MEG, ROO, SHA, SUN} \\
         & \scalebox{.8}{3} & \scalebox{.8}{$1.04{\times}10^{-4}$} & \scalebox{.8}{$[1.95{\times}10^{-5}, 3.57{\times}10^{-4}]$} & \scalebox{.8}{MEG, NAS, SHA, SUN, TIL} \\
         \bottomrule
      \end{tabular}
      \label{tbl:config_min_max}
      \end{center}
   \end{table}
}

\subsection{Geometric extremes modelling}
\label{subseq:data_GE_modelling}

A first event of interest is 
the event $E_{m,h}^{(1,\mathcal{I})}:=\{X_{j,m,h}^{\mathrm{o}}\in I_1\cup I_4, \,\forall\, j\in \mathcal{I}\}$ 
that a configuration of~$\# \mathcal{I}$ stations denoted by the index set 
$\mathcal{I}\subseteq\{1,\ldots,d\}$ does not produce any electricity at 
hour~$h$ of a given day of month~$m$.\ A second event of interest is that 
all~$\# \mathcal{I}$ stations in configuration~$\mathcal{I}$ produce maximal power, that is,
$E_{m,h}^{(2,\mathcal{I})}:=\{X_{j,m,h}^{\mathrm{o}}\in I_3, \,\forall\, j\in \mathcal{I}\}$.\ 
Both events can be interpreted in relation to hyperboxes for which 
the radial integration bounds of equation~\eqref{eq:MR_exp} 
can respectively be obtained from 
Sections~\ref{sec:outside_starbodyhyperbox} and~\ref{sec:hyperbox}.\ 

To estimate the probability of these events, we transform the hyperboxes from the original
scale of the data~$\underline{\bm x}^{\mathrm{o}}$ to the approximate standard 
Laplace scale of the data~$\underline{\bm x}$.\ 
For a general hyperbox
$\Rset:=[\Rset_{1,l},\Rset_{1,u}]\times\cdots\times[\Rset_{d,l},\Rset_{d,u}]$,
we obtain the associated hyperbox~$\Rset_{L}^{m,h}$ for month~$m$ and hour~$h$ in the approximate 
standard Laplace scale via 
\begin{IEEEeqnarray}{rCl}
\label{eq:prob_est_loc_scale}
\Rset_{L}^{m,h} = 
   \prod_{j=1}^d\left[\hat{\varphi}_j\left(\left(\Rset_{j,l}/\hat{\lambda}_{j,m,h}\right)^{\hat{\kappa}_{j,m,h}}\right),
   \hat{\varphi}_j\left(\left(\Rset_{j,u}/\hat{\lambda}_{j,m,h}\right)^{\hat{\kappa}_{j,m,h}}\right)\right].
\end{IEEEeqnarray}

We focus on model~$\M_2$ and use the same 
normalising flow architecture as that described in Section~\ref{sec:Sim_Study}, 
having demonstrated its applicability to the range of dimensions studied.\ 
We set the probability level of~$\Qq$ to $q=0.95$, and similarly to the simulation study,
we rely on mollification and early stopping, 
partitioning~$\underline{\bm x}$ into the 
training and validation sets~$\underline{\bm x}_{\mathrm{T}}$ 
and~$\underline{\bm x}_{\mathrm{V}}$ according to a 0.7--0.3 
proportion split.\ Due to the temporal autocorrelation 
in $\underline{\bm x}^{\mathrm{H}}$
(see Section~\ref{subseq:data_preprocessing}), 
and thus in $\underline{\bm x}$, we build confidence intervals for our probability estimates
via temporal block bootstrapping.\ We form blocks of $7\times24$ hours to retain sequences of 
7 days of observations which are representative of the temporal dependence 
in~$\underline{\bm x}$, and fit model~$\M_2$ to all $\binom{10}{5}=252$ 
configurations of $\#\mathcal{I}=5$ stations.\ Figure~\ref{fig:QQ_models} of 
Section~\ref{sec:Supp_case_study} displays
goodness-of-fit plots for a subset of configurations.\

Figure~\ref{fig:config_min_max} and Table~\ref{tbl:config_min_max} show,  
for the winter month of January ($m=1$) at a high electrity 
consumption time of 6:00 PM ($h=18$), the configurations of five 
stations minimising the probability of producing no electricity 
and those maximising the probability of full electricity production.\ 
In particular, this latter configuration
contains four of the five stations located at the highest altitudes where 
there are less terrain obstables and stronger winds.\ 
Figure~\ref{fig:config_worse} of Section~\ref{sec:Supp_case_study} 
shows, for the same month $m$ and hour $h$, the configuration 
with the highest estimated 
probability $0.273$ of suffering no electricity production, 
illustrating the important risk gap among different configurations.\ 
The west-east alignment of stations in this configuration reflects 
the correlation identified by \cite{Huser2017bridgingGaussian}
between stations aligned in the direction of dominant winds.\

\begin{figure}[t!]
   \centering
   \begin{overpic}[width=\textwidth]{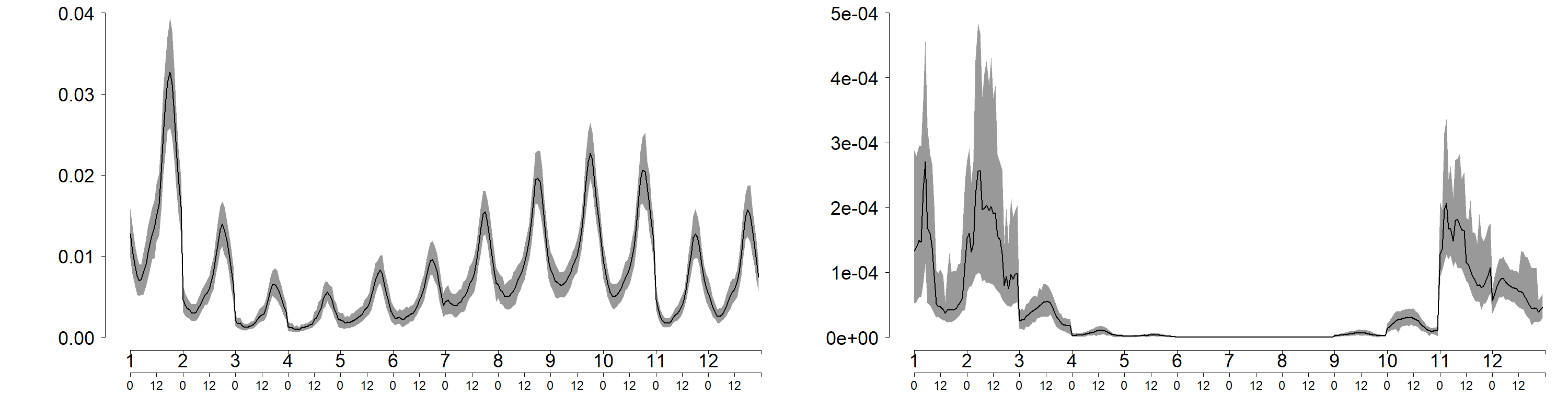}
   \put(4.1,3.3){\scalebox{.6}{$m=$}}
   \put(4.5,1.8){\scalebox{.5}{$h=$}}

   \put(54.1,3.3){\scalebox{.6}{$m=$}}
   \put(54.5,1.8){\scalebox{.5}{$h=$}}

   \put(0,12){\rotatebox{90}{\scalebox{0.8}{$\widehat{\PR}_{\bm X_J}(E_{m,h}^{(1,\mathcal{I})})$}}}
   \put(49.5,12){\rotatebox{90}{\scalebox{0.8}{$\widehat{\PR}_{\bm X_J}(E_{m,h}^{(2,\mathcal{I})})$}}}

   \put(20,0){\scalebox{0.8}{Month $m$, Hour $h$}}
   \put(70,0){\scalebox{0.8}{Month $m$, Hour $h$}}
   \end{overpic}
   \caption{Estimated probability (solid line) and 95\% block bootstrap confidence 
   intervals (shaded band) of the event $E_{m,h}^{(1,\mathcal{I})}$ of no power production (left)
   and of the event $E_{m,h}^{(2,\mathcal{I})}$ of full power production (right) as a function of the 
   month $m$ and the hour $h$ at observation stations $\mathcal{I}=\{$BID, MEG, NAS, ROO, SUN$\}$.\ }
   \label{fig:Probs_MH}
\end{figure}
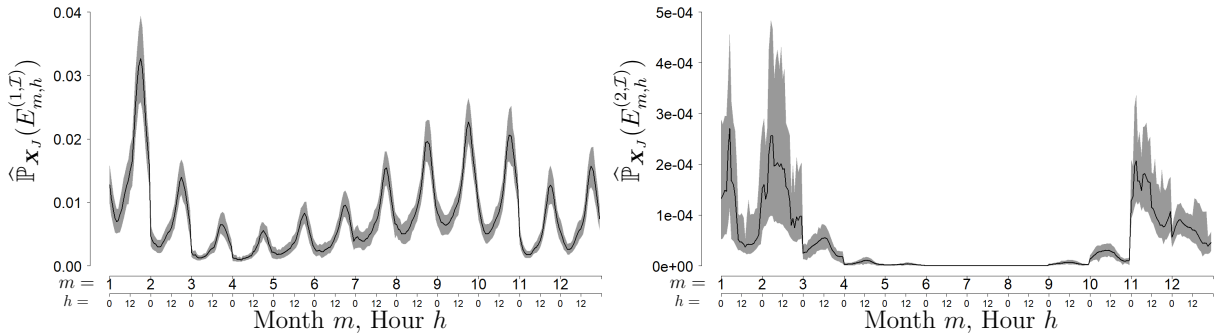

For the configuration $\mathcal{I}{=}\{$BID, MEG, NAS, ROO, SUN$\}$ of stations 
which minimises $\widehat{\PR}_{\bm X_J}(E_{1,18}^{(1,\mathcal{I})})$, we estimate 
$\{\PR_{\bm X_J}(E_{m,h}^{(e,\mathcal{I})}):m=1,\ldots,12;\,h=0,\ldots,23;\, e = 1,2\}$,
the probabilities of observing no or full electricity production 
for each month $m$ and hour $h$.\
Estimates and their associated 95\% block bootstrap confidence intervals 
are presented in Figure~\ref{fig:Probs_MH}.\ 
The results display clear periodicity along the different months and 
times of day the events defined
in relation to wind power production.\ This highlights the potential 
usefulness of our methods to guide operation management of 
electricity production.\

\section{Discussion}

Using normalising flows as models for probability densities,
we tackled the difficult task of inferring the distribution 
of directions of multivariate extreme
events projected on hyperspheres and built identifiable geometric extremes models 
by combining the shapes of structural star-body parameters.\ 
Leveraging the generative property of normalising flows, we devised a fast and accurate
method for the estimation of multivariate extreme events---made even faster via  
exact integration on radial intervals with explicit bounds whenever 
the extreme events of interest can be characterised by hyperboxes.\ As 
opposed to classical multivariate extemes frameworks such as those based on  
the multivariate GP distribution or regular
variation, our models can extrapolate in all directions of the multivariate space.\

While we adopted the more-parsimonious yet accurate exponential distribution to model 
the radial exceedances of the quantile set, our methodology 
could trivially be adapted to use the truncated gamma distribution, akin to
\cite{wadsworth2022statistical}.\
Future work could also consider the generalised Pareto model, enforcing regularity on 
a varying shape function using a penalisation method similar to that used on deformation sets,
or the use of normalising flows for conditional density estimation to model
non-stationarity.\ 

The code to use our methodology is available at \url{https://github.com/lambertdem}.\ 

\section*{Acknowledgments}
Part of this paper was written during
a visit of LDM at King Abdullah University of Science and~Technology 
(KAUST).\ LDM thanks funding from the School of 
Mathematics, University of Edinburgh~(UoE).\ We
acknowledge computing support from the Eddie (UoE) and Ibex (KAUST)
high-performance computing clusters.\ 

\bibliographystyle{Chicago}

\bibliography{bibliography}

\begin{thebibliography}{}

\bibitem[\protect\citeauthoryear{Balkema, Embrechts, and Nolde}{Balkema et~al.}{2010}]{balkema2010metadensities}
Balkema, A., P.~Embrechts, and N.~Nolde (2010).
\newblock Meta densities and the shape of their sample clouds.
\newblock {\em J. Multivar. Anal.\/}~{\em 101\/}(7), 1738--1754.

\bibitem[\protect\citeauthoryear{Balkema and Nolde}{Balkema and Nolde}{2010}]{balkema_nolde_2010}
Balkema, G. and N.~Nolde (2010).
\newblock Asymptotic independence for unimodal densities.
\newblock {\em Adv. App. Probab.\/}~{\em 42\/}(2), 411--432.

\bibitem[\protect\citeauthoryear{Balkema and Nolde}{Balkema and Nolde}{2012}]{balkema_nolde_2012}
Balkema, G. and N.~Nolde (2012).
\newblock Asymptotic dependence for light-tailed homothetic densities.
\newblock {\em Adv. App. Probab.\/}~{\em 44\/}(2), 506–527.

\bibitem[\protect\citeauthoryear{Campbell and Wadsworth}{Campbell and Wadsworth}{2024}]{campbell2024piecewiselinear}
Campbell, R. and J.~Wadsworth (2024).
\newblock Piecewise-linear modeling of multivariate geometric extremes.
\newblock {\em arXiv:2412.05195\/}.

\bibitem[\protect\citeauthoryear{Castro-Camilo, Huser, and Rue}{Castro-Camilo et~al.}{2019}]{CastroCamilo2019GenGamma}
Castro-Camilo, D., R.~Huser, and H.~Rue (2019, Sep).
\newblock A spliced gamma-generalized {Pareto} model for short-term extreme wind speed probabilistic forecasting.
\newblock {\em J. Agric. Biol. Environ. Stat.\/}~{\em 24\/}(3), 517--534.

\bibitem[\protect\citeauthoryear{Davis, Mulrow, and Resnick}{Davis et~al.}{1988}]{davis1988almost}
Davis, R.~A., E.~Mulrow, and S.~I. Resnick (1988).
\newblock Almost sure limit sets of random samples in $\mathbb{R}^d$.
\newblock {\em Adv. App. Probab.\/}~{\em 20\/}(3), 573--599.

\bibitem[\protect\citeauthoryear{Dinh, Sohl-Dickstein, and Bengio}{Dinh et~al.}{2017}]{dinh2017density}
Dinh, L., J.~Sohl-Dickstein, and S.~Bengio (2017).
\newblock Density estimation using real {NVP}.
\newblock In {\em ICLR}.

\bibitem[\protect\citeauthoryear{Durkan, Bekasov, Murray, and Papamakarios}{Durkan et~al.}{2019}]{Durkan2019NeuralSplineFlow}
Durkan, C., A.~Bekasov, I.~Murray, and G.~Papamakarios (2019).
\newblock Neural spline flows.
\newblock In H.~Wallach, H.~Larochelle, A.~Beygelzimer, F.~d\textquotesingle Alch\'{e}-Buc, E.~Fox, and R.~Garnett (Eds.), {\em Adv. Neural Inf. Process. Syst.}, Volume~32. Curran Associates, Inc.

\bibitem[\protect\citeauthoryear{Elliott, Schwartz, and Scott}{Elliott et~al.}{2004}]{Elliott2004WindWeibull}
Elliott, D., M.~Schwartz, and G.~Scott (2004).
\newblock Wind resource base.
\newblock In C.~J. Cleveland (Ed.), {\em Encyclopedia of Energy}, pp.\  465--479. New York: Elsevier.

\bibitem[\protect\citeauthoryear{Fisher}{Fisher}{1969}]{Fisher1069limitconvex}
Fisher, L. (1969).
\newblock {Limiting Sets and Convex Hulls of Samples from Product Measures}.
\newblock {\em Ann. Math. Stat.\/}~{\em 40\/}(5), 1824 -- 1832.

\bibitem[\protect\citeauthoryear{Geffroy}{Geffroy}{1958}]{geffroy1958}
Geffroy, J. (1958).
\newblock {Contribution {\`a} la th{\'e}orie des valeurs extr{\^e}mes}.
\newblock {\em {Ann. ISUP}\/}~{\em VII\/}(1), 37--121.

\bibitem[\protect\citeauthoryear{Geffroy}{Geffroy}{1959}]{geffroy1959suite}
Geffroy, J. (1959).
\newblock {Contribution {\`a} la th{\'e}orie des valeurs extr{\^e}mes (Suite)}.
\newblock {\em {Ann. ISUP}\/}~{\em VIII\/}(1), 3--65.

\bibitem[\protect\citeauthoryear{Gunturu and Schlosser}{Gunturu and Schlosser}{2012}]{Gunturu2012WindWeibull}
Gunturu, U.~B. and C.~A. Schlosser (2012).
\newblock Characterization of wind power resource in the {United States}.
\newblock {\em Atmospheric Chemistry and Physics\/}~{\em 12\/}(20), 9687--9702.

\bibitem[\protect\citeauthoryear{Hansen, Herburt, Martini, and Moszy{\'n}ska}{Hansen et~al.}{2020}]{hansen2020starshaped}
Hansen, G., I.~Herburt, H.~Martini, and M.~Moszy{\'n}ska (2020).
\newblock Starshaped sets.
\newblock {\em Aequationes {M}athematicae\/}~{\em 94}, 1001--1092.

\bibitem[\protect\citeauthoryear{Hering and Genton}{Hering and Genton}{2010}]{Hering2010Wind}
Hering, A.~S. and M.~G. Genton (2010).
\newblock Powering up with space-time wind forecasting.
\newblock {\em J. Am. Stat. Assoc.\/}~{\em 105\/}(489), 92--104.

\bibitem[\protect\citeauthoryear{Hickling and Prangle}{Hickling and Prangle}{2023}]{hickling2023flexible}
Hickling, T. and D.~Prangle (2023).
\newblock Flexible tails for normalising flows, with application to the modelling of financial return data.
\newblock {\em arXiv:2311.00580\/}.

\bibitem[\protect\citeauthoryear{Hu and Castro-Camilo}{Hu and Castro-Camilo}{2025}]{hu2025gpdflowgenerativemultivariatethreshold}
Hu, C. and D.~Castro-Camilo (2025).
\newblock {GPDF}low: Generative multivariate threshold exceedance modeling via normalizing flows.
\newblock {\em arXiv:2503.11822\/}.

\bibitem[\protect\citeauthoryear{Huser, Opitz, and Thibaud}{Huser et~al.}{2017}]{Huser2017bridgingGaussian}
Huser, R., T.~Opitz, and E.~Thibaud (2017).
\newblock Bridging asymptotic independence and dependence in spatial extremes using {G}aussian scale mixtures.
\newblock {\em Spatial Statistics\/}~{\em 21}, 166--186.

\bibitem[\protect\citeauthoryear{Kazor and Hering}{Kazor and Hering}{2015}]{Kazor2015wind}
Kazor, K. and A.~S. Hering (2015).
\newblock The role of regimes in short-term wind speed forecasting at multiple wind farms.
\newblock {\em Stat\/}~{\em 4\/}(1), 271--290.

\bibitem[\protect\citeauthoryear{Kingma and Ba}{Kingma and Ba}{2017}]{kingma2017adammethodstochasticoptimization}
Kingma, D.~P. and J.~Ba (2017).
\newblock Adam: A method for stochastic optimization.
\newblock {\em arXiv:1412.6980\/}.

\bibitem[\protect\citeauthoryear{Kinoshita and Resnick}{Kinoshita and Resnick}{1991}]{kinoshta1991convergence}
Kinoshita, K. and S.~I. Resnick (1991).
\newblock Convergence of scaled random samples in $\mathbb{R}^d$.
\newblock {\em Ann. Probab.\/}~{\em 19\/}(4), 1640--1663.

\bibitem[\protect\citeauthoryear{Kobyzev, Prince, and Brubaker}{Kobyzev et~al.}{2021}]{Kobyzev_2021}
Kobyzev, I., S.~J. Prince, and M.~A. Brubaker (2021).
\newblock Normalizing flows: an introduction and review of current methods.
\newblock {\em {IEEE} Trans. Pattern Anal. Mach. Intell.\/}~{\em 43\/}(11), 3964–3979.

\bibitem[\protect\citeauthoryear{Koenker and Bassett}{Koenker and Bassett}{1978}]{Koenker1978}
Koenker, R. and G.~Bassett (1978).
\newblock Regression quantiles.
\newblock {\em Econometrica\/}~{\em 46\/}(1), 33--50.

\bibitem[\protect\citeauthoryear{Lhaut, Rootzén, and Segers}{Lhaut et~al.}{2025}]{lhaut2025wassersteinaitchisonganangularmeasures}
Lhaut, S., H.~Rootzén, and J.~Segers (2025).
\newblock Wasserstein-{A}itchison {GAN} for angular measures of multivariate extremes.
\newblock {\em arXiv:2504.21438\/}.

\bibitem[\protect\citeauthoryear{Mackay and Jonathan}{Mackay and Jonathan}{2024}]{mackay2024modelling}
Mackay, E. and P.~Jonathan (2024).
\newblock Modelling multivariate extremes through angular-radial decomposition of the density function.
\newblock {\em arXiv:2310.12711\/}.

\bibitem[\protect\citeauthoryear{Mackay, Murphy-Barltrop, Richards, and Jonathan}{Mackay et~al.}{2024}]{mackay2024deeplearningjointextremes}
Mackay, E., C.~Murphy-Barltrop, J.~Richards, and P.~Jonathan (2024).
\newblock Deep learning joint extremes of metocean variables using the {SPAR} model.
\newblock {\em arXiv:2412.15808\/}.

\bibitem[\protect\citeauthoryear{Majumder, Shaby, Reich, and Cooley}{Majumder et~al.}{2025}]{Majumder2025semiparametric}
Majumder, R., B.~A. Shaby, B.~J. Reich, and D.~S. Cooley (2025).
\newblock {Semiparametric Estimation of the Shape of the Limiting Bivariate Point Cloud}.
\newblock {\em Bayesian Anal.\/}, 1 -- 27.

\bibitem[\protect\citeauthoryear{Murphy-Barltrop, Mackay, and Jonathan}{Murphy-Barltrop et~al.}{2024}]{murphybarltrop2024inference}
Murphy-Barltrop, C. J.~R., E.~Mackay, and P.~Jonathan (2024).
\newblock Inference for multivariate extremes via a semi-parametric angular-radial model.
\newblock {\em arXiv:2401.07259\/}.

\bibitem[\protect\citeauthoryear{Murphy-Barltrop, Majumder, and Richards}{Murphy-Barltrop et~al.}{2024}]{murphybarltrop2024deepgauge}
Murphy-Barltrop, C. J.~R., R.~Majumder, and J.~Richards (2024).
\newblock Deep learning of multivariate extremes via a geometric representation.
\newblock {\em arXiv:2406.19936\/}.

\bibitem[\protect\citeauthoryear{Ng and Zammit-Mangion}{Ng and Zammit-Mangion}{2024}]{Ng2024}
Ng, T. L.~J. and A.~Zammit-Mangion (2024).
\newblock Mixture modeling with normalizing flows for spherical density estimation.
\newblock {\em Adv. Data Anal. Classif.\/}~{\em 18\/}(1), 103--120.

\bibitem[\protect\citeauthoryear{Nolde}{Nolde}{2014}]{nolde2014geometric}
Nolde, N. (2014).
\newblock Geometric interpretation of the residual dependence coefficient.
\newblock {\em J. Multivar. Anal.\/}~{\em 123}, 85--95.

\bibitem[\protect\citeauthoryear{Nolde and Wadsworth}{Nolde and Wadsworth}{2022}]{nolde_wadsworth_2022}
Nolde, N. and J.~L. Wadsworth (2022).
\newblock Linking representations for multivariate extremes via a limit set.
\newblock {\em Adv. App. Probab.\/}~{\em 54\/}(3), 688–717.

\bibitem[\protect\citeauthoryear{Papamakarios, Nalisnick, Rezende, Mohamed, and Lakshminarayanan}{Papamakarios et~al.}{2021}]{Papamakarios2021NF}
Papamakarios, G., E.~Nalisnick, D.~J. Rezende, S.~Mohamed, and B.~Lakshminarayanan (2021).
\newblock Normalizing flows for probabilistic modeling and inference.
\newblock {\em J. Mach. Learn. Res.\/}~{\em 22\/}(1).

\bibitem[\protect\citeauthoryear{Papamakarios, Pavlakou, and Murray}{Papamakarios et~al.}{2017}]{papamakarios2017masked}
Papamakarios, G., T.~Pavlakou, and I.~Murray (2017).
\newblock Masked autoregressive flow for density estimation.
\newblock In I.~Guyon, U.~V. Luxburg, S.~Bengio, H.~Wallach, R.~Fergus, S.~Vishwanathan, and R.~Garnett (Eds.), {\em Adv. Neural Inf. Process. Syst.}, Volume~30. Curran Associates, Inc.

\bibitem[\protect\citeauthoryear{Papastathopoulos, {De Monte}, Campbell, and Rue}{Papastathopoulos et~al.}{2025}]{papastathopoulos2025statistical}
Papastathopoulos, I., L.~{De Monte}, R.~Campbell, and H.~Rue (2025).
\newblock Statistical inference for radial generalized pareto distributions and return sets in geometric extremes.
\newblock {\em arXiv:2310.06130\/}.

\bibitem[\protect\citeauthoryear{Paszke, Gross, Massa, Lerer, Bradbury, Chanan, Killeen, Lin, Gimelshein, Antiga, Desmaison, Köpf, Yang, DeVito, Raison, Tejani, Chilamkurthy, Steiner, Fang, Bai, and Chintala}{Paszke et~al.}{2019}]{paszke2019pytorch}
Paszke, A., S.~Gross, F.~Massa, A.~Lerer, J.~Bradbury, G.~Chanan, T.~Killeen, Z.~Lin, N.~Gimelshein, L.~Antiga, A.~Desmaison, A.~Köpf, E.~Yang, Z.~DeVito, M.~Raison, A.~Tejani, S.~Chilamkurthy, B.~Steiner, L.~Fang, J.~Bai, and S.~Chintala (2019).
\newblock {PyTorch}: an imperative style, high-performance deep learning library.

\bibitem[\protect\citeauthoryear{Prechelt}{Prechelt}{2012}]{Prechelt2012}
Prechelt, L. (2012).
\newblock Early stopping --- but when?
\newblock In G.~Montavon, G.~B. Orr, and K.-R. M{\"u}ller (Eds.), {\em {Neural Networks: Tricks of the Trade: Second Edition}}, Berlin, Heidelberg, pp.\  53--67. Springer Berlin Heidelberg.

\bibitem[\protect\citeauthoryear{Rezende, Papamakarios, Racaniere, Albergo, Kanwar, Shanahan, and Cranmer}{Rezende et~al.}{2020}]{rezende2020normalizing}
Rezende, D.~J., G.~Papamakarios, S.~Racaniere, M.~Albergo, G.~Kanwar, P.~Shanahan, and K.~Cranmer (2020).
\newblock Normalizing {F}lows on {T}ori and {S}pheres.
\newblock In H.~D. III and A.~Singh (Eds.), {\em Proceedings of the 37th International Conference on Machine Learning}, Volume 119 of {\em Proc. J. Mach. Learn. Res.}, pp.\  8083--8092. PMLR.

\bibitem[\protect\citeauthoryear{Rootzén and Tajvidi}{Rootzén and Tajvidi}{2006}]{Rootzén2006MGPD}
Rootzén, H. and N.~Tajvidi (2006).
\newblock Multivariate generalized {P}areto distributions.
\newblock {\em Bernoulli\/}~{\em 12\/}(5), 917--930.

\bibitem[\protect\citeauthoryear{Simpson, Rue, Riebler, Martins, and S{\o}rbye}{Simpson et~al.}{2017}]{Simpson2017PCpriors}
Simpson, D., H.~Rue, A.~Riebler, T.~G. Martins, and S.~H. S{\o}rbye (2017).
\newblock {Penalising Model Component Complexity: A Principled, Practical Approach to Constructing Priors}.
\newblock {\em Stat. Sci.\/}~{\em 32\/}(1), 1 -- 28.

\bibitem[\protect\citeauthoryear{Simpson and Tawn}{Simpson and Tawn}{2024a}]{simptawn22bivariate}
Simpson, E.~S. and J.~A. Tawn (2024a).
\newblock {Estimating the limiting shape of bivariate scaled sample clouds: With additional benefits of self-consistent inference for existing extremal dependence properties}.
\newblock {\em Electron. J. Stat.\/}~{\em 18\/}(2), 4582 -- 4611.

\bibitem[\protect\citeauthoryear{Simpson and Tawn}{Simpson and Tawn}{2024b}]{simpson2024inference}
Simpson, E.~S. and J.~A. Tawn (2024b).
\newblock Inference for new environmental contours using extreme value analysis.
\newblock {\em J. Agric. Biol. Environ. Stat.\/}.

\bibitem[\protect\citeauthoryear{Song, Sohl-Dickstein, Kingma, Kumar, Ermon, and Poole}{Song et~al.}{2021}]{song2021scorebased}
Song, Y., J.~Sohl-Dickstein, D.~P. Kingma, A.~Kumar, S.~Ermon, and B.~Poole (2021).
\newblock Score-based generative modeling through stochastic differential equations.
\newblock In {\em ICLR}.

\bibitem[\protect\citeauthoryear{Stimper, Liu, Campbell, Berenz, Ryll, Schölkopf, and Hernández-Lobato}{Stimper et~al.}{2023}]{Stimper2023}
Stimper, V., D.~Liu, A.~Campbell, V.~Berenz, L.~Ryll, B.~Schölkopf, and J.~M. Hernández-Lobato (2023).
\newblock normflows: a {PyTorch} package for normalizing flows.
\newblock {\em J. Open Source Softw.\/}~{\em 8\/}(86), 5361.

\bibitem[\protect\citeauthoryear{Tran, Franzese, Michiardi, and Filippone}{Tran et~al.}{2023}]{tran2023onelineofcode}
Tran, B.-H., G.~Franzese, P.~Michiardi, and M.~Filippone (2023).
\newblock One-line-of-code data mollification improves optimization of likelihood-based generative models.
\newblock In A.~Oh, T.~Naumann, A.~Globerson, K.~Saenko, M.~Hardt, and S.~Levine (Eds.), {\em Adv. Neural Inf. Process. Syst.}, Volume~36, pp.\  6545--6567. Curran Associates, Inc.

\bibitem[\protect\citeauthoryear{Wadsworth and Campbell}{Wadsworth and Campbell}{2024}]{wadsworth2022statistical}
Wadsworth, J.~L. and R.~Campbell (2024).
\newblock Statistical inference for multivariate extremes via a geometric approach.
\newblock {\em J. R. Stat. Soc., B\/}~{\em 86\/}(5), 1243--1265.

\bibitem[\protect\citeauthoryear{Wessel, Murphy-Barltrop, and Simpson}{Wessel et~al.}{2025}]{wessel2025comparisongenerativedeeplearning}
Wessel, J.~B., C.~J.~R. Murphy-Barltrop, and E.~S. Simpson (2025).
\newblock A comparison of generative deep learning methods for multivariate angular simulation.

\bibitem[\protect\citeauthoryear{Wood}{Wood}{2011}]{Wood2011GAM}
Wood, S.~N. (2011).
\newblock Fast stable restricted maximum likelihood and marginal likelihood estimation of semiparametric generalized linear models.
\newblock {\em J. R. Stat. Soc., B\/}~{\em 73\/}(1), 3--36.

\bibitem[\protect\citeauthoryear{Youngman}{Youngman}{2022}]{Youngman2022evgam}
Youngman, B.~D. (2022).
\newblock {evgam}: An {R} package for generalized additive extreme value models.
\newblock {\em J. Stat. Softw.\/}~{\em 103\/}(3), 1--26.

\end{thebibliography}



\appendix
\renewcommand{\thesection}{S.\arabic{section}}
\setcounter{section}{0}
\setcounter{equation}{0}

\newpage
\begin{center}
{\large\bf SUPPLEMENTARY MATERIAL}
\end{center}

\section{Multivariate radial exponential distribution}
\label{sec:MRE}

The multivariate radial exponential distribution is supported 
on the complement of a location set $\loc\in\bigstar$, 
and has a scaling set $\Sigma\in\bigstar$ 
and a directional set 
$\W\in\bigstar$ with radial function~$r_{\W}$ equal to the density of 
directions~$f_{\bm W}$.\ 

\begin{figure}[htbp]
   \centering
   \vspace{-0.5em}
   \begin{overpic}[width=0.45\textwidth]{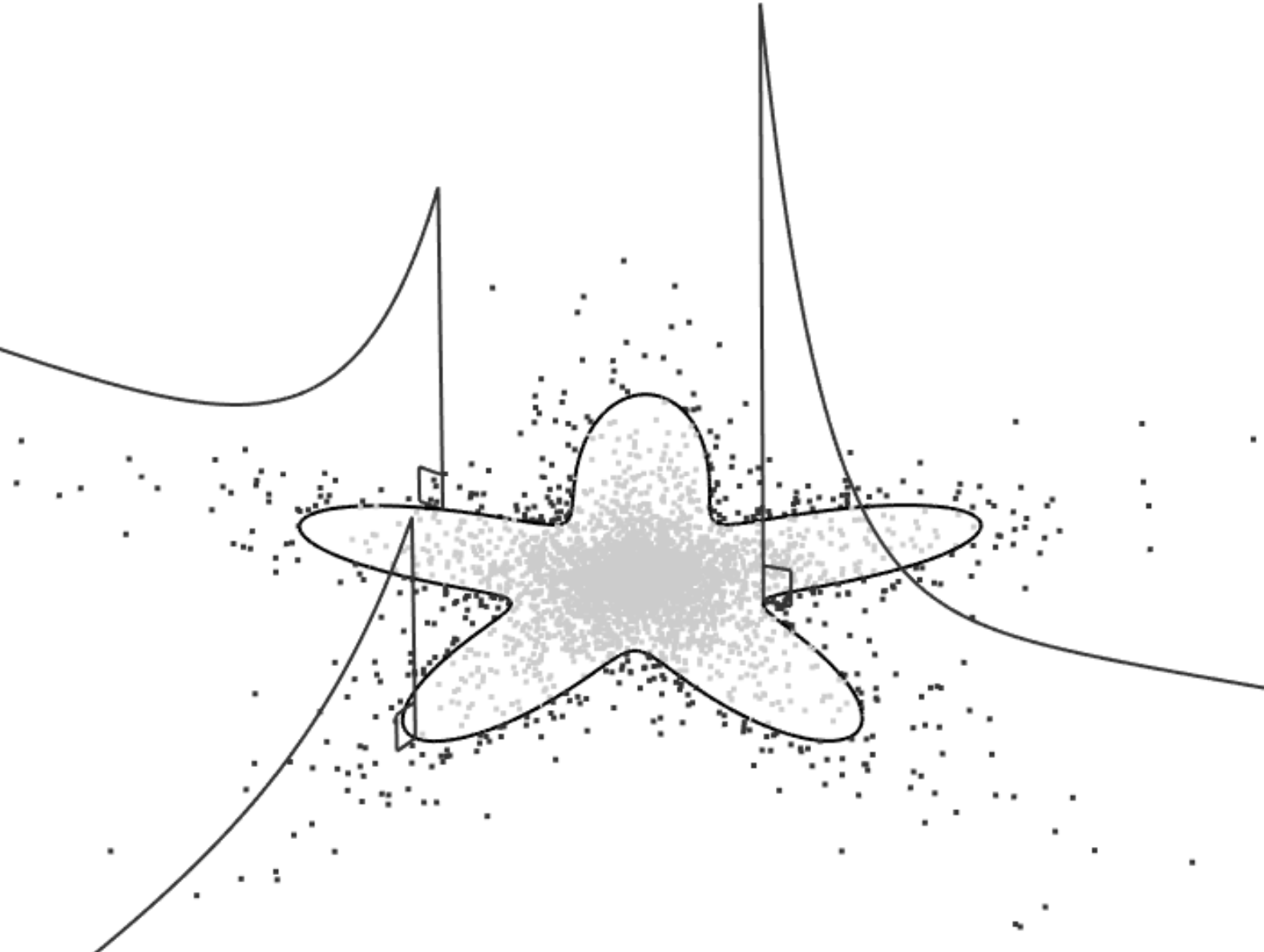}
   \put (46,28) {$\Q_{0.95}$}
   \put (6,47) {\small$f_{R_E\mid \bm W=\bm w_1}$}
   \put (0,13) {\small$f_{R_E\mid \bm W=\bm w_2}$}
   \put (81,26) {\small$f_{R_E\mid \bm W=\bm w_3}$}
   \end{overpic}
   \hspace{2em}
   \begin{overpic}[width=0.45\textwidth]{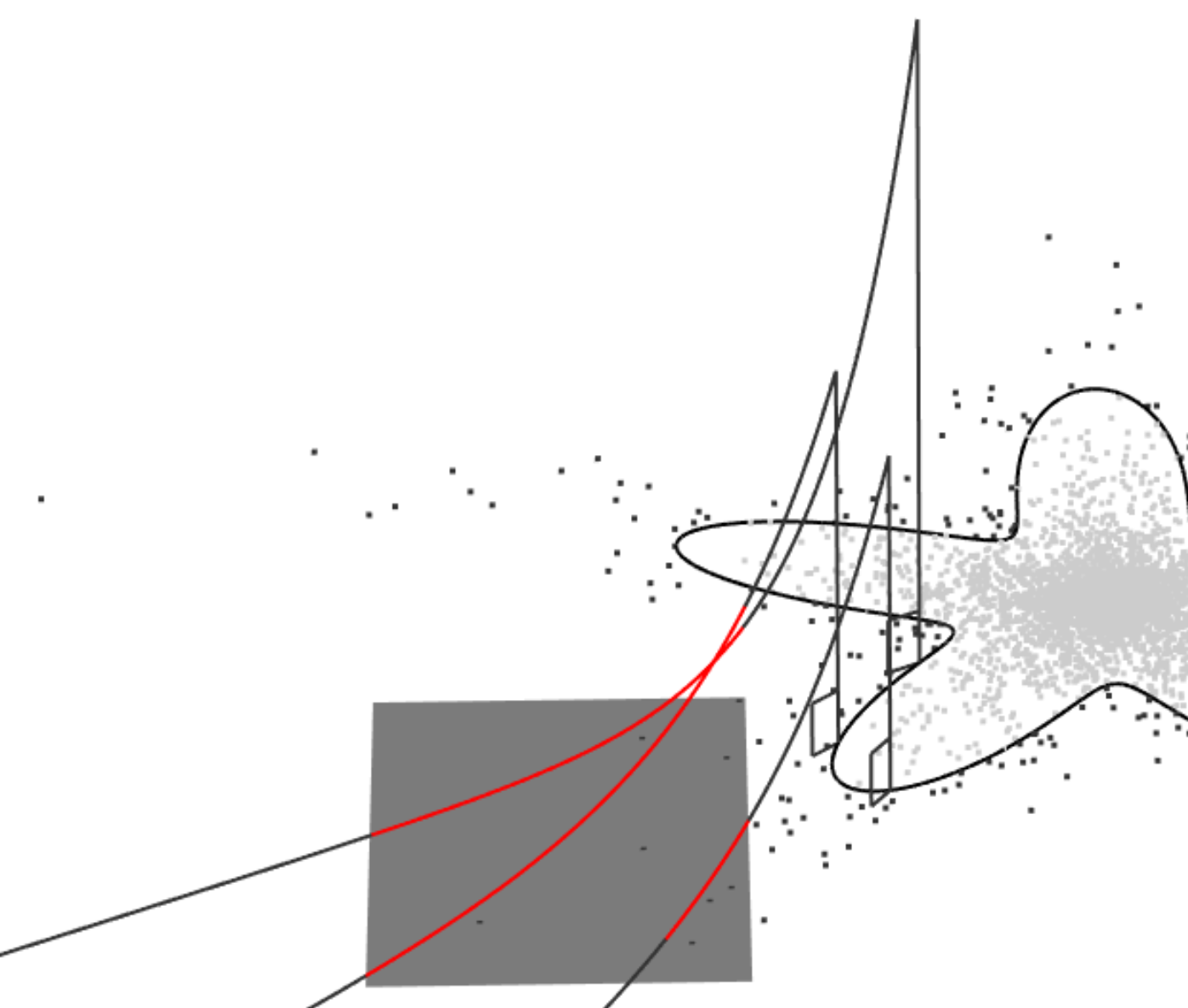}
   \put (47,8) {$\Rset$}
   \put (85,33) {$\Q_{0.95}$}
   \end{overpic}
   \caption{Samples from a bivariate radial exponential distribution 
   with scale~$\G$ given by $r_{\G}(\bm w) = 
   \exp\{\sin(5\bm w)/2\}/2$, associated 
   true quantile set~$\Q_{0.95}$, and exceedance samples (dark grey) in 
   the complement~$\Q_{0.95}^\prime$.\ 
   \textit{Left}:\ Exponential density 
   $f_{R_E\mid \bm W=\bm w}$ of radial exceedances 
   with scale~$r_{\G}$ at directions $\bm w_1,\bm w_2,\bm w_3\in\SSS^1$.\ \textit{Right}:\ Visualisation 
   of expression~\eqref{eq:MR_exp} with integration of the radial
   density along three directions intersecting with~$\Rset$.\ }
   \label{fig:Qq_MRS}
\end{figure}

Figure~\ref{fig:Qq_MRS} displays a sample from a bivariate radial exponential 
distribution with 
scale~$\G$ given by~$r_{\G}(\bm w) = 
\exp\{\sin(5\bm w)/2\}/2$ and directional set~$\W$ 
with~$f_{\bm W}\equiv1/(2\pi)$, the uniform density 
on~$\SSS^1$.\ By the memoryless property, 
$(R-u)\mid \{R>u,\bm W=\bm w\} \sim \text{Exp}(1/r_{\G}(\bm w))$ 
for any $u>0$ and $\bm w\in\SSS^{d-1}$, and hence, exceedances 
$\bm X\mid \{\bm X\in\Qq^\prime\}$ follow a radial exponential distribution with 
location~$\Qq$ and scale~$\G$ exactly.\ 

\section{Rational-quadratic spline flows formulation}
\label{sec:flow_construction}

\cite{rezende2020normalizing} proposed to apply a 
two-stage autoregressive flow to the hypercylinder 
$\CCC^{d-1}$---see their Section 2.3.1 and Appendix B
for a detailed description.\ 

The procedure performs transformations based on rational-quadratic splines 
\citep{Durkan2019NeuralSplineFlow}
to the different marginal intervals $[0,2\pi]$ and $(-1,1)$ of $\CCC^{d-1}$.\ 
A transformation $g$ applied to an interval is parameterised 
by a set of $\Phi+1$ knots 
$\{x_\phi,y_\phi\}_{\phi=0}^\Phi$ and $\Phi$ slopes $\{s_\phi\}_{\phi=1}^\Phi$
such that $x_{\phi-1}<x_\phi$, $y_{\phi-1}<y_\phi$, and $s_\phi>0$ for all $\phi=1,\ldots,\Phi$.\ 
For $x_{\phi-1}<x<x_\phi$, the transformation $g$ is defined via
\begin{IEEEeqnarray}{rCl}
\label{eq:rational_quadratic}
   g(x) = \frac{\alpha_{\phi,2}x^2+\alpha_{\phi,1}x+\alpha_{\phi,0}}{\beta_{\phi,2}x^2+\beta_{\phi,1}x+\beta_{\phi,0}},
\end{IEEEeqnarray}
where the coefficients $\{\alpha_{\phi,i},\beta_{\phi,i}\}_{i=0}^2$ are constrained so that
$g$ is strictly monotonic increasing, $g(x_{\phi-1})=y_{\phi-1}$, $g(x_{\phi})=y_{\phi}$, and 
$\partial g(t)/\partial t \mid_{t=x} = s_\phi$ for all $\phi=1,\ldots,\Phi$.\ Then,
transformations on $(-1,1)$ are defined by setting $x_0=y_0=-1$ and $x_{\Phi}={y_\Phi}=1$, and circular
transformations on $[0,2\pi]$ by setting $x_0=y_0=0$, $x_{\Phi}={y_\Phi}=2\pi$, and $s_1=s_\phi$.\ 

\section{Transformation from $\SSS^{d-1}$ to $\SSS^{1}\times(-1,1)^{d-2}$}
\label{sec:T}

Let $T^{(k)}$, for $k\geq 3$, denote the transformation mapping a direction $\bm w = (w_1,\ldots,w_k)\in\SSS^{k-1}$ to the 
hypercylinder $\SSS^{k-2}\times(-1,1)$ and define it via 
\begin{IEEEeqnarray}{rCl}
\label{eq:T_k}
   T^{(k)}(\bm w) = 
       \left(\frac{\bm w_{[1:k-1]}}{\sqrt{1-w_k^2}},w_k\right),\quad \bm w \in \SSS^{k-1}\backslash \left\{\pm\,\bm e_{k}^{(k)}\right\},\, k\geq 3,
\end{IEEEeqnarray}
where~$\bm e_{n}^{(m)}$ denotes the vector
of size~$n$ whose sole non-zero $m$th component is equal to 1.\ 
Defining the transformation 
$\widetilde{T}^{(k)}:\SSS^{k-1}\times(-1,1)^{d-k}\to \SSS^{k-2}\times(-1,1)^{d-k+1}$ via 
\begin{IEEEeqnarray}{rCl}
\label{eq:T_k_tilde}
   \widetilde{T}^{(k)}(\bm x) = \begin{cases}
       \left(T^{(k)}(\bm x_{[1:k]}),\bm x_{[k+1:d]}\right),\quad &k<d,\\
       T^{(k)}(\bm x), \quad &k=d,
   \end{cases}
\end{IEEEeqnarray}
for $\bm x \in \SSS^{k-1}\times(-1,1)^{d-k}\backslash \left\{\pm\,\bm e_{d}^{(k)}\right\}$, we have that the recursive and reversible transformation 
$T:= \widetilde{T}^{(3)}\circ \cdots \circ \widetilde{T}^{(d)}$ 
maps elements of the punctured hypersphere 
$\SSS^{d-1}\backslash \{\pm\,\bm e_{d}^{(3)},\ldots,\pm\,\bm e_{d}^{(d)}\}$ 
into the hypercylinder~$\CCC^{d-1}$.\ 

\cite{Papamakarios2021NF} show that the transformation~$T$, for a direction
$\bm w\in\SSS^{d-1}$ and associated $\bm c=T(\bm w)\in\CCC^{d-1}$, has Jacobian determinant given by
\begin{IEEEeqnarray}{rCl}
   \label{eq:Jac_T_k}
   \left\lvert\frac{\partial T(\bm w)}{\partial \bm w}\right\rvert = \prod_{i=1}^{d} (1-c_i^2)^{-\left(\frac{i+1}{2}-1\right)}.
\end{IEEEeqnarray}
Numerical instabilities in the computation of the Jacobian determinant can occur 
whenever $c_i$ is in the near vicinity of -1 or 1 for some $i\in\{1,\ldots,d\}$.\ 
However, these locations on the hypercylinder correspond to the near vicinity of the 
singularities~$S_d$ of the transformation~$T$ expressed in Section~\ref{subsec:NF},
and hence, have small or negligeable measure with respect to $\PR_{\bm X/\lVert\bm X\rVert}$
leading to negligeable impact on the general probability estimation procedure.\ 

\section{Uniform penalisation for constrained models}
\label{sec:penalisation}

The PDF of a uniform density on~$\SSS^{d-1}$ is given by 
$f_U(\bm w) = {\Gamma(d/2)}/{2\pi^{d/2}}$, $\bm w \in \SSS^{d-1}$.\ 
Hence, penalisation of~$f_{\D}$ away from~$f_U$ can be performed 
via the Kullback--Leibler divergence 
$\KLD[f_U\lVert f_{\D}] = \smash{\int_{\SSS^{d-1}}}\,\log[f_{U}(\bm w)/f_{\D}(\bm w)]f_U(\bm w)\,d\bm w$.\ 
In practice, this integral is approximated via Monte Carlo 
integration by sampling a large number~$m$ of directions 
$\bm u_1,\ldots,\bm u_m$ uniformly on~$\SSS^{d-1}$ and calculating
\begin{IEEEeqnarray}{rCl}
\label{eq:KL_unif_penalty}
    \overline{\mathrm{D}}_{\mathrm{KL}}[f_U\lVert f_{\D}] := \frac{1}{m}\sum_{i=1}^m\log[f_{U}(\bm u_i)/f_{\D}(\bm u_i)]. 
\end{IEEEeqnarray}

\section{Mollification dispersion parameter}
To perform gradually vanishing mollification, at iteration $j\in\{0,\ldots,J\}$, we set
\label{sec:sigma_j}
\begin{IEEEeqnarray}{rCl}
\label{eq:sigma_j}
      \sigma_j = \sigma\frac{\mathrm{Sigmoid}(j/(\tau J))-\mathrm{Sigmoid}(1/\tau)}
      {\mathrm{Sigmoid}(0)-\mathrm{Sigmoid}(1/\tau)},
\end{IEEEeqnarray}
where $\mathrm{Sigmoid}(x)=1/\{1+\exp(x)\}$, $\sigma>0$ is the strength of dispersion at iteration $j=0$
and $\tau>0$ controls the rate of decay of $\sigma_j$ to 0 when $j=J$.\ 

\section{Radial integration bounds of hyperboxes}
\label{sec:hyperbox}

Let $\bm a = (a_1,\ldots,a_d)\in \{\RR \cup \{-\infty,\infty\}\}^d$ 
and $\bm b = (b_1,\ldots,b_d)\in \{\RR \cup \{-\infty,\infty\}\}^d$, 
with $a_i\leq b_i$ for all $i\in\{1,\ldots,d\}$, then 
the set $\Rset=[a_1,b_1] \times \cdots \times [a_d,b_d]$ is a hyperbox.\ 
Since hyperboxes are starshaped (in the more general sense of \cite{hansen2020starshaped}), 
for $\bm w\in\SSS^{d-1}$, the intersection $[\bm 0:\bm w)\cap\Rset$ is either a 
single line segment---possibly degenerate to a singleton---or empty.\ 
For simplicity, we let 
$\s_{\Rset} := \{\bm x/\lVert \bm x \rVert \in\SSS^{d-1}: \bm x \in \Rset\}$ 
be the set of directions pointing toward~$\Rset$, so that 
$[\bm 0:\bm w)\,\cap\,\Rset = \emptyset$ if and only if $\bm w \in \s_{\Rset}$.\ 

We define the functions $\rinf{\Rset}: \SSS^{d-1}\to \RR_{\geq0}$ and 
$\rsup{\Rset}: \SSS^{d-1}\to \RR_{\geq0}$, mapping a direction 
$\bm w \in \s_{\Rset}$ respectively to the infimum and supremum 
radii of $[\bm 0:\bm w)\,\cap\,\Rset$ and a direction 
$\bm w\notin\s_{\Rset}$ to 0, through
\begin{IEEEeqnarray}{rCl}
\label{eq:rinf_sup}
   r_{\Rset}^*(\bm w) = \begin{cases}
       *\{\lVert \bm x \rVert : \bm x \in [\bm 0:\bm w)\,\cap\,\Rset\}, \quad &\bm w \in \s_{\Rset},\\
       0, &\bm w \notin \s_{\Rset},
   \end{cases} \quad \bm w \in \SSS^{d-1},
\end{IEEEeqnarray}
where $*$ is a place-holder for the $\inf$ or $\sup$ operators.\ 
Knowing~$\rinf{\Rset}$ and~$\rsup{\Rset}$ on~$\SSS^{d-1}$ is sufficient 
to evaluate the inner integral in equation~\eqref{eq:MR_exp}, but 
this requires obtaining an expression for the~$\inf$ and~$\sup$ of 
expressions~\eqref{eq:rinf_sup} 
as well as conditions to determine if a given direction~$\bm w$ belongs to~$\s_{\Rset}$ or not.\ 

We first consider the case in which~$\Rset$ lies entirely 
in one quadrant of~$\RR^d$, that is, for each $i\in\{1,\ldots,d\}$, 
either $a_i \geq 0$ or $b_i\leq 0$.\ 
Since $\rVert \bm x \rVert$ is invariant to the sign~of~$\bm x\in\RR^d$,
we map~$\Rset$ to~$\Rset_+:=[a_{1,+},b_{1,+}] \times \cdots \times [a_{d,+},b_{d,+}]$ 
where $\bm a_+ = (\min\{|a_{i,+}|,|b_{i,+}|\}:i=1,\ldots,d)$ and 
$\bm b_+ = (\max\{|a_{i,+}|,|b_{i,+}|\}:i=1,\ldots,d)$ in the~$\RR_{>0}^d$ quadrant, preserving 
that if $\bm x\in\Rset$ maps to $\bm x_+\in\Rset_+$, then 
$\lVert \bm x \rVert = \lVert \bm x_+ \rVert$.\ 
Assuming 
that~$\bm w \in \s_{\Rset}$, we have that 
\begin{IEEEeqnarray}{rCl}
\label{eq:rinf_B}
   \inf\{\lVert \bm x \rVert : \bm x \in [\bm 0:\bm w)\,\cap\,\Rset\} = \max\{\bm a_+/\bm w_+\}
\end{IEEEeqnarray}
and
\begin{IEEEeqnarray}{rCl}
   \label{eq:rsup_B}
      \sup\{\lVert \bm x \rVert : \bm x \in [\bm 0:\bm w)\,\cap\,\Rset\} = \min\{\bm b_+/\bm w_+\},
      \qquad
   \end{IEEEeqnarray}
where $\bm w_+ = (|w_i|:i=1,\ldots,d)$.\ The term 
$\max\{\bm a_+/\bm w_+\}$ in~\eqref{eq:rinf_B} is the 
smallest factor~$t$ 
so that $t \bm w$ first enters~$\Rset$.\
Analogously, the term $\min\{\bm a_+/\bm w_+\}$ in~\eqref{eq:rsup_B} 
is the biggest factor~$t$
so that $t \bm w\in\Rset$ while 
$(t+\varepsilon)\bm w\notin\Rset$ for all $\varepsilon>0$.\

It remains to determine whether a direction~$\bm w$ belongs 
to~$\s_{\Rset}$.\ Again assuming that~$\Rset$ lies 
entirely in a quadrant of~$\RR^d$, we check 
that~$\bm w\notin\s_{\Rset}$ 
if $\min\{\bm a/\bm w\}<0$: if 
$\min\{\bm a/\bm w\}<0$, there cannot exist~$r>0$ such that 
$r\bm w\in \Rset$.\ However,~$\bm w$ could still lie in the same 
quadrant as~$\Rset$ and lead to $[\bm 0:\bm w)\,\cap\,\Rset = \emptyset$: 
we must further verify that $\max\{\bm a_+/\bm w_+\}\geq\min\{\bm b_+/\bm w_+\}$.\ 
If this condition is not met, there does not exist~$r>0$ 
such that $r\bm w$ exceeds all components of~$\bm a$ 
(\textit{i.e.\ }potentially enters~$\Rset$) before it exceeds at 
least one component of~$\bm b$ (\textit{i.e.\ }potentially exits~$\Rset$).\ 

The case where~$\Rset$ does not entirely belong to a quadrant of~$\RR^d$ 
is dealt with by considering an appropriate subset of~$\Rset$ laying 
in a single quadrant to apply the same logic as before.\ Suppose 
that~$\Rset$ is a hyperbox such that $a_i<0<b_i$ for 
all $i\in J\subseteq\{1,\ldots,d\}$, \textit{i.e.}~$\Rset$ lies in more than one quadrant.\ 
Then, if $\bm w_i>0$ (respectively $\bm w_i<0$) 
and $[\bm 0:\bm w)\,\cap\,\Rset\neq\emptyset$, then each 
$\bm x\in[\bm 0:\bm w)\,\cap\,\Rset$ must have $x_i>0$ 
(respectively $x_i<0$).\ Thus, to obtain $\rinf{B}(\bm w)$ 
and $\rsup{B}(\bm w)$, it suffices to consider the hyperbox 
$\Rset_{\bm w}$ with lower- and upper-bounds~$\bm a$ and~$\bm b$ 
such that $a_i = 0$ if $w_i>0$ or $b_i = 0$ if $w_i<0$, 
for each $i\in J$.\ $\Rset_{\bm w}$ then lies entirely 
in one quadrant of~$\RR^d$, and 
$\rinf{\Rset}(\bm w) = \rinf{\Rset_{\bm w}}(\lvert\bm w\rvert)$ 
and $\rsup{\Rset}(\bm w) = \rsup{\Rset_{\bm w}}(\lvert\bm w\rvert)$ for all 
$\bm w \in \SSS^{d-1}$.\ 

Hence, if $\Rset$ is a hyperbox and $\rinf{\Rset}\geq r_{\Qq}$,
the inner integral in~\eqref{eq:MR_exp} and the integral 
in~\eqref{eq:P_B_noint_Qq} simplify according to
\begin{IEEEeqnarray}{rCl}
\label{eq:MR_exp_hyperbox}
   \PR[\bm X\in\Rset\mid \bm X \in \Qq^\prime, \bm W = \bm w] =
   \int_{\rinf{\Rset}(\bm w)}^{\rsup{\Rset}(\bm w)}
   \frac{1}{r_{\G}(\bm w)}\exp\left\{-\frac{r-r_{\Qq}(\bm w)}{r_{\G}(\bm w)}\right\}
   \dd r.
\end{IEEEeqnarray}

\section{Subset of exceedances}
\label{sec:outside_starbodyhyperbox}
In this section, we derive the radial integration bounds used to compute the probability
that at least~$\ell$ of the margins of~$\bm X$ fall below or above their associated 
negative lower- and positive upper-bounds $\bm l = (l_1,\ldots,l_d)\in\RR_{\leq 0}^d$ and 
$\bm u = (u_1,\ldots,u_d)\in\RR_{\geq 0}^d$, respectively,
for some $\ell\in\{1,\ldots,d\}$.\ That is, for some index sets~$A$ and~$B$ such that $A\cup B=\{1,\ldots,d\}$ 
and $A\cap B = \emptyset$, we assume for simplicity that 
$\Qq\subseteq \mathcal{H}:=[l_1,u_1]\times\cdots\times[l_d,u_d]$ and 
define $r^{(\ell)}\,:\,\SSS^{d-1}\to \RR_{\geq 0}$ such that 
\begin{IEEEeqnarray}{rCl}
   \label{eq:MR_exp_l_exceedances}
      \PR[\bm X_{A}\notin[\bm l_{A},\bm u_{A}],\bm X_{B}&\in&[\bm l_{B},\bm u_{B}], 
      \#A \geq \ell] = \nonumber\\
      &&\quad 
      \int_{\bm w \in \SSS^{d-1}}\int_{r^{(\ell)}(\bm w)}^{\infty}
      \frac{1}{r_{\G}(\bm w)}\exp\left\{-\frac{r-r_{\Qq}(\bm w)}{r_{\G}(\bm w)}\right\}
      \dd r\,\dd \bm w, \qquad
\end{IEEEeqnarray}
where $\bm X \in [\bm x,\bm y]$, with $\bm x,\bm y\in\RR^d$,
denotes that $l_i<X_i<u_i$ for all $i\in\{1,\ldots,d\}$.\ 
The case where $\Qq\not\subseteq\mathcal{H}$
can be dealt with by resorting to the partitioning method of 
Section~\ref{subsec:prob_est}.\ 

Given a direction $\bm w\in\SSS^{d-1}$, the sign of its $i$th component
determines whether $[\bm0:\bm w)$ points toward the face
with coordinate $l_i$ or $u_i$ of the hyperbox $\mathcal{H}$.\ Hence, the vector
\begin{IEEEeqnarray}{rCl}
   \label{eq:r_l_coordinate}
      \bm a(\bm w,\bm l, \bm u) = \left[\max\left\{\frac{l_1}{w_1},\frac{u_1}{w_1}\right\},
      \ldots ,
      \max\left\{\frac{l_d}{w_d},\frac{u_d}{w_d}\right\}\right]
\end{IEEEeqnarray}
yields in every ordinate the smallest factor $t_i$ by which $\bm w$ needs to be multiplied
so that $t_i w_i\notin[l_i,u_i]$.\ Therefore, defining 
$r^{(\ell)}(\bm w) = \bm a(\bm w,\bm l, \bm u)_{(\ell)}$ where $\bm x_{(\ell)}$ denotes the 
$\ell$th smallest component of $\bm x\in\RR^d$, we get that for $r>r^{(\ell)}(\bm w)$,
the point $r\bm w\in\RR^d$ has at least $\ell$ of its coordinates are outside their 
intervals specified by $\bm l$ and $\bm u$; that is, some index sets~$A$ and~$B$ such 
that $A\cup B=\{1,\ldots,d\}$, $A\cap B = \emptyset$, and $\#A\geq \ell$, 
$r\bm w_A\notin[\bm l_{A},\bm u_{A}]$ and $r\bm w_{B}\in[\bm l_{B},\bm u_{B}]$.\ 
Hence, $r^{(\ell)}$ satisfies equation~\eqref{eq:MR_exp_l_exceedances}.\ 



\section{Simulation study}
\label{sec:Supp_simulation_study}

\begin{figure}[H]
   \centering
   \hspace{-2em}
   \begin{overpic}[width=0.30\textwidth]{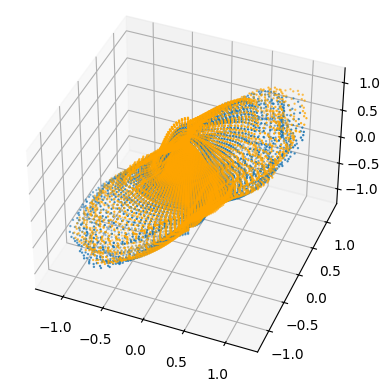}
      \put(27,0){$\widehat{\G}_1$}
      \put(90,22){$\widehat{\G}_2$}
      \put(101,62){$\widehat{\G}_3$}
   \end{overpic}
   \hspace{0.5em}
   \begin{overpic}[width=0.30\textwidth]{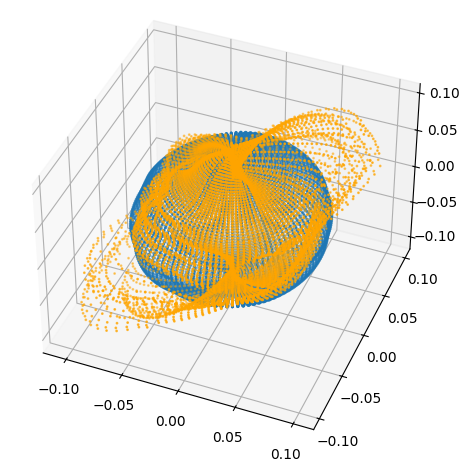}
      \put(27,0){$\widehat{\D}_1$}
      \put(90,22){$\widehat{\D}_2$}
      \put(101,62){$\widehat{\D}_3$}
   \end{overpic}
   \hspace{0.75em}
   \begin{overpic}[width=0.30\textwidth]{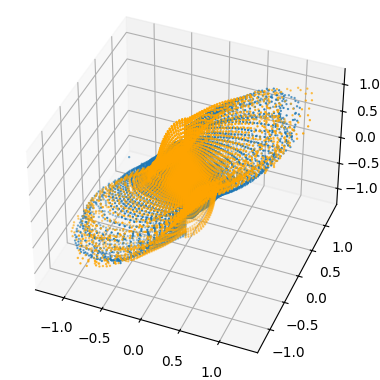}
      \put(27,0){$\widehat{\G}_1$}
      \put(90,22){$\widehat{\G}_2$}
      \put(101,62){$\widehat{\G}_3$}
   \end{overpic}
   \caption{Estimated model parameters on $10^4$ samples from a trivariate 
   normal distribution with standard Laplace margins and the same covariance matrix as the simulation study 
   of Section~\ref{sec:Sim_Study}.\ \textit{Left}:\ Estimated scaling set $\widehat{\G}$ using 
   models $\M_1$ (orange) and $\M_7$ (blue) with strong 
   penalisation $\lambda_U=100$.\ \textit{Centre}:\ 
   Estimated deformation sets $\widehat{\D}$ from model $\M_7$ 
   with $\lambda=100$ (blue) and with weak 
   penalisation $\lambda=1.0$ (orange).\ \textit{Right}:\ 
   Estimated scaling set $\widehat{\G}$ using 
   models $\M_2$ (orange) and $\M_7$ (blue) with weak 
   penalisation.\ }
   \label{fig:M1M7M2}
\end{figure}

\begin{figure}[H]
   \centering
   \begin{overpic}[width=\textwidth]{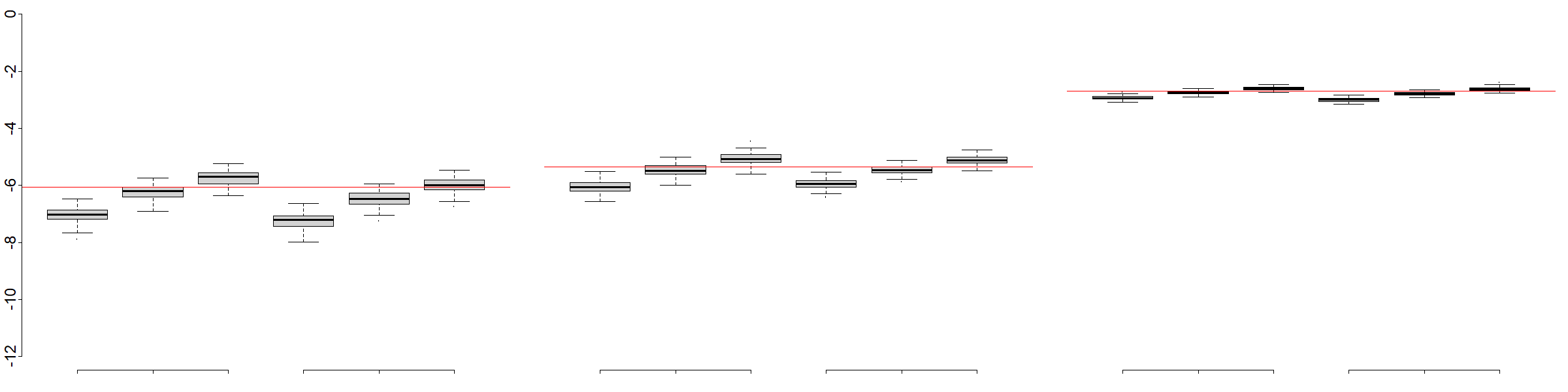}
      \put (4,-0.5) {\scalebox{0.5}{$\mathrm{CI}_{L}$}}
      \put (37.5,-0.5) {\scalebox{0.5}{$\mathrm{CI}_{L}$}}
      \put (70.5,-0.5) {\scalebox{0.5}{$\mathrm{CI}_{L}$}}
      
      \put (7.5,-0.5) {\scalebox{0.5}{$\log_{10}\widehat{\PR}_{\Rset_1}$}}
      \put (41,-0.5) {\scalebox{0.5}{$\log_{10}\widehat{\PR}_{\Rset_2}$}}
      \put (74.25,-0.5) {\scalebox{0.5}{$\log_{10}\widehat{\PR}_{\Rset_3}$}}
      
      \put (13,-0.5) {\scalebox{0.5}{$\mathrm{CI}_{U}$}}
      \put (46.5,-0.5) {\scalebox{0.5}{$\mathrm{CI}_{U}$}}
      \put (79.5,-0.5) {\scalebox{0.5}{$\mathrm{CI}_{U}$}}

      \put (18.5,-0.5) {\scalebox{0.5}{$\mathrm{CI}_{L}$}}
      \put (52,-0.5) {\scalebox{0.5}{$\mathrm{CI}_{L}$}}
      \put (85,-0.5) {\scalebox{0.5}{$\mathrm{CI}_{L}$}}
      
      \put (22,-0.5) {\scalebox{0.5}{$\log_{10}\widehat{\PR}_{\Rset_1}$}}
      \put (55.5,-0.5) {\scalebox{0.5}{$\log_{10}\widehat{\PR}_{\Rset_2}$}}
      \put (88.75,-0.5) {\scalebox{0.5}{$\log_{10}\widehat{\PR}_{\Rset_3}$}}
      
      \put (27.5,-0.5) {\scalebox{0.5}{$\mathrm{CI}_{U}$}}
      \put (61,-0.5) {\scalebox{0.5}{$\mathrm{CI}_{U}$}}
      \put (94,-0.5) {\scalebox{0.5}{$\mathrm{CI}_{U}$}}
       
      \put (8.5,2.5) {\footnotesize $\M_4$}
      \put (23,2.5) {\footnotesize $\M_7$}
      \put (41,2.5) {\footnotesize $\M_4$}
      \put (56,2.5) {\footnotesize $\M_7$}
      \put (74.5,2.5) {\footnotesize $\M_4$}
      \put (89.5,2.5) {\footnotesize $\M_7$}
      
      \put (15.5,23) {$\Rset_1$}
      \put (16.5,20.5) {\rotatebox{90}{$=$}}
      \put (14.5,19) {\tiny $[10,\infty)^3$}
      \put (48.5,23) {$\Rset_2$}
      \put (49.5,20.5) {\rotatebox{90}{$=$}}
      \put (41,19) {\tiny $[-5,5]\times[10,\infty)\times[-10,10]$}
      \put (82.5,15) {$\Rset_3$}
      \put (83.5,12.5) {\rotatebox{90}{$=$}}
      \put (75,11) {\tiny $(-\infty,5]\times[5,\infty)\times[-5,5]$}
   \end{overpic}

   \vspace{0.5em}
   \centering
   \begin{overpic}[width=\textwidth]{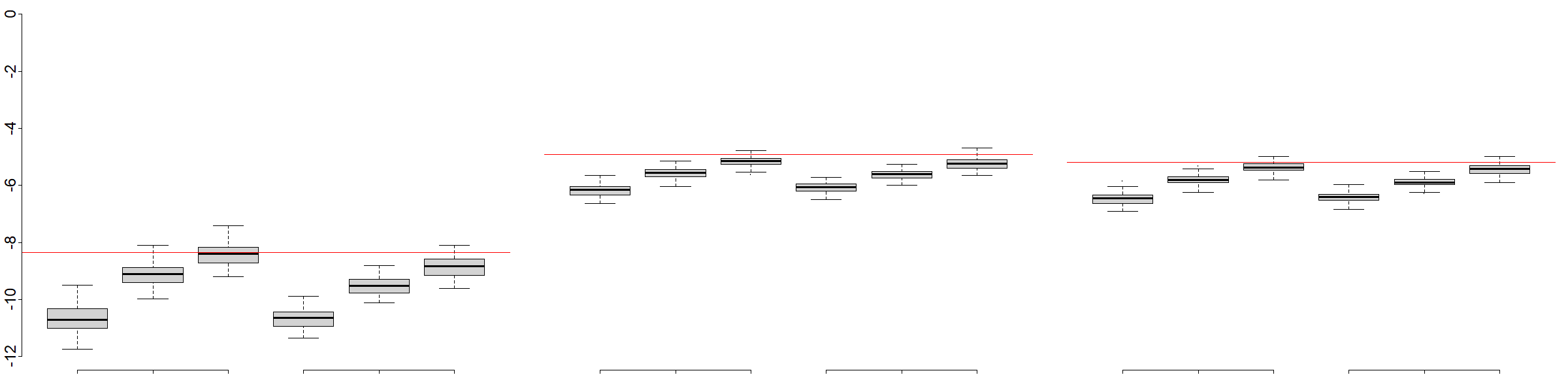}
      \put (4,-0.5) {\scalebox{0.5}{$\mathrm{CI}_{L}$}}
      \put (37.5,-0.5) {\scalebox{0.5}{$\mathrm{CI}_{L}$}}
      \put (70.5,-0.5) {\scalebox{0.5}{$\mathrm{CI}_{L}$}}
      
      \put (7.5,-0.5) {\scalebox{0.5}{$\log_{10}\widehat{\PR}_{\Rset_1}$}}
      \put (41,-0.5) {\scalebox{0.5}{$\log_{10}\widehat{\PR}_{\Rset_2}$}}
      \put (74.25,-0.5) {\scalebox{0.5}{$\log_{10}\widehat{\PR}_{\Rset_3}$}}
      
      \put (13,-0.5) {\scalebox{0.5}{$\mathrm{CI}_{U}$}}
      \put (46.5,-0.5) {\scalebox{0.5}{$\mathrm{CI}_{U}$}}
      \put (79.5,-0.5) {\scalebox{0.5}{$\mathrm{CI}_{U}$}}

      \put (18.5,-0.5) {\scalebox{0.5}{$\mathrm{CI}_{L}$}}
      \put (52,-0.5) {\scalebox{0.5}{$\mathrm{CI}_{L}$}}
      \put (85,-0.5) {\scalebox{0.5}{$\mathrm{CI}_{L}$}}
      
      \put (22,-0.5) {\scalebox{0.5}{$\log_{10}\widehat{\PR}_{\Rset_1}$}}
      \put (55.5,-0.5) {\scalebox{0.5}{$\log_{10}\widehat{\PR}_{\Rset_2}$}}
      \put (88.75,-0.5) {\scalebox{0.5}{$\log_{10}\widehat{\PR}_{\Rset_3}$}}
      
      \put (27.5,-0.5) {\scalebox{0.5}{$\mathrm{CI}_{U}$}}
      \put (61,-0.5) {\scalebox{0.5}{$\mathrm{CI}_{U}$}}
      \put (94,-0.5) {\scalebox{0.5}{$\mathrm{CI}_{U}$}}

      \put (8.5,2.5) {\footnotesize $\M_4$}
      \put (23,2.5) {\footnotesize $\M_7$}
      \put (41,2.5) {\footnotesize $\M_4$}
      \put (56,2.5) {\footnotesize $\M_7$}
      \put (74.5,2.5) {\footnotesize $\M_4$}
      \put (89.5,2.5) {\footnotesize $\M_7$}
       
      \put (15.5,22) {$\Rset_1$}
      \put (16.5,19.5) {\rotatebox{90}{$=$}}
      \put (14.5,18) {\tiny $[10,\infty)^5$}
      \put (48.5,22) {$\Rset_2$}
      \put (49.5,19.5) {\rotatebox{90}{$=$}}
      \put (34.5,18) {\scalebox{0.55}{$(-\infty,\infty)\times[6,\infty)\times[8,\infty)\times[6,\infty)\times(-\infty,\infty)$}}
      \put (82.5,22) {$\Rset_3$}
      \put (83.5,19.5) {\rotatebox{90}{$=$}}
      \put (66,18) {\scalebox{0.55}{$(-\infty,-7] \times(-\infty,0] \times (-\infty,-5] \times (-\infty,0] \times (-\infty,-7]$}}
      
   \end{overpic}

   \vspace{0.5em}
   \centering
   \begin{overpic}[width=\textwidth]{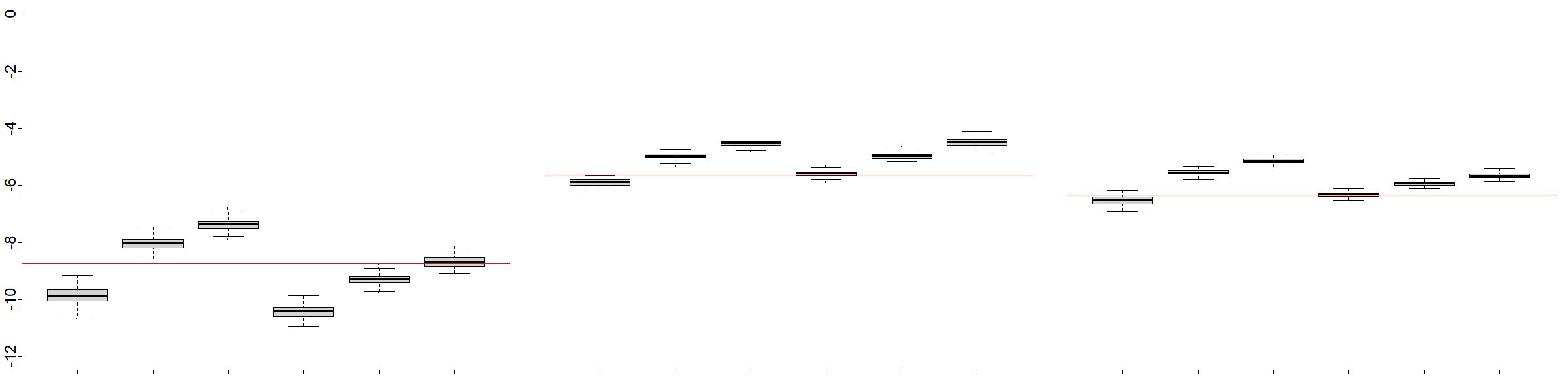}
      \put (4,-0.5) {\scalebox{0.5}{$\mathrm{CI}_{L}$}}
      \put (37.5,-0.5) {\scalebox{0.5}{$\mathrm{CI}_{L}$}}
      \put (70.5,-0.5) {\scalebox{0.5}{$\mathrm{CI}_{L}$}}
      
      \put (7.5,-0.5) {\scalebox{0.5}{$\log_{10}\widehat{\PR}_{\Rset_1}$}}
      \put (41,-0.5) {\scalebox{0.5}{$\log_{10}\widehat{\PR}_{\Rset_2}$}}
      \put (74.25,-0.5) {\scalebox{0.5}{$\log_{10}\widehat{\PR}_{\Rset_3}$}}
      
      \put (13,-0.5) {\scalebox{0.5}{$\mathrm{CI}_{U}$}}
      \put (46.5,-0.5) {\scalebox{0.5}{$\mathrm{CI}_{U}$}}
      \put (79.5,-0.5) {\scalebox{0.5}{$\mathrm{CI}_{U}$}}

      \put (18.5,-0.5) {\scalebox{0.5}{$\mathrm{CI}_{L}$}}
      \put (52,-0.5) {\scalebox{0.5}{$\mathrm{CI}_{L}$}}
      \put (85,-0.5) {\scalebox{0.5}{$\mathrm{CI}_{L}$}}
      
      \put (22,-0.5) {\scalebox{0.5}{$\log_{10}\widehat{\PR}_{\Rset_1}$}}
      \put (55.5,-0.5) {\scalebox{0.5}{$\log_{10}\widehat{\PR}_{\Rset_2}$}}
      \put (88.75,-0.5) {\scalebox{0.5}{$\log_{10}\widehat{\PR}_{\Rset_3}$}}
      
      \put (27.5,-0.5) {\scalebox{0.5}{$\mathrm{CI}_{U}$}}
      \put (61,-0.5) {\scalebox{0.5}{$\mathrm{CI}_{U}$}}
      \put (94,-0.5) {\scalebox{0.5}{$\mathrm{CI}_{U}$}}
       
      \put (8.5,2.5) {\footnotesize $\M_4$}
      \put (23,2.5) {\footnotesize $\M_7$}
      \put (41,2.5) {\footnotesize $\M_4$}
      \put (56,2.5) {\footnotesize $\M_7$}
      \put (74.5,2.5) {\footnotesize $\M_4$}
      \put (89.5,2.5) {\footnotesize $\M_7$}

      \put (15.5,22) {$\Rset_1$}
      \put (16.5,19.5) {\rotatebox{90}{$=$}}
      \put (14.5,18) {\tiny $[10,\infty)^7$}
      \put (48.5,22) {$\Rset_2$}
      \put (49.5,19.5) {\rotatebox{90}{$=$}}
      \put (39.5,18) {\scalebox{0.6}{$[0, \infty)\times[0, \infty)\times[5, \infty)\times[5, \infty)$}}
      \put (41.5,16.5) {\scalebox{0.6}{$\times[0, \infty)\times[8, \infty)\times[8, \infty)$}}
      \put (82.5,22) {$\Rset_3$}
      \put (83.5,19.5) {\rotatebox{90}{$=$}}
      \put (72.5,18) {\scalebox{0.6}{$[ 6, \infty)\times[-2, \infty)\times(-\infty,  5]\times[ 6, \infty)$}}
      \put (74.5,16.5) {\scalebox{0.6}{$\times[-2, \infty)\times(-\infty,  5]\times[ 6, \infty)$}}
   \end{overpic}
   \caption{Boxplots of 100 estimated probabilities $\widehat{\PR}_{\Rset_j}$, $j=1,2,3$, 
   and of the lower- ($\mathrm{CI}_L$) and 
   upper-bounds ($\mathrm{CI}_U$) of their 95\% bootstrap confidence intervals (on $\log_{10}$ 
   scale) for the sets 
   $\Rset_1,\Rset_2,\Rset_3$ using models $\M_4$ and $\M_7$ fitted on 
   samples of size $n=10^4$ from multivariate normal distributions with $d=3$ (top row),
   $d=5$ (middle row), and $d=7$ (bottom row).\ 
   The true $\log_{10}$-probabilities are denoted by red lines.\ }
   \label{fig:logPr_M4M7}
\end{figure}

\begin{figure}[H]
   \centering
   \begin{overpic}[width=\textwidth]{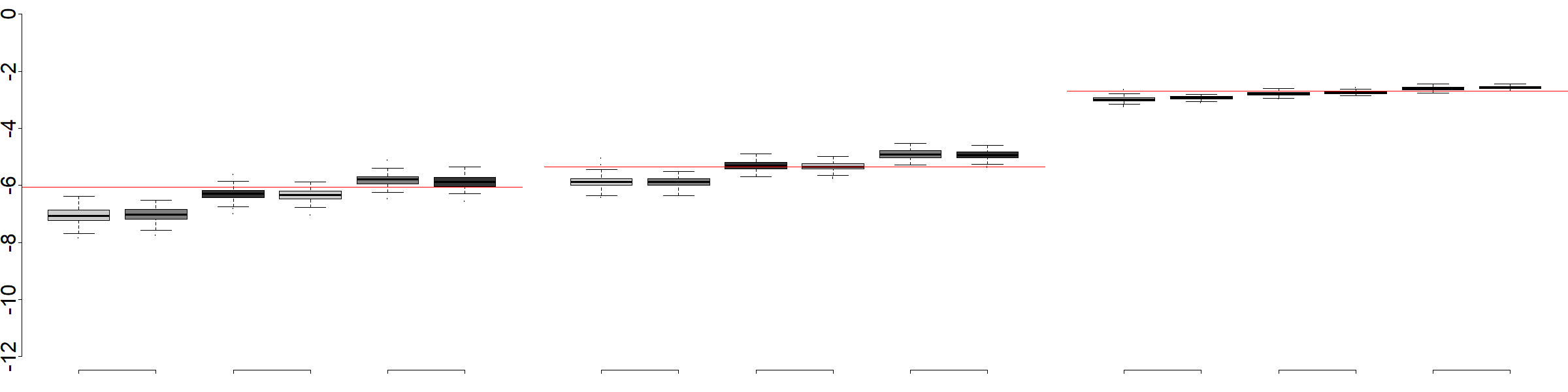}
      \put (6,-1) {\scalebox{0.5}{$\mathrm{CI}_{L}$}}
      \put (39.5,-1) {\scalebox{0.5}{$\mathrm{CI}_{L}$}}
      \put (72.5,-1) {\scalebox{0.5}{$\mathrm{CI}_{L}$}}
      
      \put (14.75,-1) {\scalebox{0.5}{$\log_{10}\widehat{\PR}_{\Rset_1}$}}
      \put (48.25,-1) {\scalebox{0.5}{$\log_{10}\widehat{\PR}_{\Rset_2}$}}
      \put (81.5,-1) {\scalebox{0.5}{$\log_{10}\widehat{\PR}_{\Rset_3}$}}
      
      \put (26,-1) {\scalebox{0.5}{$\mathrm{CI}_{U}$}}
      \put (59.5,-1) {\scalebox{0.5}{$\mathrm{CI}_{U}$}}
      \put (92.5,-1) {\scalebox{0.5}{$\mathrm{CI}_{U}$}}
       
      \put (15.5,23) {$\Rset_1$}
      \put (16.5,20.5) {\rotatebox{90}{$=$}}
      \put (14.5,19) {\tiny $[10,\infty)^3$}
      \put (48.5,23) {$\Rset_2$}
      \put (49.5,20.5) {\rotatebox{90}{$=$}}
      \put (41,19) {\tiny $[-5,5]\times[10,\infty)\times[-10,10]$}
      \put (82.5,15) {$\Rset_3$}
      \put (83.5,12.5) {\rotatebox{90}{$=$}}
      \put (75,11) {\tiny $(-\infty,5]\times[5,\infty)\times[-5,5]$}
   \end{overpic}

   \vspace{0.5em}

   \centering
   \begin{overpic}[width=\textwidth]{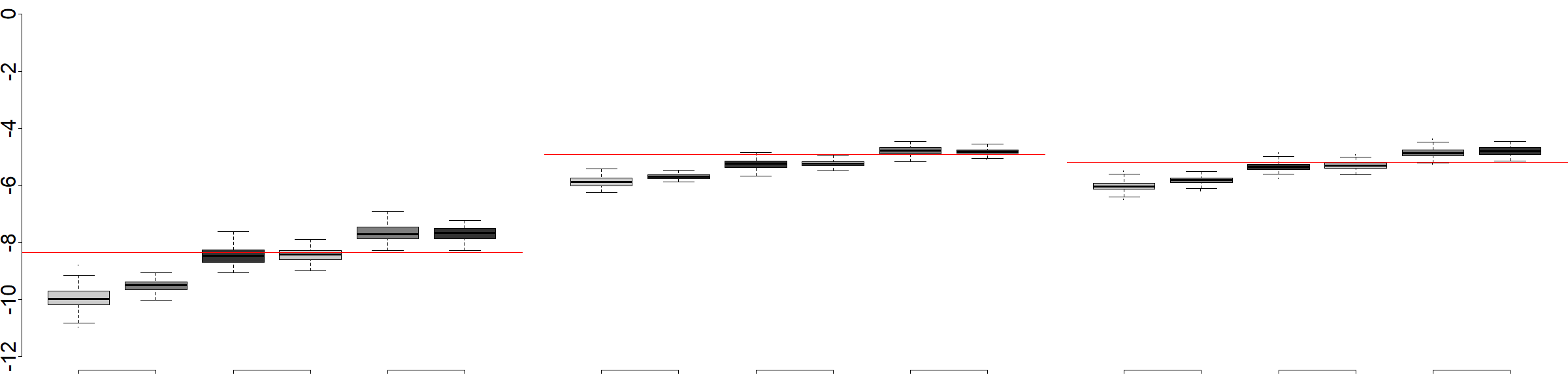}
      \put (6,-1) {\scalebox{0.5}{$\mathrm{CI}_{L}$}}
      \put (39.5,-1) {\scalebox{0.5}{$\mathrm{CI}_{L}$}}
      \put (72.5,-1) {\scalebox{0.5}{$\mathrm{CI}_{L}$}}
      
      \put (14.75,-1) {\scalebox{0.5}{$\log_{10}\widehat{\PR}_{\Rset_1}$}}
      \put (48.25,-1) {\scalebox{0.5}{$\log_{10}\widehat{\PR}_{\Rset_2}$}}
      \put (81.5,-1) {\scalebox{0.5}{$\log_{10}\widehat{\PR}_{\Rset_3}$}}
      
      \put (26,-1) {\scalebox{0.5}{$\mathrm{CI}_{U}$}}
      \put (59.5,-1) {\scalebox{0.5}{$\mathrm{CI}_{U}$}}
      \put (92.5,-1) {\scalebox{0.5}{$\mathrm{CI}_{U}$}}
      
      \put (15.5,22) {$\Rset_1$}
      \put (16.5,19.5) {\rotatebox{90}{$=$}}
      \put (14.5,18) {\tiny $[10,\infty)^5$}
      \put (48.5,22) {$\Rset_2$}
      \put (49.5,19.5) {\rotatebox{90}{$=$}}
      \put (34.5,18) {\scalebox{0.55}{$(-\infty,\infty)\times[6,\infty)\times[8,\infty)\times[6,\infty)\times(-\infty,\infty)$}}
      \put (82.5,22) {$\Rset_3$}
      \put (83.5,19.5) {\rotatebox{90}{$=$}}
      \put (66,18) {\scalebox{0.55}{$(-\infty,-7] \times(-\infty,0] \times (-\infty,-5] \times (-\infty,0] \times (-\infty,-7]$}}
   \end{overpic}

   \vspace{0.5em}

   \centering
   \begin{overpic}[width=\textwidth]{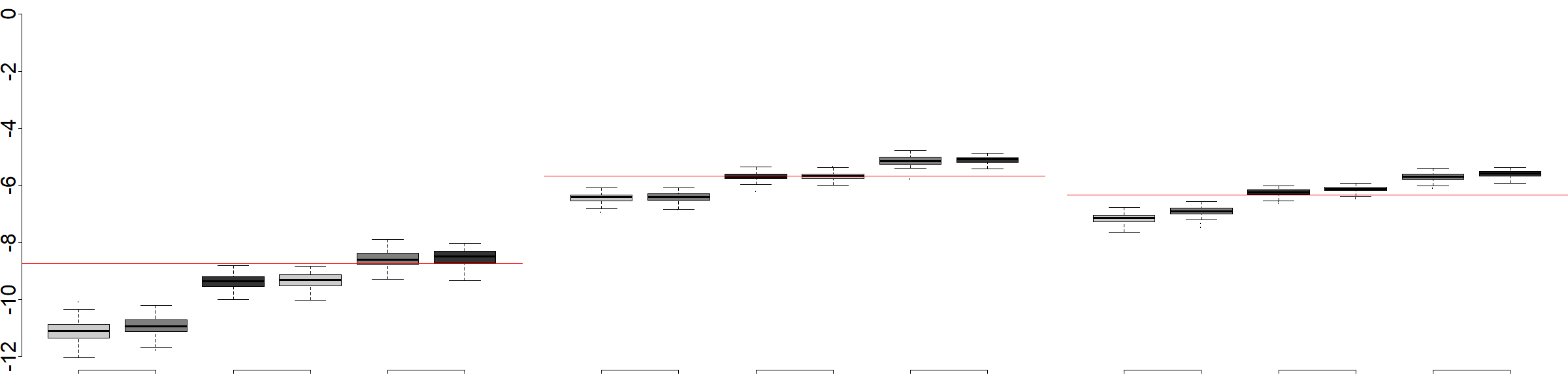}
      \put (6,-1) {\scalebox{0.5}{$\mathrm{CI}_{L}$}}
      \put (39.5,-1) {\scalebox{0.5}{$\mathrm{CI}_{L}$}}
      \put (72.5,-1) {\scalebox{0.5}{$\mathrm{CI}_{L}$}}
      
      \put (14.75,-1) {\scalebox{0.5}{$\log_{10}\widehat{\PR}_{\Rset_1}$}}
      \put (48.25,-1) {\scalebox{0.5}{$\log_{10}\widehat{\PR}_{\Rset_2}$}}
      \put (81.5,-1) {\scalebox{0.5}{$\log_{10}\widehat{\PR}_{\Rset_3}$}}
      
      \put (26,-1) {\scalebox{0.5}{$\mathrm{CI}_{U}$}}
      \put (59.5,-1) {\scalebox{0.5}{$\mathrm{CI}_{U}$}}
      \put (92.5,-1) {\scalebox{0.5}{$\mathrm{CI}_{U}$}}
      
      \put (15.5,22) {$\Rset_1$}
      \put (16.5,19.5) {\rotatebox{90}{$=$}}
      \put (14.5,18) {\tiny $[10,\infty)^7$}
      \put (48.5,22) {$\Rset_2$}
      \put (49.5,19.5) {\rotatebox{90}{$=$}}
      \put (39.5,18) {\scalebox{0.7}{$[0, \infty)\times[0, \infty)\times[5, \infty)\times[5, \infty)$}}
      \put (41.5,16.5) {\scalebox{0.7}{$\times[0, \infty)\times[8, \infty)\times[8, \infty)$}}
      \put (82.5,22) {$\Rset_3$}
      \put (83.5,19.5) {\rotatebox{90}{$=$}}
      \put (72.5,18) {\scalebox{0.7}{$[ 6, \infty)\times[-2, \infty)\times(-\infty,  5]\times[ 6, \infty)$}}
      \put (74.5,16.5) {\scalebox{0.7}{$\times[-2, \infty)\times(-\infty,  5]\times[ 6, \infty)$}}
   \end{overpic}
   \caption{
   Boxplots of 100 estimated probabilities $\widehat{\PR}_{\Rset_j}$, $j=1,2,3$, 
   and of the lower- ($\mathrm{CI}_L$) and 
   upper-bounds ($\mathrm{CI}_U$) of their 95\% bootstrap confidence intervals (on $\log_{10}$ 
   scale) for the sets 
   $\Rset_1,\Rset_2,\Rset_3$ using model $\M_2$ fitted on 
   samples of size $10^4$ (light grey) and $2\times 10^4$ (dark grey) from multivariate normal distributions with $d=3$ (top row),
   $d=5$ (middle row), and $d=7$ (bottom row).\ 
   The true $\log_{10}$-probabilities are denoted by red lines.\ 
   }
   \label{fig:logPr_d7_n20000}
   \end{figure}

\begin{figure}[H]
   \centering
   \begin{overpic}[width=\textwidth]{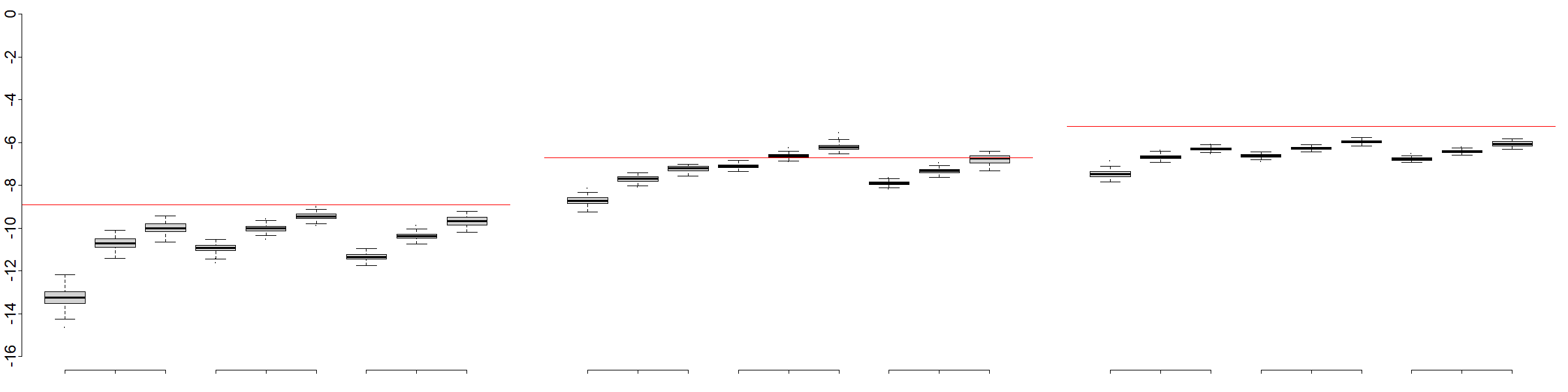}
      \put (12.5,-1) {\scalebox{0.5}{$\mathrm{CI}_{L}$}}
      \put (46,-1) {\scalebox{0.5}{$\mathrm{CI}_{L}$}}
      \put (79,-1) {\scalebox{0.5}{$\mathrm{CI}_{L}$}}
      
      \put (14.75,-1) {\scalebox{0.5}{$\log_{10}\widehat{\PR}_{\Rset_1}$}}
      \put (48.25,-1) {\scalebox{0.5}{$\log_{10}\widehat{\PR}_{\Rset_2}$}}
      \put (81.5,-1) {\scalebox{0.5}{$\log_{10}\widehat{\PR}_{\Rset_3}$}}
      
      \put (20,-1) {\scalebox{0.5}{$\mathrm{CI}_{U}$}}
      \put (53.5,-1) {\scalebox{0.5}{$\mathrm{CI}_{U}$}}
      \put (86.5,-1) {\scalebox{0.5}{$\mathrm{CI}_{U}$}}
       
      \put (6,2.5) {\footnotesize $\M_0$}
      \put (16,2.5) {\footnotesize $\M_1$}
      \put (26,2.5) {\footnotesize $\M_2$}
      \put (39,2.5) {\footnotesize $\M_0$}
      \put (49,2.5) {\footnotesize $\M_1$}
      \put (59,2.5) {\footnotesize $\M_2$}
      \put (72.5,2.5) {\footnotesize $\M_0$}
      \put (82.5,2.5) {\footnotesize $\M_1$}
      \put (92.5,2.5) {\footnotesize $\M_2$}
      
      \put (15.5,23) {$\Rset_1$}
      \put (16.5,20.5) {\rotatebox{90}{$=$}}
      \put (14.5,19) {\tiny $[10,\infty)^{10}$}
      \put (48.5,23) {$\Rset_2$}
      \put (49.5,20.5) {\rotatebox{90}{$=$}}
      \put (34,19) {\scalebox{.5}{$(-\infty,\infty)\times(-\infty,\infty)\times[8,\infty)\times[8,\infty)\times(-\infty,\infty)$}}
      \put (35.25,17.5) {\scalebox{.5}{ $\times[8,\infty)\times(-\infty,\infty)\times[8,\infty)\times[8,\infty)\times[8,\infty)$}}
      \put (82.5,23) {$\Rset_3$}
      \put (83.5,20.5) {\rotatebox{90}{$=$}}
      \put (67,19) {\scalebox{.5}{$(-\infty,-6]\times(-\infty,-6]\times(-\infty,\infty)\times(-\infty,\infty)\times(-\infty,-6]$}}
      \put (67.25,17.5) {\scalebox{.5}{$\times(-\infty,\infty)\times(-\infty,-6]\times[\infty,-6]\times(-\infty,\infty)\times(-\infty,\infty)$}}
   \end{overpic}

   \vspace{0.5em}
   \centering
   \begin{overpic}[width=\textwidth]{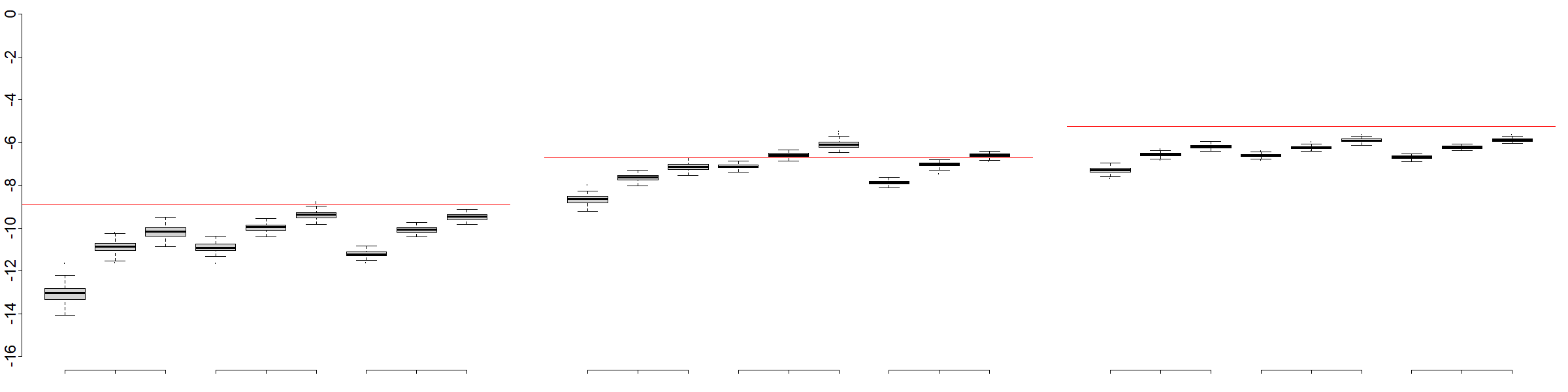}
      \put (12.5,-1) {\scalebox{0.5}{$\mathrm{CI}_{L}$}}
      \put (46,-1) {\scalebox{0.5}{$\mathrm{CI}_{L}$}}
      \put (79,-1) {\scalebox{0.5}{$\mathrm{CI}_{L}$}}
      
      \put (14.75,-1) {\scalebox{0.5}{$\log_{10}\widehat{\PR}_{\Rset_1}$}}
      \put (48.25,-1) {\scalebox{0.5}{$\log_{10}\widehat{\PR}_{\Rset_2}$}}
      \put (81.5,-1) {\scalebox{0.5}{$\log_{10}\widehat{\PR}_{\Rset_3}$}}
      
      \put (20,-1) {\scalebox{0.5}{$\mathrm{CI}_{U}$}}
      \put (53.5,-1) {\scalebox{0.5}{$\mathrm{CI}_{U}$}}
      \put (86.5,-1) {\scalebox{0.5}{$\mathrm{CI}_{U}$}}
       
      \put (6,2.5) {\footnotesize $\M_0$}
      \put (16,2.5) {\footnotesize $\M_1$}
      \put (26,2.5) {\footnotesize $\M_2$}
      \put (39,2.5) {\footnotesize $\M_0$}
      \put (49,2.5) {\footnotesize $\M_1$}
      \put (59,2.5) {\footnotesize $\M_2$}
      \put (72.5,2.5) {\footnotesize $\M_0$}
      \put (82.5,2.5) {\footnotesize $\M_1$}
      \put (92.5,2.5) {\footnotesize $\M_2$}
      
   \end{overpic}

   \vspace{0.5em}
   \centering
   \begin{overpic}[width=\textwidth]{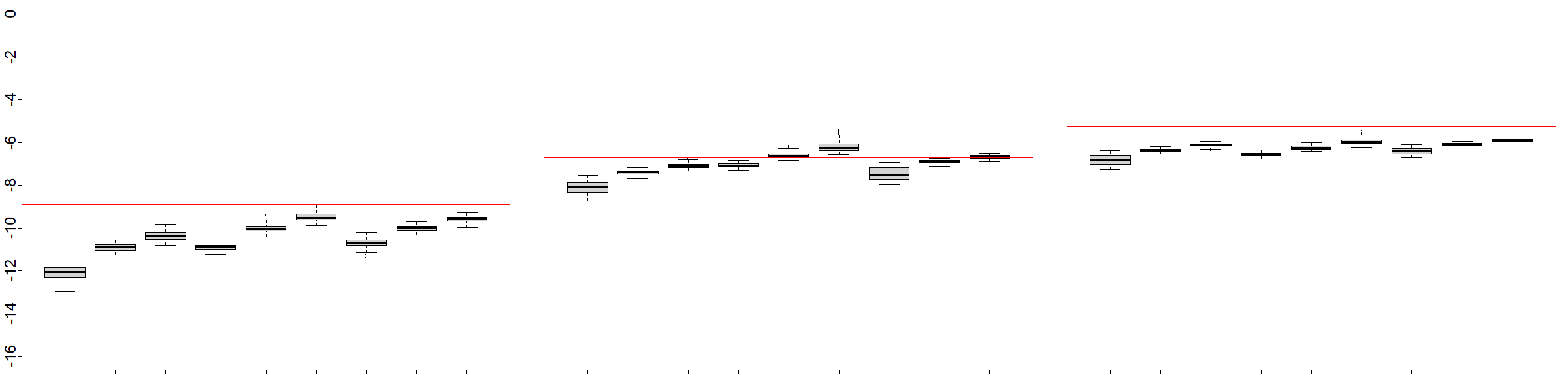}
      \put (12.5,-1) {\scalebox{0.5}{$\mathrm{CI}_{L}$}}
      \put (46,-1) {\scalebox{0.5}{$\mathrm{CI}_{L}$}}
      \put (79,-1) {\scalebox{0.5}{$\mathrm{CI}_{L}$}}
      
      \put (14.75,-1) {\scalebox{0.5}{$\log_{10}\widehat{\PR}_{\Rset_1}$}}
      \put (48.25,-1) {\scalebox{0.5}{$\log_{10}\widehat{\PR}_{\Rset_2}$}}
      \put (81.5,-1) {\scalebox{0.5}{$\log_{10}\widehat{\PR}_{\Rset_3}$}}
      
      \put (20,-1) {\scalebox{0.5}{$\mathrm{CI}_{U}$}}
      \put (53.5,-1) {\scalebox{0.5}{$\mathrm{CI}_{U}$}}
      \put (86.5,-1) {\scalebox{0.5}{$\mathrm{CI}_{U}$}}
       
      \put (6,2.5) {\footnotesize $\M_0$}
      \put (16,2.5) {\footnotesize $\M_1$}
      \put (26,2.5) {\footnotesize $\M_2$}
      \put (39,2.5) {\footnotesize $\M_0$}
      \put (49,2.5) {\footnotesize $\M_1$}
      \put (59,2.5) {\footnotesize $\M_2$}
      \put (72.5,2.5) {\footnotesize $\M_0$}
      \put (82.5,2.5) {\footnotesize $\M_1$}
      \put (92.5,2.5) {\footnotesize $\M_2$}
   \end{overpic}
   \caption{Boxplots of 100 estimated probabilities $\widehat{\PR}_{\Rset_j}$, $j=1,2,3$, 
   and of the lower- ($\mathrm{CI}_L$) and 
   upper-bounds ($\mathrm{CI}_U$) of their 95\% bootstrap confidence intervals (on $\log_{10}$ 
   scale) for the sets 
   $\Rset_1,\Rset_2,\Rset_3$ using models $\M_0$, $\M_1$, $\M_2$ fitted on 
   samples of size $5\times 10^4$ (top row), $10^5$ (middle row) and $2\times 10^5$ (bottom row) 
   from multivariate normal distributions with $d=10$.\ 
   The true $\log_{10}$-probabilities are denoted by red lines.\ }
   \label{fig:logPr_d10_n200000}
\end{figure}
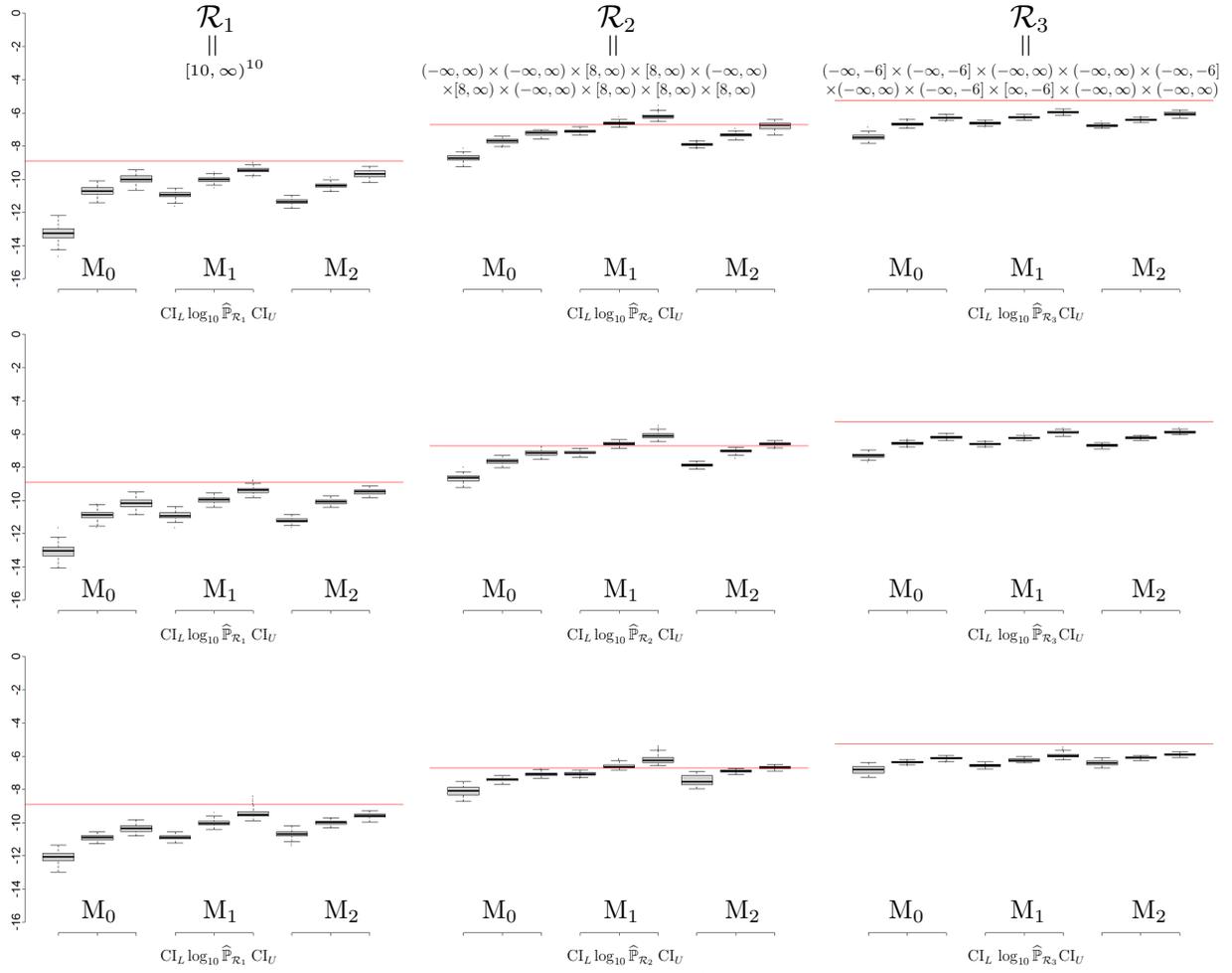

\newpage
\section{Data application to low and high wind extremes}
\label{sec:Supp_case_study}

\begin{table}[H]
   \begin{center}
   \caption{Acronyms and names of the ten numbered stations with no missing data. } 
   \begin{tabular}{cccccc}
      \toprule \toprule
      \scalebox{.8}{\textbf{Number}} & \scalebox{.8}{\textbf{Acronym}} & \scalebox{.8}{\textbf{Name}} & 
      \scalebox{.8}{\textbf{Number}} & \scalebox{.8}{\textbf{Acronym}} & \scalebox{.8}{\textbf{Name}} \\
      \midrule
      1 & MEG & Megler & 6 & HOO & Hood River \\
      2 & NAS & Naselle Ridge & 7 & SHA & Shaniko \\
      3 & TIL & Tillamook & 8 & ROO & Roosevelt \\
      4 & FOR & Forest Grove & 9 & SUN & Sunnyside \\
      5 & BID & Biddle Butte & 10 & HOR & Horse Heaven \\
      \bottomrule
   \end{tabular}
   \label{tbl:stations}
   \end{center}
\end{table}

\begin{figure}[H]
\centering
\begin{overpic}[width=\textwidth]{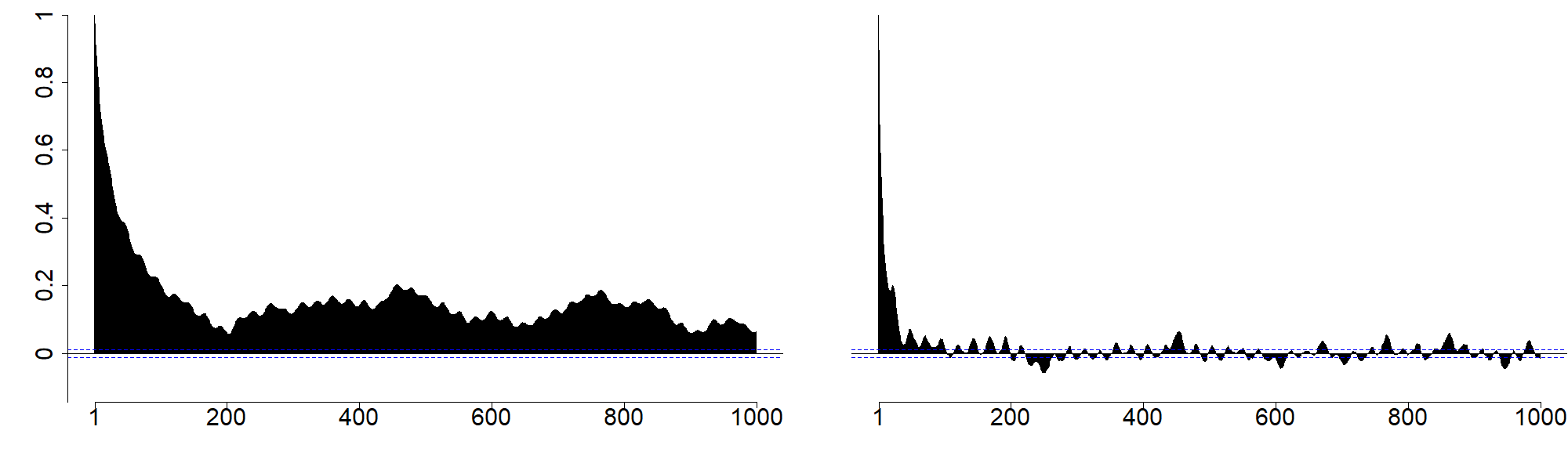}
   \put(22,0){Lag (hours)}
   \put(72,0){Lag (hours)}
   \put(0,16){{\small \rotatebox{90}{ACF}}}
\end{overpic}
\caption{Autocorrelation function (ACF) plots of the original 
   data $\underline{\bm x}_1$ (left) and of the homogenised  
   data $\underline{\bm x}_1^{\mathrm{H}}$ (right) for site 1, MEG.\ }
\label{fig:ACF}
\end{figure}

\begin{figure}[H]
\centering
\begin{overpic}[width=\textwidth]{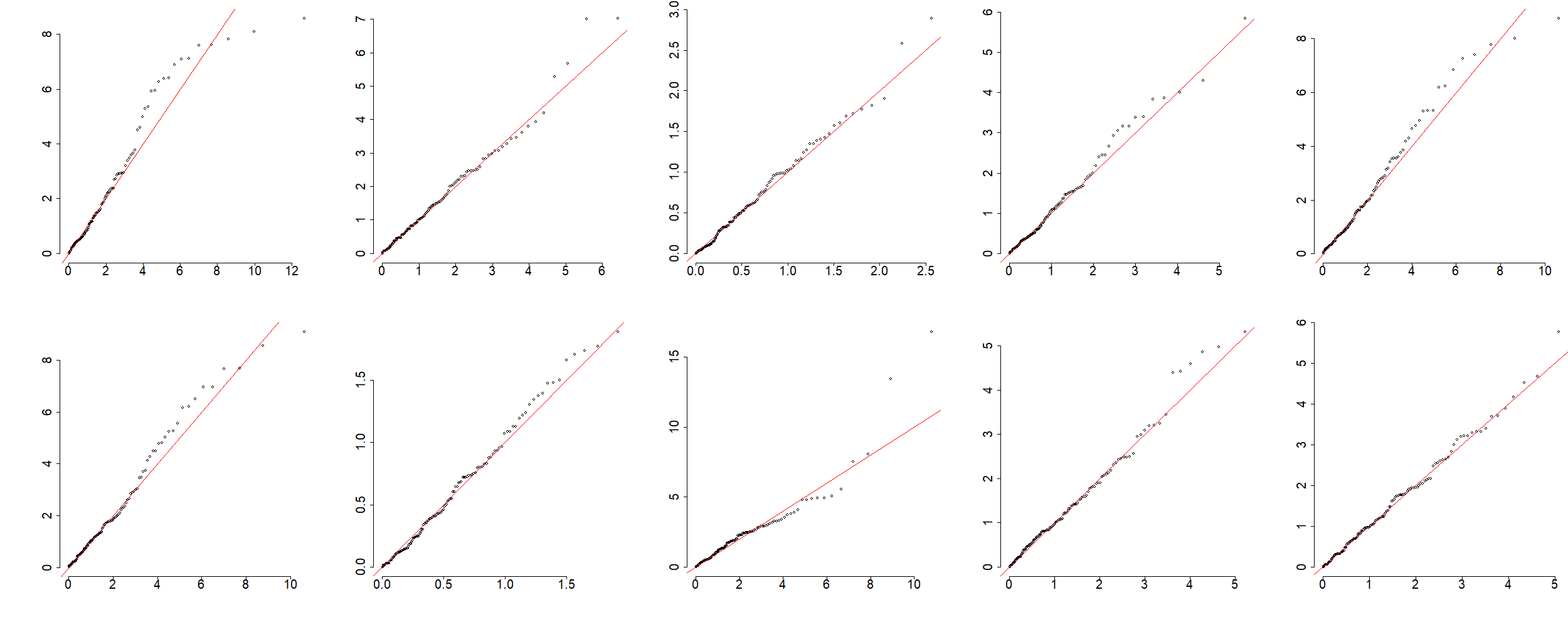}
   \put(0,12){\rotatebox{90}{{\small Empirical quantiles}}}
   \put(41,0){Model quantiles}

   \put(16.5,25){{\small $\underline{\bm x}_1^{\mathrm{H}}$}}
   \put(36.5,25){{\small $\underline{\bm x}_2^{\mathrm{H}}$}}
   \put(56.5,25){{\small $\underline{\bm x}_3^{\mathrm{H}}$}}
   \put(76.5,25){{\small $\underline{\bm x}_4^{\mathrm{H}}$}}
   \put(96.5,25){{\small $\underline{\bm x}_5^{\mathrm{H}}$}}

   \put(16.5,5){{\small $\underline{\bm x}_6^{\mathrm{H}}$}}
   \put(36.5,5){{\small $\underline{\bm x}_7^{\mathrm{H}}$}}
   \put(56.5,5){{\small $\underline{\bm x}_8^{\mathrm{H}}$}}
   \put(76.5,5){{\small $\underline{\bm x}_9^{\mathrm{H}}$}}
   \put(96.5,5){{\small $\underline{\bm x}_{10}^{\mathrm{H}}$}}
\end{overpic}
\caption{Quantile-Quantile plots for the marginal generalised Pareto
models fited above the 0.995-quantile of the
homogenised data $\underline{\bm x}_1^{\mathrm{H}},\ldots,
\underline{\bm x}_{10}^{\mathrm{H}}$.\ }
\label{fig:MarginalGP}
\end{figure}

\begin{figure}[H]
   \centering
   \begin{overpic}[width=0.35\textwidth,trim= 60 0 60 20]{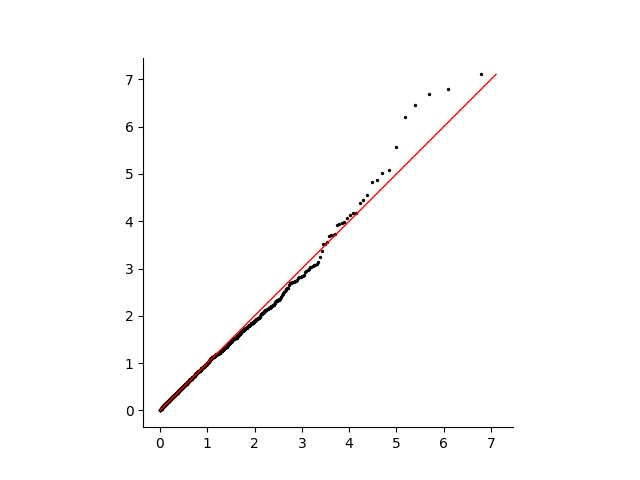}
      \put(0,45){\rotatebox{90}{\small Model}}
   \end{overpic}
   \begin{overpic}[width=0.35\textwidth,trim= 60 0 60 20]{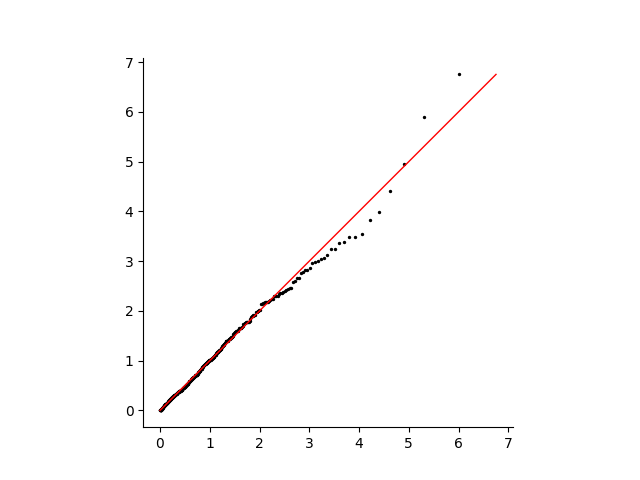}
   \end{overpic}
   \begin{overpic}[width=0.35\textwidth,trim= 60 0 60 20]{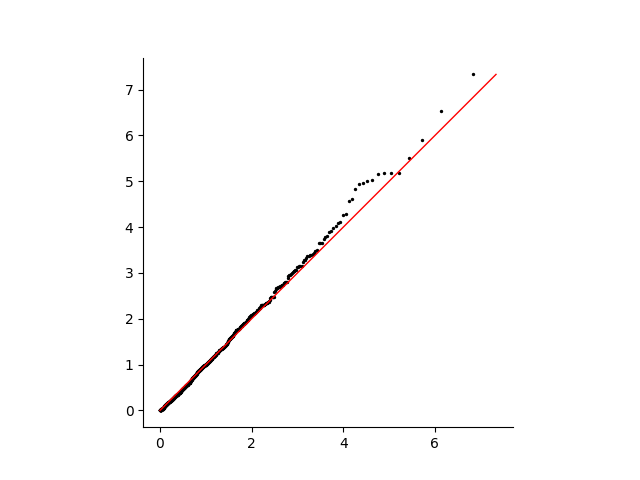}
      \put(20,0){\small Standard exponential}
      \put(0,45){\rotatebox{90}{\small Model}}
   \end{overpic}
   \begin{overpic}[width=0.35\textwidth,trim= 60 0 60 20]{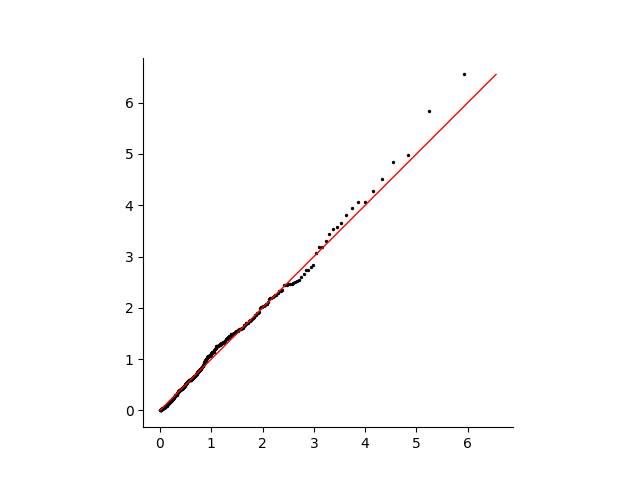}
      \put(20,0){\small Standard exponential}
   \end{overpic}
   \caption{Quantile-Quantile plots of the standard exponential quantiles against the observed radial exceedances 
   $\{[\lVert \bm x_i \rVert -r_{\Qq}(\bm x_i/\lVert \bm x_i \rVert)]/
   r_{\G}(\bm x_i/\lVert \bm x_i \rVert) \,: \,
   \lVert \bm x_i \rVert>r_{\Qq}(\bm x_i /\lVert \bm x_i \rVert), 
   i=1,\ldots,n\}$ for $m=1$ and $h=18$ at the configuration of five stations minimising 
   the probability of no production (top row) and 
   maximising the probability of full power (bottom row)
   on their training (left column) and validation (right column) data.\ }
   \label{fig:QQ_models}
\end{figure}

\begin{figure}[H]
   \centering
   \begin{overpic}[width=0.6\textwidth]{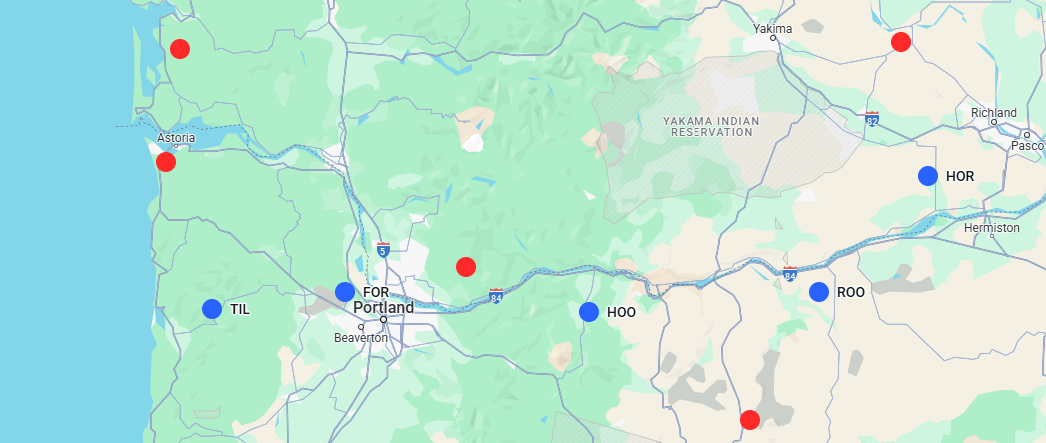}
   \end{overpic}
   \caption{Configurations of five stations (blue) maximising the 
   risk of no production for month $m=1$ and hour $h=18$.\ }
   \label{fig:config_worse}
\end{figure}

\end{document}